\documentclass[onecolumn,seceqn]{elsart3p}

\usepackage{graphicx,amssymb}
\usepackage{bm}

\newcommand \be{\begin{equation}}
\newcommand \en{\end{equation}}
\newcommand \bea{\begin{eqnarray}}
\newcommand \ena{\end{eqnarray}}
\newcommand \bib{\bibitem}

\begin{document}

\begin{frontmatter}

\title{A study on relativistic lagrangian field theories with non-topological soliton solutions}

\author[Luth,Oviedo]{J. Diaz-Alonso\corauthref{cor}},
\corauth[cor]{Corresponding author.} \ead{joaquin.diaz@obspm.fr}
\author[Oviedo]{D. Rubiera-Garcia}

\emph{\address[Luth]{LUTH, Observatoire de Paris, CNRS,
Universit\'e Paris Diderot. 5 Place Jules Janssen, 92190 Meudon, France}
\address[Oviedo]{Departamento de F\'isica, Universidad de Oviedo. Avda. Calvo Sotelo 18, E-33007 Oviedo, Asturias, Spain}}

\date{\today}

\begin{abstract}

We perform a general analysis of the dynamic structure of two
classes of relativistic lagrangian field theories exhibiting
static spherically symmetric non-topological soliton solutions.
The analysis is concerned with (multi-) scalar fields and
generalized gauge fields of compact semi-simple Lie groups. The
lagrangian densities governing the dynamics of the (multi-) scalar
fields are assumed to be general functions of the kinetic terms,
whereas the gauge-invariant lagrangians are general functions of
the field invariants. These functions are constrained by
requirements of regularity, positivity of the energy and vanishing
of the vacuum energy, defining what we call ``admissible" models.
In the scalar case we establish the general conditions which
determine exhaustively the families of admissible lagrangian
models supporting this kind of finite-energy solutions. We analyze
some explicit examples of these different families, which are
defined by the asymptotic and central behaviour of the fields of
the corresponding particle-like solutions. From the variational
analysis of the energy functional, we show that the admissibility
constraints and the finiteness of the energy of the scalar
solitons are necessary and sufficient conditions for their linear
static stability against small charge-preserving perturbations.
Furthermore we perform a general spectral analysis of the dynamic
evolution of the small perturbations around the statically stable
solitons, establishing their dynamic stability. Next, we consider
the case of many-components scalar fields, showing that the
resolution of the particle-like field problem in this case reduces
to that of the one-component case. The study of these scalar
models is a necessary step in the analysis of the gauge fields. In
this latter case, we add the requirement of parity invariance to
the admissibility constraints. We determine the general conditions
defining the families of admissible gauge-invariant models
exhibiting finite-energy electrostatic spherically symmetric
solutions which, unlike the (multi-) scalar case, are not always
stable. The variational analysis of the energy functional leads
now to supplementary restrictions to be imposed on the lagrangian
densities in order to ensure the linear stability of the solitons.
We establish a correspondence between any admissible
soliton-supporting (multi-) scalar model and a family of
admissible generalized gauge models supporting finite-energy
electrostatic point-like solutions. Conversely, for each
admissible soliton-supporting gauge-invariant model there is an
associated \emph{unique} admissible (multi-) scalar model with
soliton solutions. This shows the exhaustive character of the
admissibility and stability conditions in determining the class of
soliton-supporting generalized gauge models. The usual Born-Infeld
electrodynamic theory and its non-abelian extensions are shown to
be (very particular) examples of one of these families.

\end{abstract}


\end{frontmatter}

\large

\section{\large Introduction}

In the context of field theory the interest in extended
configurations describing fields associated to point-like
particles (particle-like solutions) dates back to the works of Mie
\cite{mie12} and Born and Infeld (B-I) \cite{born34},\cite{B-I34},
aimed to solve the problem of the divergent self-energy of the
electron field in Classical Electrodynamics. Today, this kind of
problems fall inside the large domain of Soliton Physics, whose
methods and applications concern most branches of physical
sciences.

Although there is not an universally accepted definition of the
concept of soliton, we shall adopt here minimal defining
properties which seem to be widely assumed in most contexts:
Solitons are \textit{stable, finite-energy solutions of
conservative non-linear differential equations}. However, the
accepted meaning and content of the term \textit{stable} is not
universal. In a \textit{strong sense}, it refers to the existence
of soliton entities which can be identified (if present) in field
configurations and are preserved by the dynamic evolution of the
system. With this definition, the analysis of such configurations
in terms of many solitons, interacting via radiative field
exchanges, becomes possible. This kind of stability arises in some
field theoretical models (most of them in one-space dimension)
exhibiting topological conservation laws, related to a non-trivial
structure of the vacuum \cite{scott73}. In these cases the
conserved topological charges identify the presence of the
\textit{topological solitons} and ensure their preservation.
Examples of topological solitons in three-space dimensions are the
monopole of 't Hooft and Polyakov \cite{t-p74} or the Skyrmion
\cite{skyrme}. In a \textit{weak sense}, stability is identified
with linear stability, i.e. with the preservation of the soliton
identity against a certain class of small perturbations for which
the soliton configuration is a minimum of the functional of
energy. This restricted class of perturbations is defined through
boundary conditions which amount, in general, to the preservation
of (\emph{non-topological}) charges associated with the soliton.
The conservation laws of these charges may be implicitly contained
within the structure of the field equations (as in
electrodynamics) or be consequences of constraints imposed on the
external sources, to which the field is coupled. Here we shall be only
concerned with this kind of non-topological, finite-energy,
weakly-stable soliton solutions of local relativistic lagrangian
field theories in three-space dimensions.

Non-linear field theories supporting soliton solutions have been
widely studied and applied in several contexts of Theoretical
Physics. Let us give some examples:

1) In the context of elementary particle physics, let us mention
the phenomenological description of the hadron structure and the
hadronic interactions in terms of topological solitons in the
framework of the Skyrme model \cite{skyrme}. Other descriptions of
this structure in terms of non-topological solitons have been
performed through the effective approach to the low-energy regime
of Quantum Chromodynamics (QCD) given by the Friedberg-Lee model
\cite{fried77},\cite{l-p92} and related theories. Let us also
mention the generalized chiral-invariant model of Deser, Duff and
Isham \cite{ddi}, whose lagrangian density is a rational power
(3/2) of the non-linear sigma model lagrangian, chosen in order to
circumvent Derrick's theorem \cite{derrick64}. This model and its
extensions support topological solitons \cite{nicol}. In the same
context let us also mention the toroidal solitons of Ref.
\cite{AFZ}, which might describe glueball collective states in the
low-energy limit of QCD, as suggested in Ref. \cite{fad}.

2) In nuclear physics, the analysis of high-density hadronic
matter and its chiral phase structure has been performed in terms
of systems of skyrmions in the framework of effective field
theories implemented with the large $N_c$ behaviour of QCD
\cite{vento}.

3) The glueball collective states, as soliton solutions of
non-abelian B-I gauge field models \cite{g-k00}. The introduction
of these generalized gauge models supporting soliton solutions and
their extensions to higher dimensions is suggested by string
theory, since some of them arise in the low-energy limit of
D-Branes \cite{soliton}. Moreover, it has been shown that the B-I
extension of the basic lagrangian of the Skyrme model leads to
stable soliton solutions. This extension was proposed as an
alternative to the inclusion of the quartic ``Skyrme term" in the
original lagrangian of the non-linear sigma model \cite{pavlo02}.
This ``ad hoc" term stabilizes the soliton by preventing Derrick's
scaling, but its introduction is not justified from more
fundamental reasons.

4) In the last two decades there has been an increasingly amount
of works on self-gravitating field configurations. The
aforementioned presence of B-I actions in the low-energy physics of
D-branes, whose fundamental excitation is gravity, is one of the
motivations for this renewed interest (see e.g. \cite{gib00}). But
the search for self-gravitating field configurations, as solutions
of the Einstein equations for gravity coupled to different kinds
of fields, is an older topic \cite{ortin}. Indeed, in
a four-dimensional flat space-time, Derrick's theorem \cite{derrick64}
and several non-existence theorems \cite{deser76} limit
drastically the class of lagrangian field theories supporting
soliton solutions. However, the coupling to gravity can remove
these obstructions and thus allow for particle-like solutions in
some cases. One example of this is the pure Yang-Mills theory,
which does not support glueball solutions in Minkowski space
\cite{deser76} but exhibits particle-like solutions in curved
space-time \cite{bart88}. Another example is the regular
black-hole solution of Einstein's equations coupled to a
non-linear electrodynamics \cite{ayon98}. Let us also mention that
theories supporting soliton solutions in flat space, as the
abelian and non-abelian Born-Infeld models, have been extended to
curved space leading to black-hole-like soliton solutions
\cite{v-d-g}. Finally, we cite the soliton stars introduced by
T.D. Lee as self-gravitating coherent quantum states with the
features of non-topological solitons \cite{l-p92},\cite{lee87}.

5) In the fast-evolving context of modern Cosmology let us mention
some problems for which non-linear field theories supporting
soliton solutions have been invoked. It has been suggested that
time-dependent but non-dispersive scalar solitons (Q-balls
\cite{coleman85}) may account for the behaviour of
self-interacting dark matter \cite{kusen01}. There is also the
suggestion that non-topological solitons might have been formed in
a second-order phase transition in the early Universe, and
contribute significantly to the present mass density
\cite{frieman88}. As another example, the Born-Infeld
generalization of SU(2) non-abelian gauge field theory, coupled to
tensor-scalar gravitation, has been used for the description of
dark energy \cite{darkenerg}. Also scalar field models with
lagrangian densities which are generalized functions of the
kinetic term have been used to drive inflationary evolution in the
early Universe (k-inflation) \cite{armen-thib99}. Solitonic
configurations of these k-essence fields have been used to
reproduce some properties of dark matter as well
\cite{armendariz01}.

All these considerations underline the interest of non-linear
lagrangian field theories and their eventual soliton solutions,
mainly for the cases of (one and many-components) scalar field
models and generalized gauge-invariant field theories. By
generalized gauge field theories we mean models for gauge fields
of compact semi-simple Lie groups, with lagrangian densities
defined as general functions $\varphi(X,Y)$ of the two
\textit{standard} first-order gauge invariants, namely $X =
tr(F_{\mu\nu}\cdot F^{\mu\nu})$ and $Y = tr(F_{\mu\nu}\cdot
F^{*\mu\nu})$. Aside from the already mentioned Born-Infeld-like
gauge models, defined by the very particular B-I choice of this
function, there are not in the literature systematic studies on
solitons for general gauge-invariant field theories. It is one of
the main purposes of this work to perform a study on this class of
theories in three space dimensions, by setting the conditions on
the lagrangian functions determining the families of these models
which are \textit{physically admissible} and support elementary
solutions which are non-topological solitons \footnote{In what
follows we shall call ``elementary solutions" the static
spherically symmetric solutions of scalar field equations or the
electrostatic spherically symmetric solutions of generalized gauge
field equations and we shall use the acronyms \textbf{SSS} and
\textbf{ESS}, respectively.}. The requirements for admissibility
adopted here refer to the positive-definite character of the
energy, the vanishing of the vacuum energy and the regularity,
uniqueness and definiteness in all space of the elementary
solutions. These conditions are targeted to deal with physically
meaningful theories.

In principle, generalized gauge-invariant lagrangians are
candidates to describe the dynamics of the gauge fields in gauge
theories of fundamental interactions. If one accepts the
fundamental character of string theory and the aforementioned
results, referenced in \cite{soliton}, the description of the
gauge-field dynamics through some generalized lagrangians,
regarded as effective field models of string theory, could be more
``fundamental" than the usual Maxwell-like choice $\varphi(X,Y)
\thicksim X$, a ``minimal prescription" which should be understood
as a low-energy (or weak-field) approximation limit. Nevertheless,
from the field-theory point of view, this minimal prescription is
generally assumed to describe the fundamental dynamics of the
gauge fields in the gauge-invariant lagrangians with coupling to
other (generally fermionic) sectors. In this case, when the
high-energy degrees of freedom are integrated out in the path
integral of the original action, generalized gauge-invariant
models emerge as \textit{effective lagrangians}, containing new
non-linear self-couplings of the gauge fields which account, at a
classical phenomenological level, for quantum effects and
interactions with the removed heavy-mode sector \cite{Dobado97}.
Historically, the first example of this kind of effective
lagrangians was obtained by Heisenberg and Euler
\cite{Heisenberg36} in the context of Quantum Electrodynamics
(QED). It accounts for the non-linear effects of the Dirac vacuum
on low-energy electromagnetic wave propagation, calculated to
lowest order in the fine structure constant. When higher order
operators are included we are lead to a sequence of effective
lagrangians which take the form of polynomials in the field
invariants, arranged as an expansion in operators of increasing
dimensions \cite{pirula70}. An interesting question arises here,
related to the possibility that these effective lagrangians could
exhibit soliton solutions, even though the bare lagrangian does
not. In the following sections we shall give explicit examples for
which the soliton elementary solutions of an effective model may
be interpreted as finite-energy fields of point-charges screened
by the vacuum effects, whereas the elementary field of the bare
theory is energy-divergent.

For scalar fields, Derrick's theorem \cite{derrick64} imposes
severe restrictions on the lagrangian models supporting
time-independent soliton solutions in three space dimensions. One
of the ways to circumvent the hypothesis of Derrick's theorem is
based on the choice of the lagrangian density as a function of the
kinetic term alone \cite{diaz83} (see Eq.(\ref{eq:(4-1)}) below).
This choice seems rather arbitrary from a physical point of view.
Nevertheless, it is the natural restriction for scalar fields of
the lagrangian densities of the generalized gauge-invariant
models. Moreover, as we shall see in the sequel, many of the
necessary results in the analysis of the soliton problem in
generalized (abelian and non-abelian) gauge-invariant theories
(characterization of the families of soliton-supporting
lagrangians, explicit determination of such models, analysis of
stability etc.) can be obtained from similar results more easily
established for scalar models. Consequently, the detailed study of
these scalar models will take an important place in this work.

The paper is organized as follows:

In section 2 we consider the scalar field models in detail. Many
general results concerning the properties of these families of
scalar field theories and their associated soliton solutions can
be obtained without the specification of the explicit form of the
lagrangian functions. After an initial discussion on the
admissibility conditions to be imposed on these functions, we
solve the field equations for SSS solutions of generic models.
This is achieved by obtaining the generic expression of a
first-integral, which allows the determination of the field
strength once the form of the lagrangian function is specified. We
analyze the expression of the integral of energy for these
solutions and determine the conditions that must be satisfied (at
$r=0$ and as $r \rightarrow \infty$) in order for them to be of
finite-energy. These conditions imply supplementary restrictions
to be imposed on the lagrangian functions which allow for an
exhaustive characterization of the admissible models supporting
finite-energy SSS solutions. We shall leave for section 6 the
study of the stability of the elementary solutions for these and
other models.

In section 3 we introduce explicit examples of the different
families of admissible scalar models supporting SSS soliton
solutions. The \textit{first example} is a large family of
polynomial lagrangian functions including the scalar versions of
the Euler-Heisenberg effective lagrangian of Electrodynamics and
the sequence of higher-order effective corrections. From these
examples and the results of section 5 on generalized
electromagnetic field theories it can be shown that the sequence
of effective lagrangians for Electrodynamics describing low-energy
photon-photon interaction, obtained in the perturbative expansion
\cite{pirula70}, exhibit finite-energy point-like solutions. The
\textit{second example} is a two-parameter family generalizing the
scalar version of the Born-Infeld model. Both the first and second
examples exhibit soliton solutions which are asymptotically
coulombian and differ by their behaviour at the center. The
\textit{third example} is a three-parameter family of models
behaving like the B-I one at the center, but exhibiting diverse
asymptotic behaviours. The \textit{fourth example} is a
three-parameter family that supports exponentially-damped soliton
solutions. When properly extended to the generalized
gauge-invariant case we obtain examples of gauge theories where
non-linear self-couplings lead to short-ranged interactions
without any symmetry breaking mechanism.

In section 4 we extend the results of the analysis of
one-component scalar fields to the case of $N$-components scalar
fields. For a given form of the lagrangian density as a function
of the rotationally-invariant kinetic term
($X=\sum_{i=1}^{N}\partial_{\mu}\phi_{i}\partial^{\mu}\phi_{i}$)
the $N$ components of the SSS solutions have the same form, as
functions of $r$, as the SSS solution of the one-component model
associated to the same form of the lagrangian density function.
The scalar charge of the associated one-component solution
corresponds to the mean-square of the $N$ external scalar charges
associated to the $N$-component solutions and the energy (when
finite) is the same in both cases. There is thus a degeneracy of
the energy on spheres of the charge space in the $N$-component
case. These results will be useful in the subsequent analysis of
the generalized non-abelian gauge-invariant models.

In section 5 we consider the abelian and non-abelian generalized
gauge-invariant theories. After defining the dynamic problem and
the admissibility conditions, we analyze the equations for ESS
fields. We prove that the solutions of these equations, in the
abelian and non-abelian cases, can be built from SSS solutions of
the one and many-components scalar field problems, respectively.
As a consequence there is a correspondence between scalar models
and families of generalized gauge models in such a way that the
SSS solutions and the corresponding ESS solutions have the same
functional form. Moreover, if the energy of a SSS solution is
finite, so is the energy of the ESS solutions of the corresponding
generalized gauge family. The results of section 2 characterizing
the admissible scalar models with finite-energy SSS soliton
solutions, characterize also the admissible generalized gauge
models with finite-energy ESS solutions, but the latter ones are
not always stable. Stability requires now supplementary conditions
to be satisfied by the lagrangian densities, which will be
determined in section 6.

Section 6 is devoted to a detailed analysis on the linear
stability of the elementary solutions of scalar and gauge models.
This analysis leads to necessary and sufficient conditions for the
stability of the soliton solutions, going beyond the necessary
conditions demanded by Derrick's theorem. For (multi-) scalar
models the variational study of the functional of energy and the
spectral analysis of the small perturbations around the SSS
solutions prove the static and dynamic stability of all
\textit{finite-energy} SSS solutions of \textit{admissible}
lagrangians. For admissible abelian and non-abelian generalized
gauge models, a similar analysis allows to determine supplementary
(necessary and sufficient) conditions to be satisfied by the
lagrangians in order for their finite-energy ESS solutions to
reach stability.

We conclude in section 7 by drawing some perspectives and future
developments.

Our analysis deals with fields in three-space dimensions, but most
of our results can be straightforwardly generalized to other
spatial dimensions.

Short reports of some of the main results developed here have been
already published in Refs. \cite{dr07-1} and \cite{dr07-2}.

\section{\large Scalar solitons}

We begin with lagrangian densities for scalar field
\textit{potentials} $\phi(x)$, defined in a four dimensional
Minkowski space-time as

\be L=f(\partial_{\mu}\phi\cdot\partial^{\mu}\phi),
\label{eq:(2-1)} \en
where $f(X)$ is a given {\em continuous and
derivable} function in the domain of definition ($\Omega$) which
is assumed to be open, connected and including the vacuum ($X
\equiv \partial_{\mu}\phi\cdot\partial^{\mu}\phi = 0$). For future
purposes, we also require $f(X)$ to be monotonically increasing
(more precisely, $df/dX > 0, \forall X \neq 0 \in \Omega$ and
$df/dX \geq 0$ for $X = 0$) and $df/dX$ to be continuous for $X <
0 (X \in \Omega)$. In absence of a coupling to external sources,
the associated field equations take the form of a local
conservation law

\be
\partial_{\mu}J^{\mu} = 0,
\label{eq:(2-2)}
\en
where the conserved current $J^{\mu}$ is

\be J^{\mu} = \stackrel{\bullet}{f}(X)\partial^{\mu} \phi,
\label{eq:(2-3)} \en with $\stackrel{\bullet}{f}(X) = df/dX$. In
these models, the D'Alembert linear wave equation corresponds to
\footnote{As in this linear case, plane waves of the form
$\phi=\phi(k_{\mu} \cdot x^{\mu})$, with $k^2=k_{\mu}k^{\mu}=0$,
are solutions of Eq.(\ref{eq:(2-2)}) but superpositions of such
waves are not, in general.} $f(X)=X/2$. For the SSS solutions
$\phi(r)$, Eq.(\ref{eq:(2-2)}) has the first-integral

\be r^{2}\phi^{'}\stackrel{\bullet}{f}(-\phi^{'2}) = \Lambda,
\label{eq:(2-4)} \en where $\phi^{'} = d\phi/dr$, and $\Lambda$ is
the integration constant. This is an algebraic equation which
allows, in principle, the determination of the field strength as a
function of $r$ and $\Lambda$. The positivity of
$\stackrel{\bullet}{f}(X)$ requires both $\Lambda$ and $\phi'(r)$
to be simultaneously either positive or negative. We can then
consider only the positive-sign case without loss of generality.
Strictly speaking Eq.(\ref{eq:(2-4)}) determines the field
$\phi^{'}(r)$ only for $r>0$. If we replace the solutions of
Eq.(\ref{eq:(2-4)}) in Eq.(\ref{eq:(2-2)}) we do not obtain zero,
but a Dirac $\delta$ distribution of weight $4\pi\Lambda$. We can
then identify this parameter with the central scalar charge source
of the (at rest) SSS solution, in analogy with the point-like
charges in the Maxwell theory. Alternatively, in some cases as,
for instance, the non-linear electromagnetism of Born-Infeld, this
charge may be interpreted as a continuous charge density
distribution in space. For non-linear electromagnetic models the
continuous interpretation of the charge is rather natural, owing
to the conservation of the electric charge as a consequence of the
field equations, but this is not so for the scalar models
\cite{anderson}. Nevertheless, following the analogy with the
electromagnetic case, we can define for the models
(\ref{eq:(2-1)}) the total scalar charge associated with a given
\textit{static} asymptotically vanishing field solution
$\phi(\vec{r})$ as

\be \frac{1}{4\pi}\int d_{3}\vec{r}
\left(\vec{\nabla}\cdot\left[\stackrel{\bullet}{f}(X)
\vec{\nabla}(\phi)\right]\right) = \frac{1}{4\pi}\int_{S_{\infty}}
\stackrel{\bullet}{f}(X) \vec{\nabla}(\phi)\cdot d \vec{\sigma},
\label{eq:(2-4)ter} \en which, owing to the field equation
(\ref{eq:(2-2)}), vanishes for everywhere-regular solutions. For
the SSS solutions of (\ref{eq:(2-4)}) we have
$\stackrel{\bullet}{f}(X)\vec{\nabla}(\phi) = \Lambda
\frac{\vec{r}}{r^{3}}$ and the total charge equals $\Lambda$. We
can then define the spatial charge-density distribution as
$\sigma(r) = (1/4\pi)\stackrel{\bullet}{f}(0)
\vec{\nabla}^{2}\phi$, which gives for the total scalar charge of
the SSS solutions

\be \frac{1}{4\pi} \int d_{3}\vec{r}
\stackrel{\bullet}{f}(0)\vec{\nabla}^{2}\phi =
\lim_{r\rightarrow\infty}\stackrel{\bullet}{f}(0)r^{2}\phi^{'}(r)
= \lim_{r\rightarrow\infty}\frac{\Lambda
\stackrel{\bullet}{f}(0)}{\stackrel{\bullet}{f}(X)}.
\label{eq:(2-4)quart} \en Clearly, this interpretation is only
possible if $\stackrel{\bullet}{f}(0)$ is finite, or equivalently,
if the function $r^{2}\phi^{'}(r)$ goes to a constant as $r
\rightarrow \infty$ (asymptotically coulombian fields). This
function must also vanish as $r \rightarrow 0$. As we shall see at
once, this latter condition is fulfilled for all models with
finite-energy SSS solutions, but the former defines a sub-class of
those models (see case B-2 below in this section) to which the
scalar version of the Born-Infeld model belongs.

Once the form of $f(X)$ is fixed, equation (\ref{eq:(2-4)}) gives
the expression of the field $\phi^{'}(r) \equiv\
\phi^{'}(r,\Lambda)$ in implicit form and allows the determination
of the potential $\phi(r)$ (up to an additive arbitrary constant)
through a quadrature. We also note that Eq.(\ref{eq:(2-4)})
implies that $\phi^{'}(r)$, if single-branched \footnote{In some
cases Eq.(\ref{eq:(2-4)}) can lead to discontinuities or several
branches for the function $\phi^{'}(r)$. We shall regard such
cases as ``unphysical" and rule them out from this analysis,
considering only models for which the fields of the SSS solutions
are (for $r>0$) continuous, single-branched functions defined
everywhere. We shall establish at the end of this section that the
corresponding admissibility condition for the lagrangian densities
is the strict monotonicity.}, is necessarily a monotonic function
of $r$. Moreover, the solution depends on $r$ and $\Lambda$
through the ratio $r/\sqrt{\Lambda}$. This is a straightforward
consequence of the invariance of the solutions of the field
equations (\ref{eq:(2-2)}) under space-time scale transformations.
Indeed, if $\phi(\vec{r},t)$ is a solution of this equation, so is
the modified function

\be \varphi(\vec{r},t,\lambda) =
\lambda^{-1}\phi(\lambda\vec{r},\lambda t), \label{eq:(2-4)bis}
\en $\lambda$ being a positive constant. This is a symmetry of the
solutions of the field equations without sources, but not an
invariance of the action \cite{ramond}.

The canonical energy-momentum tensor associated to the lagrangian
(\ref{eq:(2-1)}) is

\be
T_{\mu \nu} = 2\stackrel{\bullet}{f}(X)\partial_{\mu}\phi\partial_{\nu}\phi - f(X) \eta_{\mu \nu},
\label{eq:(2-5)}
\en
and the energy density in the SSS case becomes

\be
\rho=T^{00}=-f(-\phi^{'2}),
\label{eq:(2-6)}
\en
whereas the total energy is

\be \epsilon(\Lambda) =
-4\pi\int_{0}^{\infty}r^{2}f(-\phi^{'2}(r,\Lambda))dr =
\Lambda^{3/2}\epsilon(\Lambda=1), \label{eq:(2-7)} \en the last
equality being a consequence of the aforementioned scale
invariance. Using the first-integral (\ref{eq:(2-4)}) and
integrating by parts we obtain the following useful expression for
the energy in this SSS case

\be \epsilon(\Lambda) = \frac{8\pi\Lambda}{3}\left\lbrace
[\phi(r,\Lambda)]\vert_{0}^{\infty} - \left[r\phi^{'}(r,\Lambda) +
\frac{r^{3}}{2\Lambda}f(-\phi^{'2}(r,\Lambda))\right]\bigg
\vert_{0}^{\infty}\right\rbrace. \label{eq:(2-7)bis} \en
As we
shall see, if the energy of the SSS solutions is finite the second
bracket in the r.h.s. vanishes and this expression reduces to

\be \epsilon(\Lambda) =
\frac{8\pi\Lambda}{3}\left[\phi(\infty,\Lambda) -
\phi(0,\Lambda)\right], \label{eq:(2-7)ter} \en which shows that
the potential $\phi(r)$ must be a bounded function of $r$, defined
up to an arbitrary constant. Note also that Eq.(\ref{eq:(2-7)ter})
has the form of the potential energy of a point-like scalar charge
of value $2\Lambda$ placed at infinity in the soliton field.

There is another way round to obtain the expression
(\ref{eq:(2-7)ter}). Indeed, by performing the usual reescaling of
Derrick's theorem \cite{derrick64}, i.e. by defining a
uniparametric family of transformations

\be \phi_{\lambda}(x)\equiv \phi(\lambda x), \label{eq:(2-7)q} \en
it can be easily checked that the condition of extremum of the
energy against these reescalings  $\left(\frac{d
\epsilon(\phi(\lambda x))}{d \lambda}\mid_{\lambda=1}=0\right)$
leads automatically to the relation (\ref{eq:(2-7)ter}) when the
condition of finiteness of the energy is assumed. Although
Derrick's theorem is only a necessary condition for stability,
this signals the presence of a connection between stability and
finite-energy condition of the SSS solutions considered here. This
connection will be precisely established in section 6.

Going beyond the results of reference \cite{diaz83}, we shall
determine general conditions to be imposed on the functions $f(X)$
in order to obtain physically consistent field theories, whose
associated SSS solutions be stable and their energy
(\ref{eq:(2-7)}) be finite (non-topological solitons). We first
summarize some criteria of physical consistency adopted for the
purposes of the present study (defining what we shall call
\emph{``admissible"} field theories) and obtain the associated
restrictions on the lagrangian densities. Next we shall obtain the
conditions for such admissible models to support SSS soliton
solutions.

\subsection{\large Conditions on the energy functional}

If any of these models are to be used for descriptions of quantum
physical systems, the possibility of their quantization becomes
important. This implies supplementary conditions to be satisfied
by the lagrangian densities, as necessary for any quantum
extension. Thus, aside from the above mentioned continuity and
derivability conditions, we must require the function $f(X)$ to be
defined everywhere ($\Omega \equiv \Re$), in order to allow the
proper definition of the associated path integral. Here we shall
call such models \emph{\underline{class-1}} field theories
\footnote{These conditions exclude models such as the scalar
version of Born-Infeld Electrodynamics \cite{B-I34}. In such cases
it is always possible to generalize the model, by continuing the
lagrangian density function to the undefined regions through some
prescription which must preserve the classical dynamical content
of the initial model. Then, the quantum behaviour of the extended
model would depend on this prescription. But this procedure will
necessarily enlarge the space of solutions of the classical
theory. As we shall show below, these extensions introduce new
branches for the SSS solutions which become spurious at the
classical level. We shall exclude such extended models from the
present analysis since they are non-admissible according to our
physical criteria.}.

Alternatively, as discussed in the introduction, we can consider
these scalar models (and their generalized versions proposed
below) as effective classical lagrangians of more ``fundamental"
theories, including integrated high-energy and quantum effects
through new non-linear couplings. In these cases we can relax the
everywhere definiteness conditions of the admissible lagrangians
and require their regularity only within a restricted domain of
definition ($\Omega \subset \Re$), which is assumed to be open,
connected and including the vacuum ($0 \in \Omega$). We shall call
these models \emph{\underline{class-2}} field theories. Using this
criterion, models such as the Born-Infeld one become admissible
field theories belonging to this class. Although these models are
essentially classical, quantum corrections to their particle-like
solutions can be obtained by quantizing the field of small
fluctuations around these ground states. Such fluctuations obey
Euler-Lagrange linear field equations of admissible lagrangian
densities defined everywhere (see Ref. \cite{tdlee81} and section
6, Eqs.(\ref{eq:(6-7)}) and (\ref{eq:(6-7)bis}) below).

Obviously, a second condition to be imposed for admissibility in
all cases concerns the positive definite character of the energy,
which is required to hold in the entire domain of definition
($\Omega$) of the lagrangian density. The expression for the
energy density in terms of $f(X)$ is

\be \rho = T^{00} = 2\stackrel{\bullet}{f}(X)\left(\frac{\partial
\phi}{\partial t}\right)^{2} - f(X) = 2X\stackrel{\bullet}{f}(X) -
f(X) + 2\stackrel{\bullet}{f}(X)\left(\vec{\nabla}\phi\right)^{2}.
\label{eq:(2-8)} \en For the D'Alembert lagrangian we have
$f(X)=X/2$, and the energy density reduces to

\be \rho = \frac{1}{2}\left[\left(\frac{\partial \phi}{\partial
t}\right)^{2} + \left(\vec{\nabla}\phi\right)^{2}\right],
\label{eq:(2-9)} \en which is positive for any non-constant
function $\phi(t,\vec{r})$ and vanishes in vacuum. In order to
obtain a similar behaviour in the general case (\ref{eq:(2-8)})
(requiring also the energy density to vanish in vacuum) we are
lead to the necessary conditions

\bea
f(0) &=& 0\\
\nonumber \stackrel{\bullet}{f}(X) &\geq& 0 \hspace{0.3cm} \forall
X. \label{eq:(2-10)} \ena For minimal sufficient conditions on
$f(X)$ let us analyze separately the cases $X < 0$ and $X > 0$. On
the one hand for $X < 0$ the term $(\partial_{t}\phi)^{2}$ may
vanish and the positivity of the energy density (\ref{eq:(2-8)})
requires

\be f(X) < 0 \hspace{0.3cm} (\forall X < 0). \label{eq:(2-11)} \en
On the other hand, for $X>0$ the term $(\vec{\nabla}\phi)^{2}$ may
vanish and the positivity of the energy requires

\be \rho(X) \geq 2X\stackrel{\bullet}{f}(X) - f(X) =
Xf(X)\frac{d}{dX}\left[\ln\left(\frac{f^{2}(X)}{X}\right)\right]
\geq 0. \label{eq:(2-12)} \en This equation, together with the
conditions $f(0)=0$ and $\stackrel{\bullet}{f}(X)\geq 0
\hspace{0.2cm} (\forall{X})$, lead to

\be \frac{d}{dX}\left[\ln\left(\frac{f(X)}{\sqrt{X}}\right)\right]
\geq 0, \label{eq:(2-13)} \en or, equivalently, the function
$\frac{f(X)}{\sqrt{X}}$ (and hence $f(X)$ itself) must be a
positive monotonically increasing function of $X$ for $X>0$.
Equation (\ref{eq:(2-13)}), together with the boundary condition
$f(0)=0$, fix the behaviour of $f(X)$ around $X=0$ as

\be f(X \rightarrow 0^{+}) \sim X^{1+\alpha}, \label{eq:(2-14)}
\en with $\alpha > -1/2$, and the energy density behaves there as
\footnote{For values of $\alpha$ in Eqs.(\ref{eq:(2-14)}) and
(\ref{eq:(2-15)}) which lie in the interval $-1<\alpha<-1/2$ the
condition $f(0)=0$ is fulfilled, but $\stackrel{\bullet}{f}(X)$
diverges at $X = 0$ in such a way that the energy density becomes
necessarily negative in the neighbourhood of $X=0$. For $\alpha
\leq -1$, $f(X)$ diverges in vacuum. The limit case $\alpha=-1/2$
is singular.}

\be
\rho(X \rightarrow 0^{+}) \geq (1+2\alpha) X^{1+\alpha}.
\label{eq:(2-15)}
\en

\subsection{\large Conditions for finite-energy SSS solutions}

The convergence of the integral of energy (\ref{eq:(2-7)}) for the
SSS solutions is governed by the behaviour of the integrand near
the limits $r \rightarrow \infty$ and $r \rightarrow 0$. This
imposes supplementary conditions on the form of the function
$f(X)$ around the values of $X(r) = -\phi^{'2}(r)$ in these
limits. Let us assume a power law expression for the field around
these regions \footnote{Although this assumption excludes some
transcendent behaviours such as the asymptotic exponential
damping, our conclusions will remain valid for models exhibiting
these ``short-ranged" SSS solutions. In fact such models are
included in the case B-3 below. Note that damped
\textit{oscillatory} behaviour at infinity is excluded by the
monotonicity of the SSS field solutions.} ($\phi^{'}(r) \sim
r^{q}$ as $r \rightarrow \infty$ or as $r \sim 0$); from the
first-integral (\ref{eq:(2-4)}) we obtain the relation (valid for
$q \neq 0$; for $q = 0$ at $r = 0$ see the discussion of the case
\textbf{A-2} below)

\be
df/dr = -\frac{2 \Lambda \phi^{''}(r)}{r^{2}} \sim -2\Lambda q r^{q-3},
\label{eq:(2-16)}
\en
and in the limits of integration, $f(r)$ behaves as

\be
f(r) \sim \frac{2 \Lambda q}{2-q}r^{q-2} + D,
\label{eq:(2-17)}
\en
if $q \neq 2$, or as

\be f(r) \sim -4 \Lambda \ln(r) + D, \label{eq:(2-18)} \en if $q =
2$. The integration constants $D$ in these expressions are easily
related to the values of $X$ and $f(X)$ on the limits of the
integral of energy Eq.(\ref{eq:(2-7)}). Around each of these
limits the contributions to this integral take the form

\be
-4\pi\int dr\left[\frac{2 \Lambda q}{q-2}r^{q} + Dr^{2}\right],
\label{eq:(2-19)}
\en
for $q \neq 2$ and

\be -4\pi\int dr\left[4 \Lambda r^{2} \ln(r) + Dr^{2}\right],
\label{eq:(2-20)} \en for $q = 2$. Let us analyze separately the
convergence of the energy integral around $r \sim 0$ (\textbf{case
A}) and in the asymptotic limit $r \rightarrow \infty$
(\textbf{case B}). In \underline{\textbf{case A}} the convergence
of (\ref{eq:(2-19)}) requires $q > -1$ and we can distinguish
three sub-cases:

$\bullet$ \textbf{\underline{A-1})} If $-1 < q < 0$ the field
$\phi^{'}(r)$ diverges at $r \rightarrow 0$ but the integral of
energy converges there and the potential $\phi(r)$ is finite at
the origin. Then, as $r$ approaches zero, $X \rightarrow -\infty$
and $f(X)$ and $\stackrel{\bullet}{f}(X)$ diverge as

\be f(X) \sim -(-X)^{\frac{q-2}{2q}}
\hspace{0.15cm},\hspace{0.15cm} \stackrel{\bullet}{f}(X) \sim
(-X)^{-\frac{q+2}{2q}}, \label{eq:(2-20)bis} \en (see figure 1).
Such solutions can be stable and finite-energy SSS fields
(depending on their large-$r$ behaviour) and, in this sense, they
might be considered as genuine non-topological solitons.

\begin{figure}
\begin{center}
\includegraphics*[width=9.0cm,height=6.0cm]{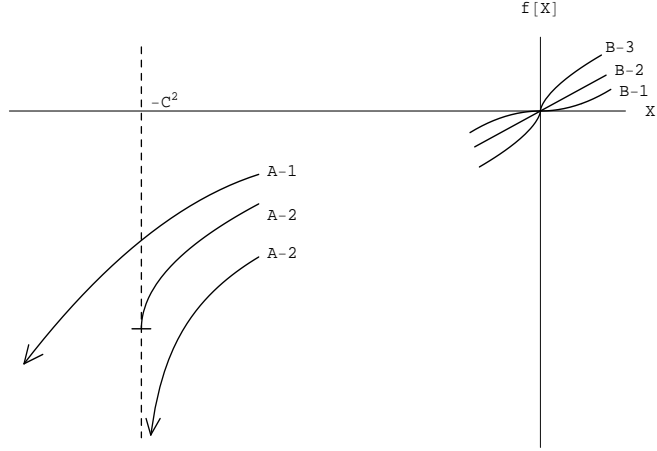}
\caption{\label{fig:epsart} Different possible behaviours of the
admissible lagrangians supporting finite-energy SSS solutions.
Note the existence of two A-2 cases, both ending at a finite value
$X = -C^2$ (corresponding to the maximum value of the field
strength) but with bounded or unbounded values of $f(X)$ there.
A-1 branch is parabolic with divergent slope as $X \rightarrow
-\infty$.}
\end{center}
\end{figure}

$\bullet$ \textbf{\underline{A-2})} If $q = 0$ the field
$\phi^{'}(r)$ goes to a constant value at the origin ($\phi^{'}(0)
= C$, which corresponds to $X+C^{2} = 0$) and can be written
around this point as

\be \phi^{'}(r) \sim C - \theta r^{\sigma}, \label{eq:(2-21)} \en
where $C, \theta$ and $\sigma$ are positive constants
\footnote{Note that the scalar version of the Born-Infeld model is
an example which belongs to this case, corresponding to $\sigma =
4$.}. Then $\stackrel{\bullet}{f}(X)$ diverges there as

\be \stackrel{\bullet}{f}(X) \sim (X+C^{2})^{- \frac{2}{\sigma}}.
\label{eq:(2-22)} \en
Consequently, for $\sigma \neq 2$, $f(X)$
behaves around $X = -C^{2}$ as

\be f(X) \sim \frac{(X+C^{2})^{1 - \frac{2}{\sigma}}} {\left(1 -
\frac{2}{\sigma}\right)} + \Delta, \label{eq:(2-23)} \en where
$\Delta$ is a constant. For $\sigma = 2$ this behaviour becomes

\be f(X) \sim \ln(X+C^{2}). \label{eq:(2-24)} \en
We see that when
$\sigma \leq 2$ these lagrangians diverge at $X = -C^{2}$ (figure
1). However they can be accepted as admissible class-2 field
theories if this point is located on the frontier of the (open)
domain of definition where the lagrangian density must be regular
(see the third example of the following section). Consequently the
set $X \leq -C^{2}$ must be excluded from the domain of definition
($\Omega$) in this case.  This requires $\sigma$ to be an
irrational number or four times the irreducible ratio between any
natural and an odd natural. When $\sigma > 2$ these values also
exclude the same region, leading again to admissible class-2 field
theories. However, in this case the lagrangians are finite in $X =
-C^{2}$ (even if $\stackrel{\bullet}{f}(X)$ diverges there). Then,
for rational values of $\sigma > 2$ which are irreducible ratios
of an odd natural and any natural \footnote{Other values of
$\sigma$, for which the lagrangian is also defined for $X <
-C^{2}$, lead to negative energy densities there and consequently
must be excluded.}, the lagrangians are defined and continuous for
any $X$ around $X = -C^{2}$, exhibiting a vertical-slope inflexion
in this point. If appropriately extended to all $X$, they lead to
class-1 field theories which satisfy the everywhere positive
definiteness condition of the energy. Nevertheless they violate
the requirement of continuity of $\stackrel{\bullet}{f}(X)$ for
$X<0 (\in \Omega)$ and this leads to associated point-like
solutions exhibiting several branches. Indeed, as we shall see at
once, this requirement for admissibility is introduced because it
endorses the single-branched character of the SSS solutions.
Consequently, in this case A-2 all admissible lagrangians must
remain undefined for $X \leq -C^{2}$ and thus belong to class-2
field theories. This implies (as happens in the Born-Infeld model
\cite{B-I34}) the existence of a maximum value of the field
strength ($\phi^{'}(r) < C$).

As a function of $r$, the energy density behaves around the center as

\be
r^{2}f(r) \sim \frac{(2C\theta)^{1-2/\sigma}}{1-2/\sigma}r^{\sigma} - \Delta r^{2},
\label{eq:(2-25)}
\en
for $\sigma \neq 2$ and as

\be r^{2}f(r) \sim 2r^{2}\ln(r) + r^{2}\ln(2C\theta),
\label{eq:(2-25)bis} \en for $\sigma = 2$. As expected, the energy
integral converges there in both cases.

$\bullet$ \textbf{\underline{A-3})} The case $q > 0$  must be
discarded. Indeed, in this case $\stackrel{\bullet}{f}(X)$ behaves
as

\be \stackrel{\bullet}{f}(X) \sim X^{- \frac{q+2}{2q}},
\label{eq:(2-26)} \en around $X = 0$. Consequently, $f(X)$ is
singular in vacuum for $0 < q \leq 2$. For $q > 2$ the energy
density for $X \rightarrow 0^{+}$ behaves as

\be \rho(X) \sim -\frac{2}{q}X^{\frac{q-2}{2q}}, \label{eq:(2-27)}
\en and becomes negative around the vacuum (see also
Eqs.(\ref{eq:(2-14)}), (\ref{eq:(2-15)}) and the footnote there).

In \underline{\textbf{case B}} the convergence of
(\ref{eq:(2-19)}) in the $r \rightarrow \infty$ limit requires $q
< -1$ (in this case $\phi^{'}(r \rightarrow \infty) = 0$ and the
integration constant $D$ in (\ref{eq:(2-19)}) vanishes). Then the
behaviour of $f(X)$ around $X=0$ must be \footnote{For the sake of
clarity we use here the parameter $p = -q$ in the exponent, in
terms of which $\phi(r \rightarrow \infty) \sim \frac{1}{r^{p}}$
with $p > 1$.}

\be f(X) \sim X^{\frac{p+2}{2p}}. \label{eq:(2-28)} \en Now the
existence of the lagrangian on both sides around $X = 0$ becomes
crucial for the consistency of the theory and this imposes
supplementary restrictions on the possible values of the parameter
$p$. Indeed, the exponent in Eq.(\ref{eq:(2-28)}) must be the
ratio of two odd naturals \footnote{If this exponent is the
irreducible ratio between an even and an odd natural numbers the
lagrangian is well defined on both sides of $X=0$, but the energy
density becomes negative for $X < 0$.} and this restricts the
possible values of $p$ to a sub-class of the rational numbers. Let
us analyze separately three possibilities:

$\bullet$ \textbf{\underline{B-1})} Consider first the case $1 <p
<2$. We define $P$ and $Q$ as two positive odd natural numbers in
such a way that $P < Q$ and the ratio $\Sigma = P/Q$ be
irreducible. Then the admissible values of the exponent in
(\ref{eq:(2-28)}) are given by

\be
\frac{p + 2}{2p} = \frac{3}{2 + \Sigma},
\label{eq:(2-29)}
\en
and the corresponding admissible values of $p$ can be written as

\be p = \frac{4 + 2 \Sigma}{4 - \Sigma}. \label{eq:(2-30)} \en Now
$\stackrel{\bullet}{f}(0)=0$ and the slope of the lagrangian
vanishes in vacuum (see figure 1).

$\bullet$ \textbf{\underline{B-2})}
For $p = 2$ the lagrangian
behaves around $X=0$ as the D'Alembert lagrangian

\be
f(X\rightarrow 0^{\pm}) \sim X,
\label{eq:(2-31)}
\en
and the soliton field becomes asymptotically coulombian (see figure 1).

$\bullet$ \textbf{\underline{B-3}) }For $p > 2$ the behaviour of
the lagrangian is also given by Eq.(\ref{eq:(2-28)}) but now the
admissible values of the exponent are constrained by

\be \frac{p + 2}{2p} = \frac{1}{1 + \Sigma} > \frac{1}{2},
\label{eq:(2-32)} \en where $\Sigma = P/Q$ must be the irreducible
ratio between an even natural $P$ and an odd natural $Q$ such that
$Q > P$. The corresponding admissible values of $p$ are

\be p = 2\frac{1 + \Sigma}{1 - \Sigma}. \label{eq:(2-33)} \en As
easily seen, the slope of the lagrangian diverges at $X=0$ in this
case (see figure 1), but the energy density remains positive
definite there.

We conclude that the set of admissible models exhibiting
finite-energy SSS solutions can be classified into six families
which are the combinations of the cases A-1 or A-2, governing the
central field behaviour and the cases B-1, B-2 or B-3, determining
the asymptotic field behaviour. Moreover, any given scalar,
monotonically decreasing, SSS function $\phi'(r)$, which satisfies
boundary conditions of A-type at the center and of B-type
asymptotically, is a finite-energy SSS solution of a particular
lagrangian model belonging to one of these families. The explicit
form of the corresponding lagrangian density can be found by
integrating Eq.(\ref{eq:(2-4)}) with respect to the variable $X =
-\phi^{'2}(r)$ with the corresponding boundary conditions (see the
fourth example of the next section).

\subsection{\large Conditions for stability}

We summarize here the main steps in the analysis of linear
stability of the scalar SSS soliton solutions (the detailed
calculations are given in section 6). The linear stability of
these solutions requires their energy (\ref{eq:(2-7)}) to be a
local minimum against small charge-preserving perturbations. We
consider finite-energy SSS solutions $\phi(r)$ and small static
perturbations $\delta\phi(\vec{r})$, finite and regular everywhere
and vanishing (as well as their radial derivatives) as $r
\rightarrow \infty$, in such a way that the scalar charge of the
perturbed fields remains unchanged at the first order in the
perturbations. For the static solutions of the field equations
(\ref{eq:(2-2)}) the first variation of the energy
(\ref{eq:(2-8)}) vanishes, while the second variation is positive
\emph{if and only if} the condition \footnote{See Eq.
(\ref{eq:(6-4)bis}) and the footnote there.}

\be \stackrel{\bullet}{f}(X) + 2X\stackrel{\bullet \bullet}{f}(X)
\geq 0, \label{eq:(2-34)} \en is satisfied in all the range of
values of $X$ covered by the solution ($X=-\phi^{'2}(r), 0 \leq r
< \infty$). As we shall see this requirement is always fulfilled
by the finite-energy SSS solutions of the \textit{admissible}
models defined in this section, proving their linear stability. A
detailed spectral analysis of the small oscillations around these
SSS finite-energy solutions has also been performed. It leads, for
admissible models, to discrete spectra of eigenvalues and
normalizable orthogonal eigenfunctions in Hilbert spaces, whose
scalar products are built as three-dimensional integrals of their
products with the functions $f(X(r))$ as kernels. In their
temporal evolutions the perturbations to the soliton solutions
remain bounded in this norm, confirming the stability (see
subsections 6-1 and 6-2).

\subsection{\large Conditions for uniqueness of the elementary solutions}

We return now to Eq.(\ref{eq:(2-4)}), which defines the SSS field
solutions $\phi^{'}(r)$, and analyze the conditions under which
they are single-branched and defined everywhere functions. As
already mentioned, owing to the positivity of
$\stackrel{\bullet}{f}(X)$ both $\Lambda$ and $\phi^{'}(r)$ must
be simultaneously either positive or negative and then we can
analyze only the case $\phi^{'}(r) > 0$ without loss of
generality. Let us write (\ref{eq:(2-4)}) under the form

\be z(y) = y \stackrel{\bullet}{f}(-y^{2}) = \frac{\Lambda}{r^{2}}
\geq 0, \label{eq:(2-35)} \en
where we have introduced the variable
$y = \phi^{'}(r) > 0$. The field strength function is given by the
values of $y$ obtained by cutting the curve $z(y)$ with horizontal
lines corresponding to the different values of $r$. Then, for the
field $\phi^{'}(r)$ to be defined in all space, $z$ must range
from $0$ to $\infty$ and the uniqueness of the solution requires a
single cut point on every $z = constant$ line. This restricts
$z(y)$ to be a continuous monotonic function. As a consequence,
the requirements of continuity (for $X < 0$) and
\underline{strict} positivity (for any $X \neq 0$) imposed on
$\stackrel{\bullet}{f}(X)$ at the beginning of this section, when
establishing the admissibility conditions, are mandatory for the
proper definition of the SSS solutions. Indeed, a glance at the
form of the function $z(y)$ reveals that the existence of a jump
in the function $\stackrel{\bullet}{f}(X)$ at a finite value $X <
0$ would lead to SSS solutions which are double-branched or
undefined in some range of values of $r$. Moreover, if
$\stackrel{\bullet}{f}(X)$ vanishes for a value $X_{0} < 0$
(horizontal-slope inflexion point for $f(X)$), then $z(y)$
vanishes at $y=(-X_{0})^{1/2}$ reaching a minimum there and the
SSS solutions become necessarily multiple-valued (the vanishing of
$\stackrel{\bullet}{f}(X)$ for a positive value $X_{0} > 0$ is
discarded by the energy-positivity condition (\ref{eq:(2-13)})).

\begin{figure}
\begin{center}
\includegraphics[width=9cm,height=6cm]{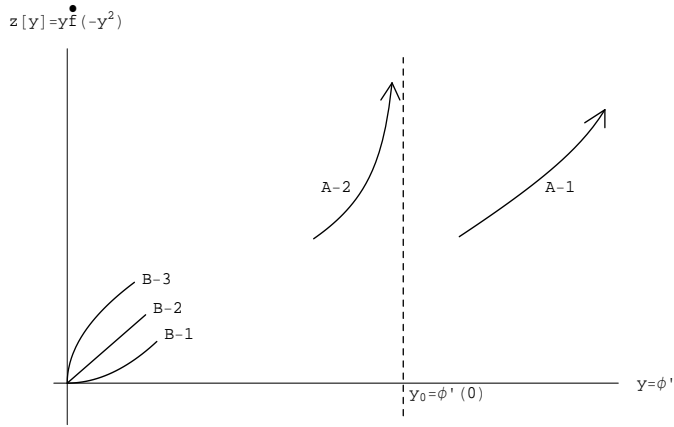}
\caption{\label{fig:epsart} Characterization in the
$y$-$z(y)=y\stackrel{\bullet}{f}(-y^2)$ plane of the admissible
models supporting finite-energy SSS solutions. Observe that all
curves start at $y=0, z=0$ ($y = \phi^{'}(r\rightarrow\infty) =
0$, cases B-1, B-2 and B-3) and grow monotonically without limit.
This guarantees the existence of a single cut point with the
$z=constant$ lines and thus a single-valued solution (compare to
figure 1).}
\end{center}
\end{figure}

For soliton solutions the finiteness of the energy requires the
origin to be a point of the curve $z(y)$ (see figure 2) and the
large-$z$ behaviour is determined by the behaviour of the field at
the center of the soliton. Then, in the cases where the field
strength diverges as $r \rightarrow 0$ (case A-1 above) the
uniqueness of the solution requires the curve $z(y)$ to start at
the origin and grow {\em monotonically} without limit as $y
\rightarrow \infty$. When the field is finite at $r = 0$ (case
A-2) the curve $z(y)$ must increase {\em monotonically} from the
origin and diverge at $y_{0} = \phi^{'}(0)$, showing a vertical
asymptote there. In this case $\stackrel{\bullet}{f}(-y^{2})$
diverges at $y = y_{0}$ and the uniqueness condition requires the
lagrangian function to remain undefined for $y > y_{0}$, for the
function $z(y)$ to exhibit an unique growing branch. Thus in all
A-2 cases the set $X < -y_{0}^{2}$ must be excluded from the
domain of definition ($\Omega$) and then, the associated models
must be necessarily class-2 field theories \footnote{It can be
shown from the analysis of the possible continuations of $z(y)$
for $y > y_{0}$ that the new branches of SSS field solutions are
pathological (non-defined everywhere, unstable, or both).}. The
monotonicity condition for the unique branch of $z(y)$ in the
admissible models with finite-energy solutions takes the form

\be \frac{dz}{dy} =  \stackrel{\bullet}{f}(-y^{2}) - 2y^{2}
\stackrel{\bullet \bullet}{f}(-y^{2}) \geq 0, \label{eq:(2-36)}
\en for any $y \geq 0$. This requirement coincides with the
condition (\ref{eq:(2-34)}) for stability which, as already
mentioned, is fulfilled by all these admissible models. To
summarize, we conclude that all \emph{finite-energy} SSS solutions
of \emph{admissible} scalar field theories considered in this
section are {\em single-branched, stable and defined everywhere}.

To close this section let us give an expression for the energy of the SSS solutions in terms of the function $z(y)$, which will be useful for the explicit calculation of the soliton energy, as we shall see in the examples of the next section. This expression can be obtained by taking into account Eqs.(\ref{eq:(2-4)}), (\ref{eq:(2-7)bis}) and (\ref{eq:(2-35)}) and reads

\be \epsilon = \frac{4\pi}{3}\Lambda^{3/2} \left\lbrace
\frac{f(-y^{2})}{z(y)^{3/2}} \bigg
\vert_{y(r=0)}^{y(r\rightarrow\infty)}
-2\int_{y(r=0)}^{y(r\rightarrow\infty)}\frac{dy}{\sqrt{z(y)}}\right\rbrace.
\label{eq:(2-37)} \en
As can be seen from the preceding analysis,
the first term in this formula vanishes for admissible models with
soliton solutions \footnote{Conversely, the conditions to be
imposed on the lagrangian densities of admissible models to
support finite-energy SSS solutions could have been directly
obtained from the requirement of cancellation of the first term in
(\ref{eq:(2-37)}).}. Once the expression of the lagrangian density
is known, the second term gives the soliton energy directly
through a quadrature.

\section{\large Some examples}

As illustrative examples of the considerations of the previous
section we shall introduce and discuss four families of admissible
models representative of the different classes analyzed there.

\subsection{\large Potential corrections to the D'Alembert Lagrangian}

The \underline{first example} is given by the two-parameter family
of field theories defined by lagrangian densities of the form

\be f(X) = \frac{X}{2} + \lambda X^{a}, \label{eq:(3-1)} \en where
$\lambda$ is a positive constant which gives the intensity of the
self-coupling. When $\lambda = 0$, $f(X)$ reduces to the usual
D'Alembert lagrangian density (see figure 3). The values of the
exponent $a$ are restricted to be irreducible ratios of two {\em
odd} natural numbers (we consider 1 as odd) $a = P/Q$ such that

\be
P > \frac{3}{2}Q \left(\Rightarrow a > \frac{3}{2}\right).
\label{eq:(3-2)}
\en

\begin{figure}[h]
\begin{center}
\includegraphics[width=17cm,height=6cm]{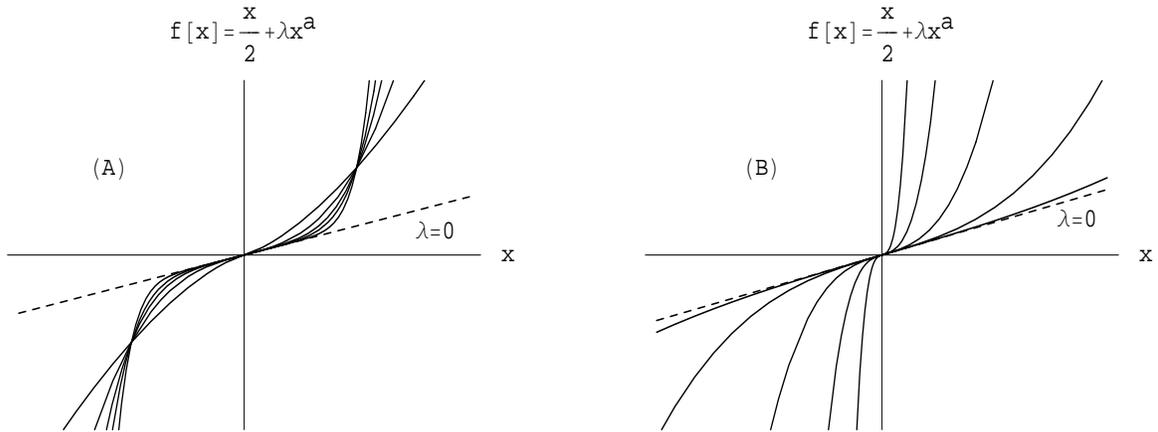}
\caption{\label{fig:epsart} Functional form of the lagrangian
densities for the family of models (\ref{eq:(3-1)}) (A) for a
fixed value of the parameter $\lambda (=1)$ and several values of
$a (=5/3, 3, 13/3, 17/3, 25/3)$ (B) for a fixed value of the
exponent $a (=3)$ and several values of $\lambda (=0.01, 0.1, 1,
10, 100)$. The dashed line corresponds in both cases to the
D'Alembert lagrangian ($\lambda=0$).}
\end{center}
\end{figure}

As easily verified these restrictions allow $f(X)$ to be defined
everywhere and the associated energy density to be positive
definite and vanishing in vacuum (admissible class-1 field
theories). The condition $a > 3/2$ is imposed to ensure the
convergence of the integral of energy at $r=0$, where the SSS
field strengths diverge.

This family of models can be extended to include rational values
of the exponent $a = P/Q> 3/2$ with $P$ being an even natural
number and $Q$ an odd one. Such models are admissible for $X<0$ if
we replace $\lambda > 0$ by $-\lambda$ in Eq.(\ref{eq:(3-1)})
\footnote{But, for complete admissibility, this function should be
matched for $X>0$ with another function satisfying the condition
(\ref{eq:(2-13)}). For example, the lagrangian obtained by
replacing $\lambda$ by $\lambda \cdot sign(X)$ in
(\ref{eq:(3-1)}). Since the structure and energy of the SSS
solutions are determined by the form of the lagrangian density for
$X<0$, the SSS solutions of these models are also solitons.}. The
following considerations are valid for the extended family. The
form of the SSS solutions is obtained from the equation
(\ref{eq:(2-36)}), which now reads

\be z(y) \equiv \frac{y}{2} + (-1)^{P-1}\lambda a y^{(2a-1)} =
\frac{\Lambda}{r^{2}}, \label{eq:(3-3)} \en with $y(r,\Lambda)
\equiv \phi^{'}(r,\Lambda)$. The function $z(y)$ shows an unique
growing branch for every value of the scalar charge $\Lambda > 0$
and, consequently, there is an unique SSS solution of
Eq.(\ref{eq:(3-3)}), which vanishes as $\phi^{'} \sim r^{-2}$ when
$r \rightarrow \infty$ (case B-2 above, asymptotically coulombian)
and diverges as $\phi^{'} \sim r^{-2/(2a-1)}$ when $r \rightarrow
0$. Thus the argument ($X$) of the lagrangian ranges from zero to
$-\infty$ in this interval and, as expected, the stability
condition (\ref{eq:(2-34)}), which now reads

\be
\stackrel{\bullet}{f}(X) + 2X\stackrel{\bullet \bullet}{f}(X) = \frac{1}{2} + (-1)^{P-1}\lambda a(2a-1)X^{a-1} > 0,
\hspace{0.3cm} \forall X < 0,
\label{eq:(3-4)}
\en
is fulfilled there.

The energy of the soliton, as a function of the model parameters,
can be explicitly obtained from the integral term of
Eq.(\ref{eq:(2-37)}). The final result is

\be \epsilon =
\frac{4\sqrt{2}\pi}{3}\frac{\Lambda^{3/2}}{(a-1)(2a\lambda)^{\frac{1}{4(a-1)}}}
B\left(\frac{1}{4(a-1)},\frac{2a-3}{4(a-1)}\right),
\label{eq:(3-5)} \en
where $B(x,y)$ is the Euler integral of first
kind

\be B(x,y) = \int_0^1dtt^{x-1}(1-t)^{y-1},
\hspace{0.3cm}\mathrm{Re}(x) > 0, \hspace{0.3cm}\mathrm{Re}(y) >
0. \label{eq:(3-5)bis} \en In figure 4 we have plotted this energy
as a function of the exponent $a$, with the coupling constant
$\lambda$ as a parameter. We see that the energy diverges, for any
value of $\lambda$, as the exponent $a$ approaches the value
$3/2$. This energy is strongly reduced in the region of values of
$a \gtrsim 3/2$ as the coupling constant $\lambda$ increases,
reaching minima which vanish as $\lambda\rightarrow\infty$ in this
region. When the exponent $a$ increases the energies of the
solitons become less dependent on $\lambda$ and approach
asymptotically the value $\epsilon/\Lambda^{3/2} =
16\pi\sqrt{2}/3$ as $a \rightarrow \infty$. An interesting feature
of these models is the existence of soliton solutions for any $a >
3/2$, no matter how small the coupling parameter $\lambda$ may be.
This implies that any small correction of this kind to the ``bare"
D'Alembert lagrangian leads to the possibility of excitation of
soliton modes. For a fixed value of the exponent, the masses of
such modes increase as the intensity of the coupling is reduced.
This behaviour is similar to the one encountered in one-space
dimensional models supporting topological soliton solutions
\cite{tdlee81}, \cite{faddeev75}.

\begin{figure}[h]
\begin{center}
\includegraphics[width=13cm,height=9cm]{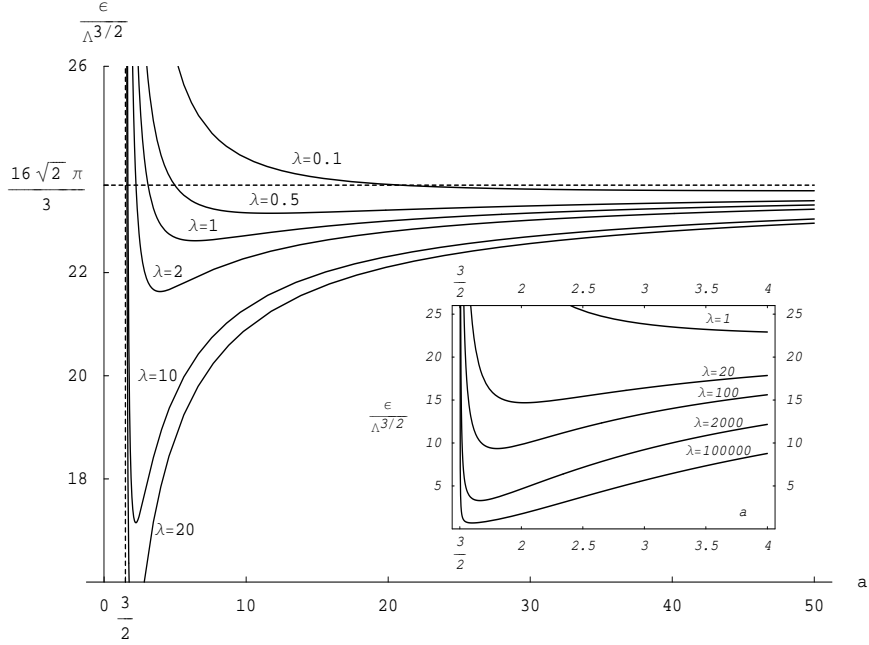}
\caption{\label{fig:epsart} Energy of the solitons of the family
(\ref{eq:(3-1)}) as a function of the exponent $a$ with the
coupling constant $\lambda$ as parameter. The dashed line
correspond to the asymptotic limit of the energy for $a
\rightarrow \infty$. The small plot shows the behaviour of the
energy for strong self-couplings. The energy is never zero for any
$\lambda < \infty$.}
\end{center}
\end{figure}

The preceding analysis can be generalized to the case of theories
whose lagrangian densities for $X<0$ take the form
\footnote{Obviously, for $X>0$ they must be extended in agreement
with the requirements of section 2.}

\be f_{N}(X) = \frac{X}{2} +
\sum_{n=2}^{N}(-1)^{P_{n}-1}\lambda_{n} X^{a_{n}},
\label{eq:(3-6)} \en where, for admissibility, the $\lambda_{n}$
are constrained to form a finite sequence of positive constants
and the exponents $a_{n} = P_{n}/Q_{n}$ to form an increasing
sequence of rational numbers, built as irreducible ratios of odd
natural numbers or of even and odd naturals, such that $a_{n} > 1$
for $n < N$ and $a_{N} > 3/2$. In these models the energy density
is positive definite for $X<0$ whereas the SSS field solutions
diverge at the origin as $\phi^{'}(r \rightarrow 0) \sim
r^{-2/(2a_{N}-1)}$ and vanish asymptotically as
$\phi^{'}(r\rightarrow \infty) \sim r^{-2}$. The associated
energies are finite and the corresponding solitons are stable.

A particular case of Eq.(\ref{eq:(3-6)}) is obtained when the
exponents are a finite sequence of consecutive naturals $a_{n} = n
(n > 1)$. Moreover, let us assume that we take the limit
$N\rightarrow\infty$ and that the infinite sequence of
$\lambda_{n} > 0$ converges to zero, in such a way that the series

\be f(X) = \frac{X}{2} + \sum_{n=2}^{\infty}(-1)^{n-1}\lambda_{n}
X^{n}, \label{eq:(3-7)} \en be convergent in an interval including
$\forall X < 0$ (case A-1), or in an interval including the range
$0 > X > -C^{2}$ but excluding the values $X < -C^{2}$ (case A-2).
We can then extend the family (\ref{eq:(3-6)}) to a large class of
analytic functions. Such functions must satisfy the conditions
established in section 2.2. for the integrals of energy associated
to the SSS solutions of such models to be finite. If we assume the
positivity of the coefficients $\lambda_{n}$ in
Eq.(\ref{eq:(3-7)}) we can explicitly check for these models the
fulfillment of the condition of stability (\ref{eq:(2-34)}), which
now takes the form

\be \stackrel{\bullet}{f}(X) + 2X\stackrel{\bullet \bullet}{f}(X)
= \frac{1}{2} + \sum_{n=2}^{\infty} n(2n-1) \lambda_{n} (-X)^{n-1}
> 0 \label{eq:(3-7)bis} \en and holds in the entire domain of
definition of the soliton. Thus the requirement of convergence of
the series (\ref{eq:(3-7)}) for any $X$, with the assumed
restrictions $\lambda_{n} > 0 \hspace{0.1cm} (\forall n)$ leads to
class-1 field theories supporting SSS solitons. As an example of
this let us mention the analytic function $f(X) =
\frac{1}{2}sh(X)$, which diverges when $X \rightarrow -\infty$
faster than $X^{\gamma}$ with $\gamma > 3/2$ (as required by
Eq.(\ref{eq:(2-20)bis}) in the A-1 case) and behaves like $X/2$
around $X=0$ (case B-2).

If we relax the requirement of positivity of the coefficients
$\lambda_{n}$, whenever the series (\ref{eq:(3-7)}) converges in
some restricted interval $X > -C^{2}$ and the sum remains a
monotonically increasing function of $X$ there, we are lead to
admissible class-2 field theories exhibiting SSS soliton
solutions. An example of this case is the lagrangian density
$f(X)=\frac{1}{2}tg(X)$ restricted to the interval $-\frac{\pi}{2}
< X < \frac{\pi}{2}$. This model supports finite-energy stable SSS
solutions belonging to cases A-2 and B-2.

Let us consider the case $\lambda_{n} \geq 0, \forall n$. The
partial sums in (\ref{eq:(3-7)}) give rise to an infinite sequence
of admissible lagrangian models of the form (\ref{eq:(3-6)}) with
natural exponents, all of them supporting SSS soliton solutions
belonging to cases A-1 and B-2. The explicit forms
$\phi_{N}(r,\Lambda)$ of these solutions can be obtained by
solving the equation (\ref{eq:(2-35)}) for each lagrangian in the
sequence. These equations take the form

\be \phi_{N}^{'}\left(\frac{1}{2} + \sum_{n=2}^{N} n \lambda_{n}
(\phi_{N}^{'})^{2(n-1)}\right) = \frac{\Lambda}{r^{2}}.
\label{eq:(3-7)s} \en If the series (\ref{eq:(3-7)}) converges in
$X < 0$, defining an analytic lagrangian density function there,
the sequence of SSS soliton solutions of the partial-sum models in
the expansion, corresponding to the same value of the scalar
charge $\Lambda$ for all $N$, must converge to the SSS soliton
solution (with the same charge) associated to this lagrangian
density ($\phi_{N \rightarrow \infty}(r,\Lambda) \rightarrow
\phi(r,\Lambda)$). This can be directly established from
Eq.(\ref{eq:(3-7)s}), which defines the forms of the SSS
solutions. Then the limit solution can be written as a functional
series expansion in terms of the members of the sequence as

\be \phi(r,\Lambda) = \phi_{1}(r,\Lambda) +
\sum_{N=2}^{\infty}\delta_{N}(r,\Lambda), \label{eq:(3-7)ter} \en
where $\delta_{N}(r,\Lambda) = \phi_{N}(r,\Lambda) -
\phi_{N-1}(r,\Lambda)$. Using Eqs.(\ref{eq:(2-7)bis}) and
(\ref{eq:(2-7)ter}) it is easy to show that the sequence of
energies of the equal-charge solitons associated to the
partial-sum lagrangian densities converges towards the energy of
the equal-charge soliton associated to the full series lagrangian
(\ref{eq:(3-7)}), which can be written as the series expansion:

\be \epsilon(\Lambda) = \epsilon_{1}(\Lambda) +
\sum_{N=2}^{\infty}\Delta_{N}\epsilon(\Lambda),
\label{eq:(3-7)quart} \en
where $\Delta_{N}\epsilon(\Lambda) =
\epsilon_{N}(\Lambda) - \epsilon_{N-1}(\Lambda)$ is the difference
between the energies of two consecutive solitons in the sequence.
The first two terms of this expansion are energy-divergent. The
first one corresponds to the divergent self-energy of the Coulomb
field whereas the first correction $\Delta_{2}\epsilon(\Lambda)$
cancels this divergence and ``renormalizes" the self-energy to a
finite value. The subsequent terms are all finite and the series
converges towards the energy of the limit soliton. The energy
associated with the soliton of order $N$ in the sequence can be
obtained making use of the expression (\ref{eq:(2-37)}) and reads

\be
\epsilon_{N}(\Lambda) = \frac{8\pi}{3}\sqrt{2}\Lambda^{3/2}\int_{0}^{\infty} \frac{dy}{y^{1/2}\sqrt{1+2\sum_{n=2}^N n\lambda_{n}y^{2(n-1)}}},
\label{eq:(3-7)q}
\en
which can be numerically calculated once the coefficients are fixed.

\begin{figure}[h]
\begin{center}
\includegraphics[width=10cm,height=7cm]{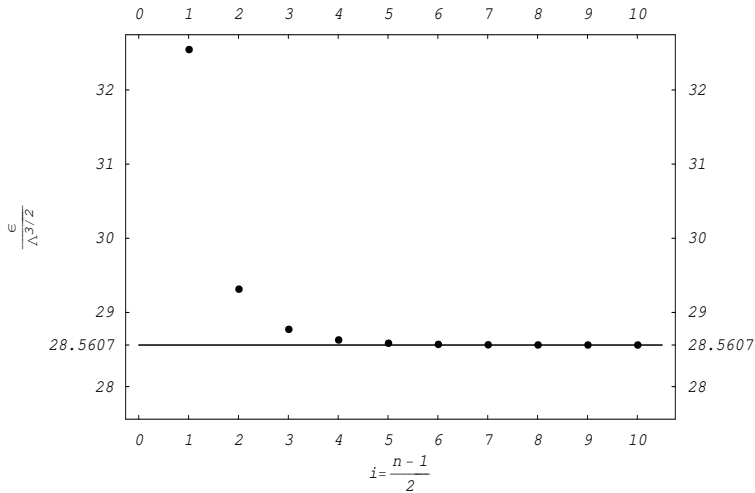}
\caption{\label{fig:epsart} Energy of the sequence of SSS soliton
solutions (with the same scalar charge, $\Lambda$) of the
partial-sum lagrangian models obtained from the McLaurin expansion
of the lagrangian density $f(X)=\frac{1}{2}sh(X)$, as functions of
the integer parameter $i=\frac{N-1}{2}$, $N$ being the odd
exponents of the surviving terms in the expansion. As $N$
increases these energies approach asymptotically the value
$\epsilon(\Lambda) \simeq 28.5607 \Lambda^{3/2}$, corresponding to
the soliton energy of the exact theory.}
\end{center}
\end{figure}

To illustrate this procedure let us consider the above mentioned
analytic lagrangian $f(X) = \frac{1}{2}sh(X)$. The energy
associated to the soliton solutions of this model, obtained from
Eq.(\ref{eq:(2-37)}), is $\epsilon(\Lambda) \simeq 28.5607
\Lambda^{3/2}$. The partial sums of the McLaurin expansion of this
lagrangian function are admissible models supporting a sequence of
SSS soliton solutions. Their energies, obtained from
Eq.(\ref{eq:(3-7)q}), are plotted in figure 5 as functions of
$i=\frac{N-1}{2}$, for the same value of the scalar charge
$\Lambda$ of each solution ($N=2i+1$ being the exponents of the
surviving terms in the expansion which, in this case, are the
sequence of odd naturals). Obviously the energy of the first-order
term ($N=1, i=0$), which corresponds to the Coulomb field,
diverges but, as expected, the first correction already
``renormalizes" this coulombian divergent energy and the
subsequent orders reduce the (now finite) energy, which approaches
asymptotically the energy of the soliton of the exact model as $i$
increases. The convergence in this example is related to the
analytic character of the sum (\ref{eq:(3-7)}) but the
``renormalization" of the divergent self-energy is due to the
first correction to the pure D'Alembert lagrangian and would arise
even if the series were not convergent.

These results can be useful in the analysis of particle-like
solutions in effective models of gauge-invariant interactions.
Indeed, effective lagrangians arise frequently in perturbative
schemas which lead to polynomial expressions in certain field
invariants. For example, in the case of QED the perturbative
expansion of the photon effective action, which is obtained by
integrating out the high-energy degrees of freedom of the electron
sector, defines a sequence of lagrangians which take this
polynomial form in the field invariants (Euler-Heisenberg
lagrangians \cite{Heisenberg36} and the higher-order corrections
\cite{pirula70}). On the other hand, as we shall establish in the
following sections, the solution of the electrostatic spherically
symmetric problem for a generalized gauge-invariant lagrangian
model can be reduced to that of an associated scalar field model,
whose lagrangian density is univocally defined from the
gauge-invariant one. If the sequence of gauge-invariant effective
lagrangians are of polynomial forms in the field invariants, the
associated scalar lagrangians are also polynomials in the kinetic
term, taking the form of partial sums of a series as
(\ref{eq:(3-7)}). In this way we have established that the
sequence of effective lagrangians describing the low-energy
photon-photon interaction in QED support electrostatic point-like
finite-energy solutions \cite{dr08-1}.

Let us give a \textit{tentative} physical interpretation of these
results. The non-linear terms in these effective lagrangians
describe, at a classical level, a self-interaction of the gauge
field mediated by the Dirac vacuum. The point-like solution of the
``bare" Maxwell lagrangian is the Coulomb field, which has a
divergent self-energy. The first non-linear correction term of the
effective lagrangian (Euler-Heinsenberg) incorporates polarization
effects of the vacuum on the classical field of the point charge,
calculated to lowest order in a perturbative expansion. These
screening effects ``renormalize" the charge field, which becomes
finite-energy. The subsequent corrections in the expansion
describe higher-order approximations to the behaviour of the
screening, but the finite-energy character of the screened fields
is preserved to all orders. Unfortunately, the validity of this
effective approach is limited to energies much lower than the
electron mass \cite{Dobado97} and is not accurate to describe the
strong fields arising near the center of the particle-like
solutions. Consequently, this tentative interpretation can not be
maintained only on these grounds. For a more rigorous
investigation this question should be considered starting from a
different effective approach incorporating the vacuum polarization
effects in presence of the strong fields of point-like charges.
The analysis of this approach is currently in progress (see also
the comments on this point in the last section).

\subsection{\large B-I-like models}

The \underline{second example} is a two-parameter family of field
theories defined by lagrangian densities of the form

\be f(X) = \frac{(1+\mu^{2}X)^{\alpha} - 1}{2\alpha\mu^{2}},
\label{eq:(3-8)} \en where $\mu$ is a real constant. The
admissibility conditions require the values of the parameter
$\alpha$ to be restricted to the range $1/2 \leq \alpha < 1$.
Indeed, if $1 \leq \alpha \leq 3/2$ the energy of the SSS
solutions diverges around $r=0$. On the other hand if $\alpha>3/2$
the solution $\phi^{'}(r)$ is multi-valued and the different
branches are either unstable or energy-divergent. We also discard
the models with $0 < \alpha < 1/2$, since the energy density in
this case is not positive definite for $X > 0$. Moreover, if
$\alpha$ is a rational number built as the irreducible ratio of an
even natural and an odd natural numbers, the function $f(X)$ is
defined everywhere, but $\stackrel{\bullet}{f}(X)$ changes sign in
$X=-1/\mu^{2}$ and the energy becomes negative for large negative
values of $X$. Finally, if $\alpha$ is the irreducible ratio of
two odd naturals we are lead to models which exhibit
multi-branched SSS solutions \footnote{In fact, the function
$z(y)$ in (\ref{eq:(2-36)}) has now two separated branches. The
field associated to the first branch ranges in the interval $0
\leq \phi^{'}(r) < 1/\mu$ and satisfies the condition
(\ref{eq:(2-34)}) there, leading to a stable and finite-energy SSS
solution, finite and defined everywhere (these branches fall
inside the cases A-2 and B-2, with $\phi^{'}(0) = 1/\mu$ and
coulombian asymptotic behaviour). The remaining branch of $z(y)$
ranges in the interval $1/\mu<y<\infty$ and exhibits a minimum at
$y=(\mu\sqrt{2\alpha-1})^{-1}$. Consequently there are two
additional solutions $\phi^{'}(r)$ defined only inside the
interval $0 \leq r \leq \sqrt{2\mu\Lambda}
(2-2\alpha)^{(1-\alpha)/2}(2\alpha-1)^{(2\alpha-1)/4}$.}.
Consequently, we must exclude the models with these values of the
parameter $\alpha$ and restrict the family to the lagrangian
densities which result from \textit{irrational values of $\alpha$
or rational values which are irreducible ratios of an odd and an
even natural} (always within the range $1/2 \leq \alpha < 1$). In
these cases the lagrangian densities are defined only for $X >
-1/\mu^{2}$ and behave as $X/2$ around $X=0$, corresponding to
class-2 field theories and cases A-2 and B-1. There are now
unique, stable and finite-energy SSS solutions for each model,
which are defined everywhere and fall inside the cases A-2 and B-2
(with the maximum field strength $\phi^{'}(0)=1/\mu$, and
coulombian behaviour at infinity). In the limits $\mu \rightarrow
0$ or $\alpha \rightarrow 1$ Eq.(\ref{eq:(3-8)}) reduces to the
D'Alembert lagrangian density. The scalar Born-Infeld model is a
member of this family, corresponding to the frontier value
$\alpha=1/2$.

In calculating the energy of these soliton solutions as a function
of the model parameters we evaluate the integral in
(\ref{eq:(2-37)}), as in the previous example. The final
expression is

\be
\epsilon=\frac{4\sqrt{2}\pi}{3}\frac{\Lambda^{3/2}}{\vert\mu\vert^{1/2}}B\left(\frac{1}{2},\frac{3-\alpha}{2}
\right). \label{eq:(3-9)} \en We see that, as a function of $\mu$
the soliton energy behaves like $\epsilon \sim
\frac{1}{\sqrt{\mu}}$ and diverges as $\mu \rightarrow 0$
(D'Alembert limit), whereas it vanishes in the strong-coupling
limit $\mu \rightarrow \infty$. This energy is not very sensitive
to the exponent $\alpha$ in the range of admissible values and the
behaviour for the whole family is similar to that of the scalar
Born-Infeld model.

\subsection{\large A three-parameter family}

The \underline{third example} is the three-parameter family of
models defined by the lagrangian densities

\be f(X) = \frac{1}{2} \frac{X^{\alpha}}{(1+\mu^{2}X)^{\beta}},
\label{eq:(3-10)} \en where $\alpha$ is chosen as the irreducible
ratio of two positive odd naturals (in order for the lagrangian to
be well defined on both sides of $X=0$). The exponent $\beta$ must
be chosen as a positive irrational number or as the irreducible
ratio of an odd and an even natural numbers. In this way the
lagrangian is defined only for $X>-1/\mu^{2}$, thus avoiding a
non-positive definite character of the energy as well as a
singularity inside the domain of definition \footnote{We emphasize
that, as mentioned in section 2, we regard as acceptable
singularities of the lagrangian density only those lying on the
boundary of the (open and connected) domain of definition. In
fact, one of the motivations in introducing this example is to
show how models with such a kind of singularities can also lead to
physically reasonable results.}. These restrictions lead to a
family of class-2 field theories belonging to the A-2 case, being
examples of the sub-cases whose lagrangians diverge at $X
\rightarrow -1/\mu^{2}$ in the boundary of the domain of
definition. The behaviours of the lagrangians around $X=0$
($r\rightarrow\infty$ for the SSS solutions) belong respectively
to the cases B-1 ($\alpha>1$), B-2 ($\alpha=1$) or B-3
($\alpha<1$), corresponding to asymptotic dampings of the soliton
field strengths which are slower than coulombian, coulombian or
faster than coulombian, respectively (see figure 6).

\begin{figure}[h]
\begin{center}
\includegraphics[width=13cm,height=8.5cm]{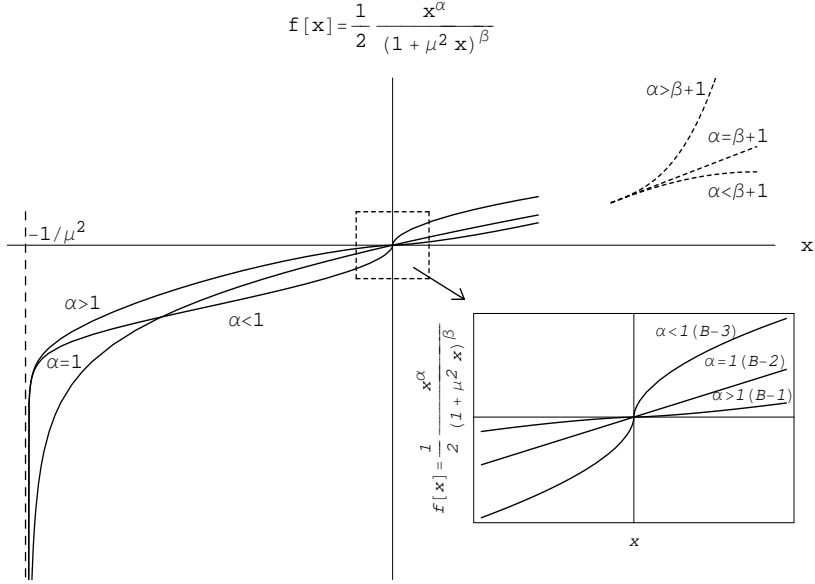}
\caption{\label{fig:epsart} Form of the lagrangian functions
corresponding to models of the family (\ref{eq:(3-10)}) for three
sets of values of the parameter $\alpha \left(\lesseqqgtr
1\right)$. All lagrangians diverge at $X \rightarrow -1/\mu^2$,
which corresponds to the maximum value of the field strength. At
the point $X=0$, which determines the asymptotic behaviour of the
solitons, the three sets of values of $\alpha$ give the three
different behaviours: case B-1 ($\alpha > 0$), case B-2 ($\alpha =
0$) and case B-3 ($\alpha < 0$). The dashed lines show the
behaviour of the lagrangian function at large positive values of
$X$ for the admissible models corresponding to different relations
among the parameters.}
\end{center}
\end{figure}

We also impose the condition $\alpha > \beta + 1/2$, necessary to
ensure the positivity of the energy density for any $X \in
\Omega$. Moreover, the convergence of the integral of energy for
the SSS solutions as $ r \rightarrow \infty$ requires $\alpha <
3/2$, as can be easily verified from the analysis of the field
equation (\ref{eq:(2-4)}) and the integral of energy in this
limit. Let us summarize in the following equations the
restrictions imposed on the parameters of the models
(\ref{eq:(3-10)}) in order to obtain admissible models with
soliton solutions:

\be \frac{1}{2} < \alpha \equiv \frac{odd}{odd} <
\frac{3}{2}\hspace{0.5cm} ; \hspace{0.5cm} \beta \equiv
\frac{odd}{even} \hspace{0.1cm} or \hspace{0.1cm} irrational
\hspace{0.5cm} ; \hspace{0.5cm} 0 < \beta < \alpha - \frac{1}{2} <
1. \label{eq:(3-11)} \en where the terms ``odd" and ``even" are
implicitly understood to apply for natural numbers. The D'Alembert
lagrangian is a limit member of this family obtained as $\alpha
\rightarrow 1$ and $\beta \rightarrow 0$ or as $\alpha \rightarrow
1$ and $\mu \rightarrow 0$.

As results from the analysis of the A-2 cases, near the center the
SSS solutions behave as

\be
\phi^{'}(r \rightarrow 0) \sim \frac{1}{\mu} - \lambda r^\sigma,
\label{eq:(3-12)}
\en
where the exponent $\sigma$ is given by

\be
1 < \sigma = \frac{2}{1+\beta} < 2,
\label{eq:(3-13)}
\en
and $\lambda$ is a positive constant, which is the solution of the equation

\be 2\alpha\mu\lambda + \beta =
\Lambda\mu^{(\alpha+\beta)}(2\lambda)^{\beta+1}. \label{eq:(3-13)}
\en As in the preceding examples, the energy of the soliton
solutions can be explicitly obtained from (\ref{eq:(2-37)}). The
final expression is

\be
\epsilon=\frac{4\sqrt{2}\pi}{3}\frac{\Lambda^{3/2}}{\alpha^{1/2}}\vert\mu\vert^{\frac{2\alpha-3}{2}}
B\left(\frac{3-2\alpha}{4},\frac{\beta+3}{2}\right)
F_1^2\left(\frac{1}{2},\frac{3-2\alpha}{4},\frac{9-2(\alpha-\beta)}{4},\frac{\alpha-\beta}{\alpha}\right),
\label{eq:(3-14)} \en where $B(x,y)$ is again the Euler
beta-function and $F_1^2(a,b,c,z)$ is the hypergeometric function
defined as

\be F_1^2=F(a,b,c,z)=\frac{1}{B(b,c-b)}\int_{0}^{1}
t^{b-1}(1-t)^{c-b-1}(1-tz)^{-a}dt, \hspace{0.1cm}\mathrm{Re}(b) >
0, \hspace{0.1cm}\mathrm{Re}(c) > 0. \label{eq:(3-15)} \en In
figure 7 we have plotted the energy as a function of $\alpha$ with
$\mu$ and $\beta$ as parameters. For a given value of $\mu$ the
energy is rather insensitive to the parameter $\beta$. As $\alpha$
approaches the limit $3/2$ the energy diverges for all values of
$\mu$ and $\beta$. As a function of $\mu$ the energy decreases as
the power $1/ \mu^{(\frac{3}{2}-\alpha)}$.

\begin{figure}[h]
\begin{center}
\includegraphics[width=13cm,height=8.0cm]{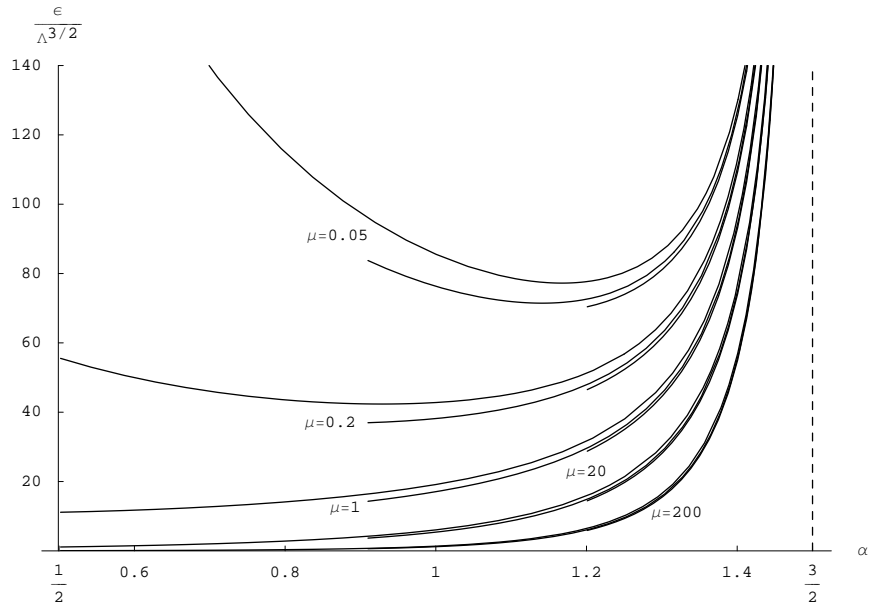}
\caption{\label{fig:epsart} Behaviour of the energy as a function
$\alpha$ for five values of $\mu$ and three of
$\beta(0.01,0.4,0.7)$. Note that the lower branches for each value
of $\mu$ do not cover all the range $1/2<\alpha<3/2$ since the
constraint $\alpha>\beta+1/2$ must be always fulfilled for
admissibility.}
\end{center}
\end{figure}

\subsection{\large Short-ranged solutions}

As a \underline{fourth example} let us look for a family of models
whose SSS solutions are exponentially damped for large $r$
\footnote{As already mentioned, these kinds of theories allow to
describe short-range interactions through the exchange of
self-coupled scalar fields. Their generalizations to the case of
gauge fields, performed in section 5, lead to effective non-linear
lagrangians which also describe short-range interactions and
preserve the explicit gauge-invariance. From this point of view
such models provide alternatives to the usual symmetry breaking
mechanism in the description of weak interactions.}. In obtaining
these models we shall proceed backwards, looking for lagrangians
whose associated field equations have \textit{prescribed} SSS
solutions. In this way we shall look for a family of lagrangian
density functions of the form (\ref{eq:(2-1)}) whose associated
SSS field solutions have the simple exponentially damped form
\footnote{More complex choices of exponentially damped SSS fields
(as, for example, $\phi^{'}(r) = a(r) \exp\left(-\sigma
\frac{r^{n}}{\Lambda^{n/2}}\right)$, where $a(r)$ is assumed to be
a bounded function) may be analyzed in a similar way, but in the
present example the calculations can be performed in terms of
elementary functions. With this choice the SSS field will be a
soliton, but the method works also in obtaining models with
exponentially-damped SSS solutions which are energy-divergent.}

\be \phi^{'}(r,\Lambda) = A \exp \left(-\sigma
\frac{r^{n}}{\Lambda^{n/2}}\right), \label{eq:(3-16)} \en where
$A$, $\sigma$ and $n$ are positive constants determining the
different models within this family \footnote{The constant $A$ is
a parameter of the model and not an integration constant of the
solutions. It plays the role of the maximum field strength and is
shared by all SSS solutions of a given model, but differs for the
various models in the family.}. The constant $\Lambda$ is the
integration constant of the first-integral (\ref{eq:(2-4)}) of the
field equation (whose solution is required to be
(\ref{eq:(3-16)})) and parameterizes all SSS solutions of a given
model. It is explicitly introduced in (\ref{eq:(3-16)}) by
implementing the scale law (\ref{eq:(2-4)bis}). These fields
belong to the cases A-2 (finite field strength at the center) and
B-3 (asymptotic damping faster than coulombian), and the
corresponding models are class-2 field theories.

By eliminating $r$ between (\ref{eq:(3-16)}) and the
first-integral (\ref{eq:(2-4)}) we obtain the form of the first
derivative of the lagrangian density

\be \stackrel{\bullet}{f}(X) = \frac{\Lambda}{r^{2}\phi^{'}(r)} =
\frac{(2\sigma)^{2/n}}{\sqrt{-X}\ln^{2/n}\left(\frac{-A^{2}}{X}\right)},
\label{eq:(3-17)} \en which holds in the interval $-A^{2} < X < 0$
(where the SSS solution (\ref{eq:(3-16)}) is defined) and diverges
at the boundaries. The lagrangian density in this interval is

\be f(X) = 2\sigma^{2/n} \int_{\sqrt{-X}}^{0}
\frac{dy}{\ln^{2/n}(A/y)} = -2A\sigma^{2/n}
\int_{\ln(A/\sqrt{-X})}^{\infty} \frac{e^{-z}}{z^{2/n}} dz.
\label{eq:(3-18)} \en

\begin{figure}[h]
\begin{center}
\includegraphics[width=13cm,height=9cm]{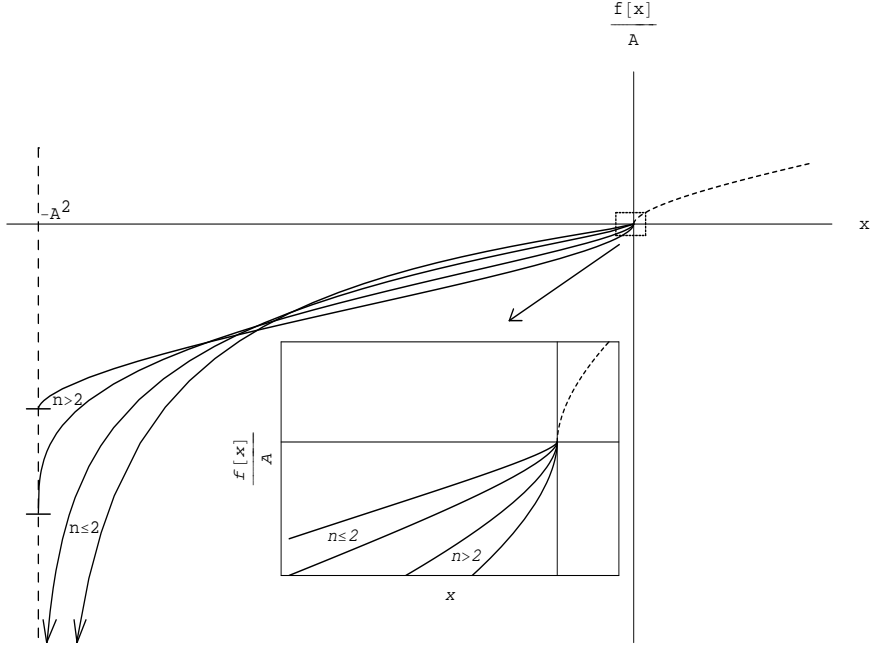}
\caption{\label{fig:epsart} Functional form of the lagrangian
densities for the family (\ref{eq:(3-18)}), exhibiting
short-ranged soliton solutions. The dashed line indicates possible
continuations of the lagrangian density for $X>0$.}
\end{center}
\end{figure}
As easily verified, $f(0) = 0$ for any set of positive values of
the parameters. In the lower boundary of the interval we have $f(X
= -A^{2}) = -2A \sigma^{2/n} \Gamma\left(1-\frac{2}{n}\right)$ for
$n > 2$ (with $\Gamma(t) = \int_0^{\infty}z^{t-1}e^{-z}dz, t>0$
being the usual Euler gamma function) and $f(X \rightarrow -A^{2})
\rightarrow -\infty$ for $n \leq 2$ (see figure 8). This
expression of the lagrangian density could be continued to the
region $X > 0$ by matching (\ref{eq:(3-18)}) to any function
satisfying the admissibility conditions there, but such
continuations do not affect the structure of the solitons, which
is completely determined by the part (\ref{eq:(3-18)}) of the
lagrangian density \footnote{Nevertheless, the requirements of
positivity of the energy and vanishing vacuum energy of the
complete lagrangians are still necessary for the stability of the
solitons (see section 6).}.

We calculate now the energy of these soliton solutions starting by
convenience from Eq.(\ref{eq:(2-7)}) (although formula
(\ref{eq:(2-37)}) would also work). After a partial integration we
obtain

\be \epsilon = \frac{4\pi}{3} \left[-r^{3}
f\left(-\phi^{'2}(r)\right) \vert_{0}^{\infty} + \int_{0}^{\infty}
r^{3}\frac{df}{dr} dr\right]. \label{eq:(3-19)} \en The integrated
part in this equation can be shown to vanish for the prescribed
solutions. In calculating the integral in the second term we use
the first-integral field equation (\ref{eq:(2-4)}) and the
expression of the field (\ref{eq:(3-16)}), which leads to

\be
\frac{df}{dr} = -2\stackrel{\bullet}{f}(X) \phi^{'}(r) \phi^{''}(r) = \frac{-2\Lambda}{r^{2}}\phi^{''}(r) =
2\Lambda^{(2-n)/2} A n \sigma r^{n-3} \exp\left(-\sigma \frac{r^{n}}{\Lambda^{n/2}}\right),
\label{eq:(3-20)}
\en
and the final expression for the energy of the solitons reads

\be \epsilon = \frac{8 \pi A}{3 \sigma^{1/n}}
\Lambda^{3/2}\Gamma\left(\frac{1+n}{n}\right), \label{eq:(3-21)}
\en which is proportional to the maximum amplitude of the field
strength and decreases as the range of the field is reduced. In
figure 9 we have plotted the energy of the unit
maximum-field-strength as a function of $n$ for several values of
the constant $\sigma$. As we see the energy diverges as $n
\rightarrow 0$ and becomes less dependent on the constant $\sigma$
for large values of $n$, approaching asymptotically a fixed value
$\frac{\epsilon}{A\Lambda^{3/2}}=\frac{8\pi}{3}$.

\begin{figure}[h]
\begin{center}
\includegraphics[width=11cm,height=7cm]{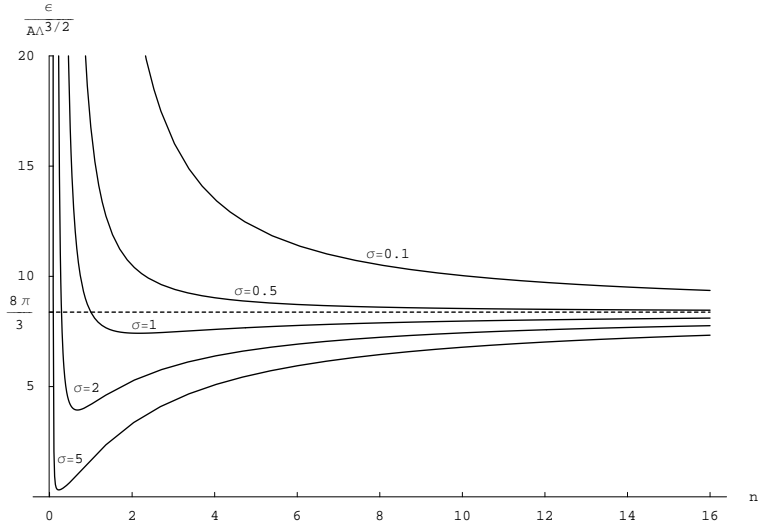}
\caption{\label{fig:epsart} Energy for short-ranged solitons
(\ref{eq:(3-16)}) of unit maximum-field strength as a function of
the exponent $n$, for several values of the parameter $\sigma$
(note that we have plotted $\frac{\epsilon}{A \Lambda^{3/2}}$
instead of $\frac{\epsilon}{\Lambda^{3/2}}$ as in the previous
examples).}
\end{center}
\end{figure}

\section{\large The multicomponent scalar field}

We shall extend the results of section 2 to the case of a set of
scalar fields $\phi_{i}(x) (i=1...N)$. In many cases the covariant
lagrangians including $N$ scalar fields and their first-order
derivatives are constrained by conditions which allow to implement
some internal symmetries. Such conditions manifest themselves in
the structure of the manifold where the field takes its values. A
well-known example is the non-linear sigma model where this
manifold is a Riemann space implementing chiral symmetry
\cite{ddi},\cite{tdlee81}. Here we shall restrict ourselves to the
case where the field manifold is the $N$-dimensional Euclidean
space and the $SO(N)$ invariant lagrangian density depends only on
derivative terms

\be L(\phi_{i}, \partial_{\mu}\phi_{i}) = f\left(\sum_{i=1}^{N}
\partial_{\mu}\phi_{i}\partial^{\mu}\phi_{i}\right),
\label{eq:(4-1)} \en where, as in section 2, $f(X)$ is a given
\textit{continuous, derivable ($C^{1}$ for $X<0$) and
monotonically increasing} ($\frac{df}{dX} > 0, \forall X \neq 0;
\frac{df}{dX} \geq 0, X = 0$) function defined in a open and
connected domain ($\Omega \subseteq \Re$) which includes the
vacuum ($(X=0) \in \Omega$). Besides the fact that these models
are the natural generalizations of the scalar field theories
studied so far, there is another motivation for their analysis.
Indeed, as we shall see in the following sections, a class of
soliton solutions arising in generalized gauge field theories of
some compact semi-simple Lie groups of dimension $N$ reduce to
multiscalar ($N$ components) solitons of some of the models
(\ref{eq:(4-1)}).

The field equations associated to the lagrangians (\ref{eq:(4-1)}) take the form of $N$ local conservation laws

\be
\partial_{\mu}J_{i}^{\mu} = 0,
\label{eq:(4-2)}
\en
where the conserved currents $J_{i}^{\mu}$ are

\be
J_{i}^{\mu} = \stackrel{\bullet}{f}(X)\partial^{\mu} \phi_{i},
\label{eq:(4-3)}
\en
with $X = \sum_{i=1}^{N}\partial_{\alpha}\phi_{i}\cdot\partial^{\alpha}\phi_{i}$. The canonical energy-momentum tensor is

\be
T_{\mu \nu} = 2\stackrel{\bullet}{f}(X)\sum_{i=1}^{N}\partial_{\mu}\phi_{i}\partial_{\nu}\phi_{i} - f(X) \eta_{\mu \nu},
\label{eq:(4-4)}
\en
and the corresponding energy density

\be
\rho(x) = 2\stackrel{\bullet}{f}(X)\sum_{i=1}^{N}\left(\frac{\partial \phi_{i}}{\partial t}\right)^{2} - f(X),
\label{eq:(4-5)}
\en
is positive definite under the same conditions established for the lagrangian function $f(X)$ of the one-component case.

For the SSS solutions $\phi_{i}(r)$, equations (\ref{eq:(4-2)}) have $N$ first-integral field equations of the form

\be
r^{2}\phi_{i}^{'}\stackrel{\bullet}{f}\left(-\sum_{j=1}^{N}\phi_{j}^{'2}\right) = \Lambda_{i},
\label{eq:(4-6)}
\en
where $\phi_{i}^{'} = d\phi_{i}/dr$ and $\Lambda_{i}$ are the integration constants. Now the signs of every component of the scalar field and of the corresponding integration constant are the same, but may differ for different components. In order to solve the system (\ref{eq:(4-6)}) let us introduce the functions $X_{i}(r) = -\phi_{i}^{'2}(r)$, in such a way that $X(r) = \sum_{i=1}^{N}X_{i}(r)$. By squaring and adding Eqs.(\ref{eq:(4-6)}) we obtain

\be
r^{4}X\stackrel{\bullet}{f}^{2}(X) = -\sum_{i=1}^{N}\Lambda_{i}^{2},
\label{eq:(4-7)}
\en
or, equivalently,

\be
r^{2}\sqrt{-X}\stackrel{\bullet}{f}(X) = \Lambda,
\label{eq:(4-8)}
\en
where

\be \Lambda = \sqrt{\sum_{i=1}^{N}\Lambda_{i}^{2}}.
\label{eq:(4-9)} \en Equation (\ref{eq:(4-8)}) has the same form
as the first-integral of the one-component scalar case
(\ref{eq:(2-4)}). Consequently, if the function $f(X)$ is the same
in both cases, we can associate to any SSS solution of the
one-component case, of the form $\phi^{'}(r,\Lambda)$, a set of
sequences of $N$ functions which are SSS solutions of the
multicomponent scalar equations. Such functions take the form

\be \phi_{i}^{'}(r,\Lambda_{j}) =
\frac{\Lambda_{i}}{\Lambda}\phi^{'}(r,\Lambda), \label{eq:(4-10)}
\en and, owing to Eq.(\ref{eq:(4-9)}), there is a one-to-one
correspondence between such sequences and the points of the sphere
of radius $\Lambda$ in the $N$-dimensional Euclidean space
($\Re^{N}$). Obviously, this is a straightforward consequence of
the invariance of the lagrangian (\ref{eq:(4-1)}) under rotations
in the internal space. The constants $\Lambda_{i}$ can now be
identified as the ``source point-charges" associated to the
different components of the SSS field, namely

\be \Lambda_i=\frac{1}{4\pi}\int
d^3x\vec{\nabla}\cdot\left(\stackrel{\bullet}{f}(X)\vec{\nabla}\phi_i\right).
\label{eq:(4-10)bis} \en The field potentials obtained by
integration of (\ref{eq:(4-10)}) read

\be \phi_{i}(r,\Lambda_{j}) =
\frac{\Lambda_{i}}{\Lambda}\phi(r,\Lambda) + \Delta_{i},
\label{eq:(4-11)} \en
where $\Delta_{i}$ are integration constants
and $\Lambda$ is the mean-square scalar charge (\ref{eq:(4-9)}).

If we consider now the energy associated to these SSS solutions we find from (\ref{eq:(4-5)})

\be
\epsilon = -4\pi\int_{0}^{\infty}r^{2}f\left(\sum_{i=1}^{N}X_{i}(r)\right) dr =
-4\pi\int_{0}^{\infty}r^{2}f\left( X(r)\right) dr.
\label{eq:(4-12)}
\en
This is the energy of the \textit{one-component} SSS solution corresponding to the integration constant (\ref{eq:(4-9)}). Thus the set of SSS solutions of the multicomponent scalar field associated to the points of the sphere of radius $\Lambda$ in $\Re^{N}$ is degenerate in energy. Moreover, the search for conditions to be imposed on $f(X)$ for the existence of finite-energy SSS solutions (as well as the admissibility constraints) in the multicomponent case reduces to the analysis of section 2 for the one-component case.

Concerning the conditions for stability of the solutions (\ref{eq:(4-11)}), the analysis of the one-component case can be straightforwardly generalized to the present situation (see subsection 6.3. for details). The final conclusion is that the multicomponent soliton solutions of \textit{admissible} models are also linearly stable against charge-preserving perturbations.

\section{\large Gauge fields}

We shall now extend the analysis developed in the previous sections to \textit{generalized} gauge field theories of compact
semi-simple Lie groups. To start with, we study the simpler case of generalized ($U(1)$-invariant) electromagnetic fields before
proceeding further to the case of non-abelian gauge fields.

\subsection{\large Abelian case}

We define lagrangian densities for \textit{generalized electromagnetic} fields defined as arbitrary functions of the two quadratic field invariants, built from the Maxwell tensor and its dual. Following the conventions of Ref. \cite{landau75}, these
tensors are defined as

\bea
F_{\mu\nu}&=&\partial_{\mu}A_{\nu}-\partial_{\nu}A_{\mu}
\\ \nonumber
F^{*}_{\mu\nu}&=&\frac{1}{2}\varepsilon_{\mu\nu\alpha\beta}F^{\alpha\beta},
\label{eq:(5-0)} \ena where
$\varepsilon^{0123}=-\varepsilon_{0123}=1$. The electric and
magnetic fields are defined as $E^i=-F^{0i}$ and
$H^i=-\frac{1}{2}\varepsilon^{ijk}F_{jk}$ while the quadratic
invariants $X$ and $Y$ are

\bea
X &=& -\frac{1}{2}F_{\mu\nu}F^{\mu\nu} = \vec{E}^{2} - \vec{H}^{2} \\
\nonumber
Y &=& -\frac{1}{2}F_{\mu\nu}F^{*\mu\nu} = 2\vec{E}\cdot\vec{H}.
\label{eq:(5-1)}
\ena

We define the general form of the lagrangian density as

\be
L = \varphi(X,Y),
\label{eq:(5-2)}
\en
where $\varphi$ is a given \textit{continuous and derivable} function on its domain of definition ($\Omega$) of the $X-Y$ plane ($\Re^{2}$). As in the scalar case we assume $\Omega$ to be open and connected and including the vacuum ($(X=0,Y=0) \in \Omega$). As a minimal extension of the assumptions of section 2 concerning the regularity properties of the scalar lagrangian functions, which is necessary for future purposes (see Eq.(\ref{eq:(5-11)}) below), we shall assume $\varphi(X,Y)$ to be of class $C^{1}$ on the line $(X>0, Y=0) \bigcap \Omega$ and $\partial \varphi/\partial X$ to be strictly positive there. By generalizing the definitions of section 2 we shall call ``class-1 field theories" the models defined and regular everywhere ($\Omega \equiv \Re^{2}$) and ``class-2 field theories" those with $\Omega \subset \Re^{2}$. We also require $\varphi(X,Y)$ to be symmetric in the second argument, in order to implement parity invariance

\be \varphi(X,Y) = \varphi(X,-Y). \label{eq:(5-2)bis} \en
This
implies that the odd partial derivatives of $\varphi(X,Y)$ with
respect to $Y$ must vanish on $Y=0$. The Maxwell lagrangian
density corresponds to $\varphi(X,Y)=\frac{X}{8\pi}$, while the
Born-Infeld electrodynamics is given by the lagrangian density

\be L_{B-I} = \varphi_{B-I}(X,Y) = \frac{1 - \sqrt{1 - \mu^{2}X -
\frac{\mu^{4}}{4}Y^{2}}}{4\pi\mu^{2}}, \label{eq:(5-3)} \en
where
$\frac{1}{\mu}$ is the maximum field strength, attained at the
center of the solution. The lagrangian (\ref{eq:(5-3)}) reduces to
the Maxwell one in the limit $\mu \rightarrow 0$.

The symmetric (gauge-invariant) energy-momentum tensor obtained from the lagrangian density (\ref{eq:(5-2)}) takes the form

\be
T^{s}_{\mu\nu} = 2\left(\frac{\partial \varphi}{\partial X}F_{\mu\alpha}F^{\alpha}_{\nu} + \frac{\partial \varphi}{\partial Y} F_{\mu\alpha}F^{*\alpha}_{\nu}\right) - \varphi \eta_{\mu\nu} =
\newline
2\frac{\partial \varphi}{\partial X}F_{\mu\alpha}F^{\alpha}_{\nu} + \left( Y\frac{\partial \varphi}{\partial Y} - \varphi\right) \eta_{\mu\nu},
\label{eq:(5-4)}
\en
and the associated energy density is

\be
\rho^{s} = T^{s}_{00} = 2\frac{\partial \varphi}{\partial X}\vec{E}^{2} + 2\frac{\partial \varphi}{\partial Y}\vec{E}\cdot
\vec{H} - \varphi(X,Y) = 2X\frac{\partial \varphi}{\partial X} - \varphi(X,Y) + Y\frac{\partial \varphi}{\partial Y} +
2\frac{\partial \varphi}{\partial X}\vec{H}^{2}.
\label{eq:(5-5)}
\en
We assume the symmetric energy-momentum tensor (\ref{eq:(5-4)}) to give the correct space-time energy density distribution. Let us analyze the conditions for the positivity of the energy density of any field configuration. The inspection of Eq.(\ref{eq:(5-5)}), together with the requirement of vanishing of the vacuum energy, lead to the set of \textit{necessary} conditions

\be
\varphi(0,0) = 0 \hspace{0.5cm};\hspace{0.5cm} \varphi(X,0) < 0 \hspace{0.3cm} \forall (X < 0,Y=0) \in \Omega \hspace{0.5cm}; \hspace{0.5cm}\frac{\partial \varphi}{\partial X} > 0 \hspace{0.3cm} \forall (X,Y) \in \Omega,
\label{eq:(5-6)}
\en
to be satisfied by the lagrangian densities. However, it is possible to obtain a minimal set of \textit{necessary and sufficient} conditions of admissibility for a satisfactory energetic behaviour. Solving Eqs.(\ref{eq:(5-0)}) for the fields we obtain

\bea
E^{2} &=& \frac{1}{2} \left(\sqrt{X^{2} + \frac{Y^{2}}{\cos^{2}(\vartheta)}} + X\right) \geq \frac{1}{2} \left(\sqrt{X^{2} + Y^{2}} + X\right)  \\
\nonumber
H^{2} &=& \frac{1}{2} \left(\sqrt{X^{2} + \frac{Y^{2}}{\cos^{2}(\vartheta)}} - X\right) \geq \frac{1}{2} \left(\sqrt{X^{2} + Y^{2}} - X\right),
\label{eq:(5-6)bis}
\ena where $\vartheta$ is the angle between $\vec{E}$ and $\vec{H}$. From these expressions the energy density can be written as

\be \rho^{s} = \frac{\partial \varphi}{\partial
X}\left(\sqrt{X^{2}+\frac{Y^{2}}{\cos^{2}(\vartheta)}} + X\right)
+ Y\frac{\partial \varphi}{\partial Y} - \varphi(X,Y).
\label{eq:(5-6)ter} \en
Consequently, the requirement of the
positive definite character of the energy leads to the minimal
\textit{necessary and sufficient} condition

\be
\rho^{s} \geq \left(\sqrt{X^{2}+Y^{2}} + X\right) \frac{\partial \varphi}{\partial X} +
Y\frac{\partial \varphi}{\partial Y} - \varphi(X,Y) \geq 0,
\label{eq:(5-7)}
\en
to be satisfied in the entire domain of definition ($\Omega$). Generalizing the criteria of section 2, we only regard as ``admissible" those models whose lagrangian densities satisfy the condition (\ref{eq:(5-7)}), aside from the vanishing of the vacuum energy and the regularity and parity-invariance conditions stated above \footnote{Obviously, excepting the vanishing of the vacuum energy, the remaining conditions in (\ref{eq:(5-6)}) are consequences of (\ref{eq:(5-7)}).}. Consequently, the admissible lagrangians must be solutions of the first-order linear inhomogeneous partial differential equation

\be
\left(\sqrt{X^{2}+Y^{2}} + X\right) \frac{\partial \varphi}{\partial X}+ Y\frac{\partial \varphi}{\partial Y} -
\varphi(X,Y) = \Psi(X,Y),
\label{eq:(5-7)bis}
\en
where $\Psi(X,Y)$ is any function being positive definite in $\Omega$ and vanishing in vacuum ($\Psi(0,0)=0$). Such solutions must also satisfy the regularity and parity-invariance requirements, as supplementary conditions.

Let us now consider the trace of the symmetric energy-momentum
tensor

\be T^{s} = 4\left[\frac{\partial \varphi}{\partial X}(\vec{E}^{2}
- \vec{H}^{2}) + 2\frac{\partial \varphi}{\partial Y}
\vec{E}\cdot\vec{H} - \varphi(X,Y)\right]  =
4\left[X\frac{\partial \varphi}{\partial X} + Y\frac{\partial
\varphi}{\partial Y} - \varphi(X,Y)\right]. \label{eq:(5-8)} \en
From the last expression we see that the sub-class of models with
traceless symmetric energy-momentum tensors is given by the
lagrangian densities $\varphi(X,Y)$ which are solutions of the
first-order linear homogeneous partial differential equation

\be
X\frac{\partial \varphi}{\partial X} + Y\frac{\partial \varphi}{\partial Y} - \varphi(X,Y) = 0.
\label{eq:(5-8)bis}
\en
The general solution of this equation is the family of all conic surfaces in the ($X,Y,\varphi$)-space having the origin as a vertex. Clearly the set of planes of the form $\varphi=aX+bY$ ($a$ and $b$ being constants) are particular solutions of this equation (which violate parity invariance if $b\neq0$). The simple case $b = 0$ with $a = \frac{1}{8\pi}$ corresponds to the Maxwell theory.

The field equations obtained from the lagrangian (\ref{eq:(5-2)}) are

\be
\partial_{\mu}\left( \frac{\partial \varphi}{\partial X}F^{\mu\nu} + \frac{\partial \varphi}{\partial Y} F^{*\mu\nu}\right) =0,
\label{eq:(5-9)}
\en
to which we add the Bianchi identities

\be
\partial_{\mu} F^{*\mu\nu} = 0.
\label{eq:(5-9)ter}
\en
In terms of the fields these equations read

\bea
&\vec{\nabla}\cdot&\left(\frac{\partial \varphi}{\partial X} \vec{E} + \frac{\partial \varphi}{\partial Y} \vec{H}\right) = 0
\nonumber \\
&-\frac{\partial}{\partial t}&\left(\frac{\partial \varphi}{\partial X} \vec{E} + \frac{\partial \varphi}{\partial Y} \vec{H}\right) + \vec{\nabla} \times \left(\frac{\partial \varphi}{\partial X} \vec{H} - \frac{\partial \varphi}{\partial Y}
\vec{E} \right) = 0,
\label{eq:(5-9)bis}
\ena
and

\bea
\vec{\nabla} \times \vec{E} &=&  -\frac{\partial \vec{H}}{\partial t}
\nonumber \\
\vec{\nabla}\cdot\vec{H} &=& 0.
\label{eq:(5-9)q}
\ena
For electrostatic fields we have $Y=0$ and in the ESS case these equations can be written in terms of the electrostatic potential $A^{0}(r)$ ($\vec{A}=0, \vec{E}(r) = -\vec{\nabla}A^{0}(r)$) in such a way that the first group of equations in (\ref{eq:(5-9)bis}) leads to the first-integral \footnote{The remaining three equations in (\ref{eq:(5-9)bis}) are identically satisfied for arbitrary electrostatic fields, owing to Eq.(\ref{eq:(5-2)bis}). On the other hand, the set of equations (\ref{eq:(5-9)q}) are trivially satisfied by the ESS solutions.}

\be
r^{2}\frac{dA^{0}}{dr} \frac{\partial \varphi}{\partial X}(X,Y=0) = q
\label{eq:(5-10)}
\en
where now $X=\left(\frac{dA^{0}}{dr}\right)^{2}$ and $q$ is an integration constant. Using the identification $\phi(r,\Lambda) \equiv A^{0}(r,q=\Lambda)$ this equation coincides with the first-integral (\ref{eq:(2-4)}) for the SSS solutions of a scalar field model with a lagrangian density defined by

\be
L_{scalar} = f(\partial_{\mu}\phi\cdot\partial^{\mu}\phi) \equiv f(X) = - \varphi(-X,Y=0),
\label{eq:(5-11)}
\en
which leads to

\be
\stackrel{\bullet}{f}(X) =  \frac{\partial\varphi}{\partial X}(-X,Y=0).
\label{eq:(5-12)}
\en
Conversely we can associate to each scalar model defined by a lagrangian density $f(X)$, a family of electromagnetic field models defined by lagrangian densities $\varphi(X,Y)$ satisfying Eq.(\ref{eq:(5-11)}) as well as the admissibility (\ref{eq:(5-7)}) and stability constraints (see Eq.(\ref{eq:(5-18)}) below). The ESS field solutions of all electromagnetic generalizations ($\vert\vec{E(r,q)}\vert$) have the same form, as functions of $r$, as the SSS field solutions ($\phi^{'}(r,\Lambda)$) of the original scalar model, $q$ and $\Lambda$ being the integration constants, which should be identified as the electric and scalar point-like charges associated to the solution, respectively. Indeed, in the generalized electromagnetic case the definition of the electric charge associated to a given field is

\be
\frac{1}{4\pi} \int d^{3}r \vec{\nabla}\cdot\left(\frac{\partial \varphi}{\partial X} \vec{E} +
\frac{\partial \varphi}{\partial Y} \vec{H}\right),
\label{eq:(5-12)bis}
\en
which now is conserved as a consequence of the field equations. Substituting in this equation the ESS field coming from the solution of (\ref{eq:(5-10)}) we obtain $q$ as the value of its total electric charge

\be
\frac{1}{4\pi} \vec{\nabla}\cdot\left[\frac{\partial \varphi}{\partial X} \vec{E}\right] = q \delta_{3}(\vec{r}).
\label{eq:(5-12)ter}
\en
As pointed out in Ref.\cite{B-I34} for the Born-Infeld model this charge can be interpreted as a source point-like charge at the center of the ESS solution or, alternatively, as a continuous charge-density distribution associated with the field and given by

\be
\frac{1}{4\pi}\frac{\partial \varphi}{\partial X}(X=0,Y=0)\vec{\nabla}\cdot\vec{E}.
\label{eq:(5-12)q}
\en
This interpretation, as already discussed for scalar models, requires the function $r^{2}E(r)$ to vanish at the origin (this condition is always fulfilled for the finite-energy ESS solutions) and the field $E(r)$ to be asymptotically coulombian (B-2 case models).

In calculating the total energy of these electrostatic central
fields from the energy density (\ref{eq:(5-5)}) we are lead to

\bea \epsilon_{e}(q) &=&
8\pi\int_{0}^{\infty}r^{2}\frac{\partial\varphi}{\partial X}
\left[X = \vec{E}^{2}(r,q), Y = 0\right] \vec{E}^{2}(r,q) dr -
\nonumber \\ &-& 4\pi\int_{0}^{\infty}r^{2}\varphi\left[X =
\vec{E}^{2}(r,q), Y = 0\right] dr, \label{eq:(5-13)} \ena (the
index \textbf{e} stands for electric field). The energy associated
with the corresponding SSS scalar field solutions, obtained from
Eqs.(\ref{eq:(2-7)}) and (\ref{eq:(5-11)}) reads

\be
\epsilon_{s}(\Lambda) = 4\pi \int_{0}^{\infty}r^{2} \varphi\left[\vec{E}^{2}(r,\Lambda),0\right] dr,
\label{eq:(5-14)}
\en
(the index \textbf{s} stands for scalar field). If the total energy (\ref{eq:(5-14)}) associated to the scalar field is finite, so is
$\epsilon_{e}(q)$. Indeed, using Eq.(\ref{eq:(5-10)}) this energy becomes

\be
\epsilon_{e}(q) = 8\pi q\left[A^{0}(\infty,q) - A^{0}(0,q)\right] - \epsilon_{s}(q),
\label{eq:(5-15)}
\en
which, owing to (\ref{eq:(2-7)ter}), is related with the scalar soliton energy through

\be \epsilon_{e}(q) = 2\epsilon_{s}(q), \label{eq:(5-16)} \en and
must be also finite. Thus the energy of a ESS solution is twofold
the energy of the corresponding SSS soliton when the integration
constants take the same value ($\Lambda=q$). Equivalently,
Eq.(\ref{eq:(2-7)}), which gives the scaling of energies, leads to
the following relation between the integration constants (the
electric ($q$) and the scalar ($\Lambda$) charges) of an
electrostatic soliton and the associated scalar soliton of
equal-energy

\be
q = (2)^{2/3}\Lambda.
\label{eq:(5-17)}
\en
We then conclude that the ESS solutions for the families of electromagnetic models which generalize (through Eq.(\ref{eq:(5-11)})) the different classes of scalar models with soliton solutions, have the same functional forms as the corresponding SSS scalar solutions and are also of finite-energy. Moreover, the classification of the admissible models with soliton solutions in the scalar case according to the central and asymptotic behaviours of the fields, immediately induces, through Eq.(\ref{eq:(5-11)}), a similar classification of the finite-energy ESS solutions in the electromagnetic case.

Let us point out an immediate consequence of this analysis (which is a corollary of the non-existence theorems established in Refs. \cite{deser76},\cite{coleman77}): \emph{there are not ESS soliton solutions for admissible generalized electromagnetic field theories with traceless energy-momentum tensor}. Indeed, as mentioned above the lagrangian densities of such theories (see Eq.(\ref{eq:(5-8)})) are given by conic surfaces in the $(X,Y,\varphi)$ space and the associated scalar field lagrangian densities $f(X) = -\varphi(-X,Y=0)$ are straight lines in the $(X,f)$ plane for $X<0$. Consequently the associated SSS solutions are \textit{coulombian} in form as well as energy-divergent, and so are the ESS solutions of these generalized electromagnetic models.

Although finite-energy SSS solutions of admissible scalar models are always linearly stable against charge-preserving perturbations, this is not so for the finite-energy ESS solutions of admissible generalized electromagnetic theories. Indeed, the analysis of the linear stability of the electrostatic solitons leads to a generalization of the criteria obtained in the scalar case (see section 6 for details). As a result of this analysis, the electrostatic finite-energy central field solutions of admissible generalized electromagnetic field models, with their lagrangian densities $\varphi(X,Y)$ satisfying the supplementary condition

\be \frac{\partial \varphi}{\partial X}-2X\frac{\partial^2
\varphi}{\partial Y^2} > 0, \label{eq:(5-18)} \en in the entire
domain of existence of the ESS solutions in the plane $(X,Y=0)$,
can be shown to be local minima of the energy functional against
small charge-preserving perturbations. Consequently,
\textit{admissibility}, \textit{finite-energy} condition of the
ESS solutions aside from Eq.(\ref{eq:(5-18)}) are
\textit{necessary and sufficient conditions} for linear static
stability. Moreover, the linear analysis of dynamics of the small
perturbations of the soliton solutions performed for the scalar
models can be generalized to the associated families of
electromagnetic models which satisfy (\ref{eq:(5-18)}) (see
section 6).

Finally, the conditions for univoque and everywhere defined ESS
solutions are straightforwardly deduced from those of the scalar
case, aside from Eq.(\ref{eq:(5-11)}).

\subsection{\large Non-abelian case}

The results obtained for generalized electromagnetic fields can be
extended to non-abelian generalized gauge field theories of
compact semi-simple Lie groups of dimension $N$. As usually, in
this case the tensor field strength components in the algebra and
their duals are defined from the gauge fields $A_{a\mu}$ and the
structure constants $C_{abc}$ as

\bea
F_{a\mu\nu}&=&\partial_{\mu}A_{a\nu}-\partial_{\nu}A_{a\mu} - g\sum_{bc} C_{abc} A_{b\mu} A_{c\nu}
\nonumber \\
F^{*}_{a\mu\nu}&=&\frac{1}{2}\varepsilon_{\mu\nu\alpha\beta}F_{a}^{\alpha\beta},
\label{eq:(5-19)} \ena whose components define the fields
$\vec{E}_{a}$, $\vec{H}_{a}$ in the usual form. In order to
introduce the lagrangian densities governing the generalized
dynamics of these fields we must define pertinent field
invariants. However there is now an ambiguity in the calculation
of the traces over the group indices, leading to different
possibilities in the definition of these invariants. Although at
this regard different prescriptions have been introduced, mainly
in the context of string theory, where B-I-like actions arise as a
low-energy effective field limit \cite{soliton},\cite{abou87},
here we shall restrict our analysis to the case of actions built
from the two simplest first-order field invariants, defined from
the ordinary prescription for the calculation of the traces as

\bea
X &=& -\frac{1}{2} \sum_{a}(F_{a\mu\nu}F_{a}^{\mu\nu}) = \sum_{a}\left(\vec{E}_{a}^{2} - \vec{H}_{a}^{2}\right)
\nonumber \\
Y &=& -\frac{1}{2} \sum_{a}(F_{a\mu\nu}F_{a}^{*\mu\nu}) = 2\sum_{a}\left(\vec{E}_{a}\cdot\vec{H}_{a}\right),
\label{eq:(5-20)}
\ena
where $1 \leq a \leq N$. The generalized lagrangian density is now assumed to be a given function $\varphi(X,Y)$ which (again for parity invariance) must be symmetric in the second argument ($\varphi(X,Y) = \varphi(X,-Y)$) and satisfy the same admissibility constraints of definition, continuity and derivability (as well as the distinction between class-1 and class-2 field theories) as in the electromagnetic case.

The associated field equations read now

\be
\sum_{c}D_{ac\mu} \left[\frac{\partial \varphi}{\partial X} F_{c}^{\mu\nu} +
\frac{\partial \varphi}{\partial Y} F_{c}^{*\mu\nu}\right] = 0,
\label{eq:(5-21)}
\en
where $D_{ac\mu} \equiv \delta_{ac}\partial_{\mu} - g\sum_{b} C_{abc} A_{b\mu}$ is the gauge-covariant derivative.

The symmetric energy-momentum tensor is

\be
T^{s}_{\mu\nu} = 2\sum_{a}F_{a \mu\alpha}\left(\frac{\partial \varphi}{\partial X}F_{a \nu}^{\alpha} + \frac{\partial \varphi}{\partial Y}F_{a \nu}^{*\alpha}\right)- \varphi \eta_{\mu\nu},
\label{eq:(5-22)}
\en
and the energy density takes the form

\be \rho^{s} = T^{s}_{00} = 2\frac{\partial \varphi}{\partial
X}\sum_{a}\vec{E}_{a}^{2} + 2\frac{\partial \varphi}{\partial Y}
\sum_{a}\vec{E}_{a}\cdot\vec{H}_{a} - \varphi(X,Y).
\label{eq:(5-23)} \en The admissibility conditions to be imposed
on the lagrangian density, for the energy functional to be
positive definite and vanishing in vacuum, take the same form as
in the electromagnetic case (see Eqs.(\ref{eq:(5-6)}) and
(\ref{eq:(5-7)})). Also the trace of the energy-momentum tensor
has the same expression (see the last member of
Eq.(\ref{eq:(5-8)})), and vanishes under the same conditions
(\ref{eq:(5-8)bis}). The subclass of models with non-vanishing
trace energy-momentum tensor breaks the scale invariance and thus
circumvent the non-existence theorems
\cite{deser76},\cite{coleman77} allowing for soliton-like
solutions.

Let us consider the ESS solutions of these models. We consider fields of the form

\be
\vec{E}_{a}(\vec{r}) = -\vec{\nabla}\left(A^{0}_{a}(r)\right) = -A^{'0}_{a}(r)\frac{\vec{r}}{r}\hspace{0.5cm} ; \hspace{0.5cm} \vec{H}_{a} = 0,
\label{eq:(5-24)}
\en
where the functions $A^{0}_{a}(r)$ are the time-like components of the gauge potential in the Lorentz gauge ($\vec{A}_{a} = 0$) and $A^{'0}_{a} = \frac{dA^{0}_{a}}{dr}$. When replaced in the field equations (\ref{eq:(5-21)}) we are lead to

\bea \vec{\nabla}\cdot\left(\frac{\partial \varphi}{\partial X}
\vec{E}_{a}\right) = -\vec{\nabla} &\cdot& \left(\frac{\partial
\varphi}{\partial X}\vec{\nabla}A^{0}_{a}(r)\right) =
-\vec{\nabla}\cdot\left(\frac{\partial \varphi}{\partial
X}A^{'0}_{a}(r)\frac{\vec{r}}{r}\right) = 0
\nonumber \\
\frac{\partial \varphi}{\partial X}\sum_{bc} C_{abc}A^{0}_{b}(r)
\vec{E}_{c} &=& -\frac{\partial \varphi}{\partial X}\sum_{bc}
C_{abc}A^{0}_{b}(r) \vec{\nabla}A^{0}_{c}(r) = \\
\nonumber &=&-\frac{\partial \varphi}{\partial X}\sum_{bc}
C_{abc}A^{0}_{b}(r) A^{'0}_{c}(r)\frac{\vec{r}}{r} = 0,
\label{eq:(5-25)} \ena where $X = \sum_{a} \vec{E}^{2}_{a}$. The
first group of equations has a set of first-integrals of the form

\be
r^{2}\frac{\partial \varphi}{\partial X}A^{'0}_{a}(r) = Q_{a},
\label{eq:(5-25)bis}
\en
where the $Q_{a}$ are integration constants which will be identified below as ``source color charges". With the identification $\phi_{a}(r) \equiv A^{0}_{a}(r)$ and $\Lambda_{a} \equiv Q_{a}$, these equations coincide with the field equations (\ref{eq:(4-6)}) for a multicomponent SSS scalar field theory whose lagrangian density is given by

\be
L = f\left(\sum_{a}\partial_{\mu}\phi_{a}\cdot\partial^{\mu}\phi_{a}\right) \equiv f(X) = -\varphi(-X,Y=0).
\label{eq:(5-26)}
\en
Thus the solutions of equations (\ref{eq:(5-25)bis}) are obtained from Eq.(\ref{eq:(4-10)}) as

\be
\vert\vec{E}_{a}(r,Q_{b})\vert = A^{'0}_{a}(r,Q_{b}) = \frac{Q_{a}}{Q}\phi^{'}(r,Q)\hspace{0.5cm}; \hspace{0.5cm} \vec{H_{a}} = 0,
\label{eq:(5-26bis)}
\en
where $Q = \sqrt{\sum_{a}Q_{a}^{2}}$ is the mean-square color charge and $\phi^{'}(r,Q)$ is the SSS solution of the associated one-component scalar model defined by a lagrangian density of the form (\ref{eq:(5-26)}). These equations can be integrated once, leading to

\be
A^{0}_{a}(r,Q_{b}) = \frac{Q_{a}}{Q}\phi(r,Q) + \chi_{a},
\label{eq:(5-26ter)}
\en
where $\chi_{a}$ are integration constants. These functions must also satisfy the second set of equations (\ref{eq:(5-25)}). Owing to the antisymmetry of the structure constants these equations lead to the supplementary restriction

\be
\chi_{a} = \frac{Q_{a}}{Q}\chi,
\label{eq:(5-27)}
\en
and the final solution of (\ref{eq:(5-25)}) is

\be
A^{0}_{a}(r,Q_{b}) = \frac{Q_{a}}{Q}\left(\phi(r,Q) + \chi\right),
\label{eq:(5-28)}
\en
where $\chi$ is an arbitrary constant, being now the same for all components of the potential. In terms of the fields the final solution is still given by Eq.(\ref{eq:(5-26bis)}). The integration constants $Q_{a}$ in this solution must be interpreted as source point-like color charges associated with the different components of the gauge field. Indeed, color charges are defined in general as

\be
Q_{a} = \frac{1}{4\pi}\int d_{3} \vec{r} \vec{\nabla}\cdot\left(\frac{\partial \varphi}{\partial X}\vec{E}_{a} +
\frac{\partial \varphi}{\partial Y} \vec{H}_{a}\right),
\label{eq:(5-29)}
\en
which, owing to the field equations (\ref{eq:(5-21)}), include now external source charges and charges carried by the field itself. The latter ones come from the integration of the term

\be
g\sum_{bc} C_{abc} A_{b\mu}\left[\frac{\partial \varphi}{\partial X} F_{c}^{\mu\nu} +
\frac{\partial \varphi}{\partial Y} F_{c}^{*\mu\nu}\right],
\label{eq:(5-30)}
\en
which vanishes for the ESS solutions. On the other hand the former are associated to Dirac distributions of weight $4\pi Q_{a}$, as can be easily seen from the substitution of Eqs.(\ref{eq:(5-25)bis}) in the first set of Eqs.(\ref{eq:(5-25)}).

The calculation of the energy associated to these solutions proceeds in the same way as in the electromagnetic case. The integration of Eq.(\ref{eq:(5-23)}) gives

\bea
\epsilon_{gf}(\Lambda_{a}) &=& 8\pi\int_{0}^{\infty}r^{2}\frac{\partial\varphi}{\partial X} \left[X = \sum_{b} \vec{E}_{b}^{2}(r,\Lambda_a), Y = 0\right] \sum_{b}\vec{E}_{b}^{2}(r,\Lambda_a)dr -
\nonumber\\
&-& 4\pi\int_{0}^{\infty}r^{2}\varphi\left[X = \sum_{b}
\vec{E}_{b}^{2}(r,\Lambda_a), Y = 0\right] dr, \label{eq:(5-31)}
\ena where the index \textbf{gf} stands for gauge field. From this
expression it is now straightforward to show that, as in the
corresponding multiscalar case, this energy is finite if the
energy of the associated scalar solitons is finite, and  depends
on the charges only through the constant $Q$, having the same kind
of degeneration on spheres of radius $Q$ in the $N$-dimensional
color-charge space. The relation between the finite energies of
the solitons with equal mean-square charges in the gauge models
and in the associated multiscalar models is given by the same
equation (\ref{eq:(5-16)}) relating the finite energies in the
cases of the abelian models and their associated scalar models.

Concerning the stability of the gauge field solitons we shall show in the next section that the finite-energy ESS solutions of admissible generalized non-abelian models are linearly stable if (and only if) the lagrangian density functions satisfy the same criterion (\ref{eq:(5-18)}) obtained in the abelian case.

Let us summarize the main conclusions of this section: The set of
generalized gauge field theories of compact semi-simple Lie
groups, whose lagrangian densities are functions $\varphi(X,Y)$ of
the field invariants (\ref{eq:(5-20)}), satisfying the
\textit{admissibility conditions} and the stability criterion
(\ref{eq:(5-18)}), and supporting finite-energy ESS
non-topological soliton solutions, can be split in equivalence
classes. Two models belong to the same class if their respective
lagrangian densities satisfy the condition $\varphi_{1}(X,0) =
\varphi_{2}(X,0)$. The forms of the ESS soliton solutions and
their energies coincide for all models belonging to the same
class. There is a one-to-one correspondence, given by
Eq.(\ref{eq:(5-11)}), between the set of these classes and the set
of \textit{admissible} scalar field models defined by
Eq.(\ref{eq:(2-1)}) and supporting finite-energy SSS
non-topological solitons. The forms and energies of the gauge
solitons are obtained from those of the corresponding scalar
solitons through Eqs.(\ref{eq:(5-26bis)}), (\ref{eq:(5-26ter)})
and (\ref{eq:(5-16)}). The analysis and classification of scalar
solitons performed in section 2 can be immediately generalized to
gauge solitons through this correspondence. Furthermore, the
explicit examples supporting scalar solitons, introduced in
section 3, can also be extended to the gauge field case simply by
including the $Y$ invariant in such a way that the admissibility
(\ref{eq:(5-7)}) and stability (\ref{eq:(5-18)}) constraints be
fulfilled by the extended models.

\section{\large Stability}

Let us now analyze the stability of the finite-energy solutions of
the different models introduced so far. We make a distinction
between ``strong" stability, defined as the ability of a soliton
to maintain its identity under any perturbation or in closed
many-soliton configurations \cite{scott73},\cite{Zabuski65}, and
``weak" stability, identified with the usual linear stability
under small perturbations. Rigorous analysis of stability in the
strong sense has been performed for a few field theories in
one-space dimension which exhibit conserved discrete topological
charges associated with the soliton solutions. In three-space
dimensions similar topological conservation laws are responsible
for the stability of the 't Hooft-Polyakov monopole solution
\cite{t-p74} or the chiral soliton solution of Deser et al.
\cite{ddi}. But satisfactory \emph{general methods} for the
analysis of interactions between non-topological solitons and
strong external fields in three space dimensions are still lacking
and only numerical analysis of the evolution of the solutions can
give some insight on this issue for most models. In our context, a
tentative approach to this question has been developed by
Chernitsky for the Born-Infeld model \cite{chern98}. It is based
on the use of the discontinuity of the field strength at the
center of \textit{static} B-I solitons as a marker of the presence
and location of the \textit{dynamic} soliton evolving in
interaction with strong external fields, or in many-soliton
configurations. Since all SSS soliton solutions of the models
considered here exhibit similar central field singularities, this
procedure might be extended to these cases, but such an extension
lies beyond the scope of the present work.

In the case of interactions between solitons and weak external
fields (or for widely separated soliton configurations) linear
stability \textit{ensures} the identity preservation of the
solitons and becomes a basic condition for the consistency of the
\textit{low-energy} analysis. The results of this analysis may be
interpreted in terms of particle-field (or particle-particle)
force laws and describe the radiative behaviour in these processes
\cite{chern98}.

\subsection{\large Static stability of scalar solitons.}

We shall begin with the study of the \textit{static} linear stability for the soliton solutions of the scalar models of section 2, by analyzing the behaviour of their energy, which must be a minimum against appropriate small perturbations \footnote{As emphasized in Ref. \cite{jackiw80} this criterion is a sufficient condition for stability, but is not a necessary one. Here we shall not consider the problem of stability of solutions which do not correspond to minima of the energy, a complicated task which deserves a study in itself.}. We shall show that, for these models, the conditions of admissibility guarantee this kind of stability for all finite-energy SSS solutions. We start with the SSS potential $\phi(r)$ and introduce a set of small \underline{static} perturbations $\delta\phi(\vec{r})$, finite and regular (as well as their first order spatial derivatives) everywhere. We also require the perturbation to leave unchanged the scalar charge associated to the solution \footnote{This condition is essential for the energy of the soliton to be a minimum. Indeed, without such a condition the perturbation does not necessarily lead to an increase of the energy of the soliton (as can be easily seen by differentiating Eq.(\ref{eq:(2-7)}) with respect to small variations of $\Lambda$) and the perturbed soliton might evolve towards less energetic states.}. To first order in the perturbations the modification of this scalar charge, obtained from Eq.(\ref{eq:(2-4)ter}), reads

\bea \Delta \Lambda = \frac{1}{4\pi} \int d_{3}\vec{r}
\vec{\nabla}\cdot\left[\stackrel{\bullet}{f}(X_{0}) (\vec{\nabla}
\delta\phi) -
2\stackrel{\bullet\bullet}{f}(X_{0})\left(\vec{\nabla}\phi\cdot\vec{\nabla}\delta\phi\right)
\vec{\nabla}\phi \right] = 0, \label{eq:(6-1)} \ena where now
$X_{0} = -(\vec{\nabla}(\phi))^{2} = -\phi^{'2}(r)$. The condition
$\Delta \Lambda = 0$ imposes restrictions on the behaviour of the
admissible perturbations at $r=0$ and as $r\rightarrow\infty$. In
particular, $\delta\phi(\vec{r})$ must satisfy

\be \lim_{r \rightarrow \infty}
\frac{\delta\phi(\vec{r})}{\phi(\infty) - \phi(r)} = 0.
\label{eq:(6-1)bis} \en In this manner the perturbed fields remain
inside the space of functions defined by the prescribed boundary
conditions (on $S_{\infty}$ in this case) which determine uniquely
the solution associated to a given value of the charge. At the
center of the soliton $\delta\phi(\vec{r})$ must be regular (see
Eq.(\ref{eq:(6-21)}) and the analysis of the dynamic stability
below).

The first-order perturbation of the energy, calculated by expanding (\ref{eq:(2-7)}) becomes

\bea \Delta_{1}\epsilon &=& 2\int d_{3}\vec{r}
\stackrel{\bullet}{f}(X_{0})
\vec{\nabla}(\phi)\cdot\vec{\nabla}(\delta\phi) = \\ \nonumber &=&
2\int d_{3}\vec{r}
\vec{\nabla}\cdot\left(\stackrel{\bullet}{f}(X_{0})\delta\phi\vec{\nabla}(\phi)\right)
- 2\int d_{3}\vec{r} \delta\phi
\vec{\nabla}\cdot\left(\stackrel{\bullet}{f}(X_{0})\vec{\nabla}(\phi)\right),
\label{eq:(6-2)} \ena where a partial integration has been
performed. Owing to Eq.(\ref{eq:(2-4)}) and the assumed asymptotic
behaviour of the perturbation $\delta\phi(\vec{r})$, the two
integrals in the last equation converge and cancel each other so
that the first variation of the energy vanishes. This is the
necessary condition for the energy of the soliton to be an
extremum. The second variation reads

\be
\Delta_{2}\epsilon = \int d_{3}\vec{r} \stackrel{\bullet}{f}(X_{0}) (\vec{\nabla} \delta\phi)^{2} - 2\int d_{3}\vec{r} \stackrel{\bullet\bullet}{f}(X_{0})\left(\vec{\nabla}\phi\cdot\vec{\nabla}\delta\phi\right)^{2},
\label{eq:(6-3)}
\en
where, owing to the boundary behaviours of the perturbation and the SSS field itself, both integrals are also convergent. From the arbitrariness of $\delta\phi$, the positivity of $\stackrel{\bullet}{f}(X)$ and the minimum condition of the energy $\Delta_{2}\epsilon > 0$, we see that static stability is reached \textit{if} the requirement

\be
\stackrel{\bullet \bullet}{f}(X) < 0,
\label{eq:(6-4)}
\en
is
fulfilled in all the range of values of $X = X_{0} = -\phi^{'2}(r)$ covered by the solution. However, if we rewrite Eq.(\ref{eq:(6-3)}) as

\bea\label{eq:(6-3)bis} \Delta_{2}\epsilon &=& \int d_{3}\vec{r}
\left[\stackrel{\bullet}{f}(X_{0}) +
2X_{0}\stackrel{\bullet\bullet}{f}(X_{0})\right]
\left(\frac{\partial \delta\phi}{\partial r}\right)^{2} + \\
\nonumber &+& \int d_{3}\vec{r}
\stackrel{\bullet}{f}(X_{0})\left[\frac{1}{r^{2}}\left(\frac{\partial
\delta\phi}{\partial \theta}\right)^{2} +
\frac{1}{r^{2}\cos^{2}(\theta)}\left(\frac{\partial
\delta\phi}{\partial \varphi}\right)^{2}\right], \ena we are lead
to the less restrictive \textit{static stability criterion}
\footnote{The only model for which (\ref{eq:(6-4)bis}) vanishes
everywhere is singular and corresponds to the lagrangian $f(X) =
\lambda\sqrt{X}$}

\be
\stackrel{\bullet}{f}(X_{0}) + 2X_{0}\stackrel{\bullet\bullet}{f}(X_{0}) \geq 0,
\label{eq:(6-4)bis}
\en
which is a necessary and sufficient condition for linear stability, as opposed to Eq.(\ref{eq:(6-4)}) which is only a sufficient one. This criterion is always fulfilled for admissible models with finite-energy SSS solutions. Indeed, by deriving the first-integral equation (\ref{eq:(2-4)}) with respect to $r$ we obtain

\be
\stackrel{\bullet}{f}(X_{0}) - 2\phi^{'2}(r)\stackrel{\bullet \bullet}{f}(X_{0}) = -\frac{2\Lambda}{r^{3}\phi^{''}(r)},
\label{eq:(6-4)ter}
\en
which, owing to the monotonicity of $\phi^{'}(r)$, is positive in all the range of values of $X_{0}$ covered by the solution. We conclude that \textbf{the finite-energy SSS solutions of admissible scalar models are statically stable}.

Against perturbations which modify the charge the solitons are
unstable, but these instabilities are blocked if charge
conservation is implicit in the model (as in the case of
generalized gauge field theories considered below) or if it is a
consequence of the nature of the external sources.

\subsection{\large Dynamic stability of scalar solitons.}

Let us now consider the \textit{dynamic stability} of the SSS solutions. The initial perturbation defines the following Cauchy conditions

\be
\Phi(\vec{r},t=0) = \phi(r) + \delta\phi(\vec{r})\hspace{0.3cm};\hspace{0.3cm}\frac{\partial \Phi}{\partial t}(\vec{r},t=0) = 0,
\label{eq:(6-5)}
\en
for a dynamical problem determining the temporal evolution of the perturbed field $\Phi(\vec{r},t)$, which is governed by the hyperbolic field equations (\ref{eq:(2-2)}). At the first order, the evolution of the perturbation

\be
\delta\phi(\vec{r},t) = \Phi(\vec{r},t) - \phi(r),
\label{eq:(6-6)}
\en
is given by the linearized scalar field equation

\be
\frac{\partial}{\partial t}\left(\stackrel{\bullet}{f}(X_{0})\frac{\partial \delta\phi}{\partial t}\right) -
\vec{\nabla}\cdot\left[\stackrel{\bullet}{f}(X_{0})\vec{\nabla}(\delta\phi) - 2\stackrel{\bullet\bullet}{f}(X_{0}) \left(\vec{\nabla}\phi\cdot\vec{\nabla}\delta\phi\right)\vec{\nabla}\phi\right] = 0.
\label{eq:(6-7)}
\en
This is the Euler-Lagrange equation associated with the lagrangian density

\be L = \frac{1}{2}\left[\stackrel{\bullet}{f}(X_{0})\partial_{\mu} \delta\phi\cdot\partial^{\mu} \delta\phi - 2\stackrel{\bullet \bullet}{f}(X_{0})\left(\vec{\nabla}\phi\cdot\vec{\nabla}\delta\phi\right)^{2}\right],
\label{eq:(6-7)bis}
\en
which is defined everywhere. Equation (\ref{eq:(6-7)}) has the form of a local conservation law for a charge density $\eta =  \stackrel{\bullet}{f}(X_{0})\frac{\partial \delta\phi}{\partial t}$ which, in integral form, becomes

\be
\frac{d}{dt}\int d_{3}\vec{r} \stackrel{\bullet}{f}(X_{0})\frac{\partial \delta\phi}{\partial t} =
\int d_{3}\vec{r} \vec{\nabla}\cdot\left[\stackrel{\bullet}{f}(X_{0})\vec{\nabla}(\delta\phi) -
2\stackrel{\bullet\bullet}{f}(X_{0})\left(\vec{\nabla}\phi\cdot\vec{\nabla}\delta\phi\right) \vec{\nabla}\phi\right] = 0.
\label{eq:(6-8)}
\en
The r.h.s. of this equation is proportional to the first-order perturbation of the scalar charge of the soliton (see Eq.(\ref{eq:(6-1)})) which, consequently, remains conserved as time evolves. Moreover, for solutions satisfying the initial conditions (\ref{eq:(6-5)}) the quantity

\be \int d_{3}\vec{r}
\stackrel{\bullet}{f}(X_{0})\delta\phi(\vec{r},t),
\label{eq:(6-8)bis} \en
remains constant in time.

Equation (\ref{eq:(6-7)}), together with the conditions
(\ref{eq:(6-5)}) and (\ref{eq:(6-1)bis}), outline a spectral
problem to which we can apply standard methods. We look for
solutions separating time and spatial variables under the form

\be
\delta\phi(\vec{r},t,\Gamma)=T(t,\Gamma)\psi(\vec{r},\Gamma),
\label{eq:(6-9)}
\en
where $\Gamma$ is the separation constant and the eigenfunction $\psi(\vec{r},\Gamma)$ is assumed to satisfy the boundary condition (\ref{eq:(6-1)bis}). Replacing this expression in (\ref{eq:(6-7)}) and using the second of the initial conditions (\ref{eq:(6-5)}) we are lead to

\be
T(t,\Gamma) = \cos(\sqrt{\Gamma}t),
\label{eq:(6-10)}
\en
and

\be
\Gamma \stackrel{\bullet}{f}(X_{0})\psi = -\vec{\nabla}\cdot\left[\stackrel{\bullet}{f}(X_{0})\vec{\nabla}(\psi) - 2\stackrel{\bullet \bullet}{f}(X_{0}) \phi^{'2}\left(\frac{\vec{r}}{r}\cdot\vec{\nabla}\psi\right)\frac{\vec{r}}{r}\right].
\label{eq:(6-11)}
\en
The sign of the eigenvalue $\Gamma$ is crucial for stability. Multiplying this equation by $\psi(\vec{r},\Gamma)$ and integrating over all space we are lead (after an integration by parts of the right-hand-side) to

\be
\Gamma \int d_{3}\vec{r} \stackrel{\bullet}{f}(X_{0})\psi^{2} = \int d_{3}\vec{r} \left[\stackrel{\bullet}{f}(X_{0}) (\vec{\nabla}\psi)^{2} - 2\stackrel{\bullet \bullet}{f}(X_{0})\phi^{'2}\left(\frac{\partial \psi}{\partial r}\right)^{2}\right].
\label{eq:(6-12)}
\en
Owing to Eq. (\ref{eq:(6-4)ter}), together with the admissibility and boundary conditions, both the r.h.s. of this equation and the coefficient of $\Gamma$ are finite and positive. Then so is for $\Gamma$, and the evolution is oscillatory and bounded in time \footnote{Note that the r.h.s. of (\ref{eq:(6-12)}) coincides with the second variation of the energy associated to the eigenfunction $\psi(\vec{r},\Gamma)$ and has the same sign as $\Gamma$. This establishes a strict correspondence between static and dynamic stabilities of the SSS solutions.}. Moreover, if we consider two different eigenvalues ($\Gamma_{i}, i=(1,2)$) and their associated eigenfunctions ($\psi_{i}=\psi(\vec{r},\Gamma_{i}), i=(1,2)$) equation (\ref{eq:(6-11)}) leads to

\be
(\Gamma_{2} - \Gamma_{1})\int d_{3}\vec{r} \stackrel{\bullet}{f}(X_{0})\psi_{1}\psi_{2} = \int d_{3}\vec{r}
\left(\psi_{2}\vec{\nabla}\cdot\vec{\Sigma}_1 - \psi_{1}\vec{\nabla}\cdot\vec{\Sigma}_2\right),
\label{eq:(6-13)}
\en
where

\be
\vec{\Sigma}_i = \stackrel{\bullet}{f}(X_{0}) \vec{\nabla}\psi_{i} - 2\stackrel{\bullet \bullet}{f}(X_{0})
\left(\vec{\nabla}\phi\cdot\vec{\nabla}\psi_{i}\right)\vec{\nabla}\phi.
\label{eq:(6-14)}
\en
After a partial integration and making use of the boundary conditions we see that the right-hand-side of (\ref{eq:(6-13)}) vanishes and thus we are lead to the orthogonality relation

\be
\int d_{3}\vec{r} \stackrel{\bullet}{f}(X_{0})\psi_{1}\psi_{2} = 0.
\label{eq:(6-15)}
\en
These results outline a Sturm-Liouville problem for each admissible scalar model of the form (\ref{eq:(2-1)}) supporting finite-energy SSS solutions and lead to the following conclusions \cite{cou89}: \emph{i)} The analysis of the dynamics of the small oscillations around these solutions leads in all cases to discrete spectra of eigenvalues $\Gamma_{i}$, \emph{ii)} The associated eigenfunctions are orthogonal and finite-norm with respect to the scalar product

\be
<\psi_{i},\psi_{j}> = \int d_{3}\vec{r} \stackrel{\bullet}{f}(X_{0})\psi_{i}(\vec{r})\psi_{j}(\vec{r}) = \frac{\Delta_{2}\epsilon_{i}}{\Gamma_{i}} \delta_{ij},
\label{eq:(6-16)}
\en
defined with the kernel $\stackrel{\bullet}{f}(X_{0}) > 0$. Such functions generate a complete Hilbert space in which any perturbation can be expanded.

On the other hand, we can now separate the spatial eigenfunctions in radial and angular coordinates as

\be
\psi(r,\vartheta,\varphi,\Gamma,l) = R(r,\Gamma,l)\textit{Y}_{l}(\vartheta,\varphi),
\label{eq:(6-17)}
\en
where the angular components are the usual spherical harmonics satisfying

\be
\sin\vartheta\frac{\partial}{\partial \vartheta}\left(\sin\vartheta \frac{\partial \textit{Y}_{l}}{\partial \vartheta}\right) + \frac{\partial^{2} \textit{Y}_{l}}{\partial \varphi^{2}} + l(l+1) \sin^{2}\vartheta \textit{Y}_{l} = 0.
\label{eq:(6-18)}
\en
The radial components obey to the equation

\be
\frac{d}{d r}\left(\frac{1}{r\phi^{''}}\frac{d R}{d r}\right) + \frac{l(l+1) - \Gamma r^{2}}{2r^{2}\phi^{'}} R = 0,
\label{eq:(6-19)}
\en
which is obtained from (\ref{eq:(6-11)}), (\ref{eq:(6-17)}) and using the first-integral (\ref{eq:(2-4)}). These equations have the standard Sturm-Liouville form \cite{cou89}

\bea
L y &+& \lambda \mu(x) y = 0
\nonumber \\
L y &=& \frac{d}{d x}\left[k(x)\frac{d y}{d x}\right] - q(x) y
\hspace{0.5cm} \mathrm{with} \hspace{0.2cm} k(x) > 0, \mu(x) > 0.
\label{eq:(6-20)}
\ena
The asymptotic behaviour of $R(r)$ is obtained from the asymptotic form of the admissible solitons
($\phi^{'}(r \rightarrow \infty) \sim 1/r^{p}; p>1$) through

\be
\frac{d^{2} R}{d r^{2}} + \frac{p}{r} \frac{d R}{d r} - p\frac{(l(l+1) - \Gamma r^{2})}{2r^{2}} R = 0,
\label{eq:(6-21)}
\en
which is a Lommel equation and can be solved in terms of Bessel functions \cite{wat44}. For large $r$ we can assume for the solution the asymptotic form

\be
R(r \rightarrow \infty) \sim \frac{\varrho(r)}{r^{q}},
\label{eq:(6-22)}
\en
where $\varrho(r)$ is a bounded function and, owing to the boundary condition (\ref{eq:(6-1)bis}), $q$ is restricted to be $q > p > 1$. By neglecting the higher-order terms in $1/r$ in the resulting equation for $\rho$ we are lead to

\be \varrho^{''} + \frac{p\Gamma}{2} \varrho = 0.
\label{eq:(6-23)} \en
Owing to the positivity of $\Gamma$ the
solution of this equation is oscillatory and the asymptotic
behaviour of the eigenfunctions is given by \footnote{The value of
$q$ as a function of $p$ and $l$ can be explicitly obtained from
recursion formulae methods \cite{som64} applied to
Eq.(\ref{eq:(6-21)}), which also allow for the approximate
determination of the eigenvalue spectrum of $\Gamma$.}

\be
R(r \rightarrow \infty) \sim \frac{\cos\left(\sqrt{\frac{p\Gamma}{2}}r + \chi\right)}{r^{q}},
\label{eq:(6-24)}
\en
where $\chi$ is a constant phase. This asymptotic form of the eigenfunctions makes the integral of the second variation of the energy (\ref{eq:(6-3)bis}) to converge in the $r \rightarrow \infty$ limit.

To determine the behaviour of $R(r)$ around the center of the soliton we must consider separately cases A-1 and A-2. Let us assume in both cases a form

\be
R(r \rightarrow 0) \sim \alpha - \beta r^{q}.
\label{eq:(6-25)}
\en
In \underline{case A-1} ($\phi^{'}(r \rightarrow 0) \sim 1/r^{p}, 0<p<1$) equation (\ref{eq:(6-19)}) becomes

\be
2\beta q \frac{p + q - 1}{p} r^{q} + \left(l(l+1) - \Gamma r^{2}\right) \left(\alpha - \beta r^{q}\right) \approx 0.
\label{eq:(6-26)}
\en
For the first surface spherical harmonic ($l = 0$) we are lead to \cite{som64}

\be
\alpha = 0 \hspace{0.3cm};\hspace{0.3cm} q = 1 - p,
\label{eq:(6-27)}
\en
or

\be
\frac{\beta}{\alpha} = \frac{\Gamma p}{4(p + 1)}\hspace{0.3cm};\hspace{0.3cm} q = 2.
\label{eq:(6-27)bis}
\en
For $l \neq 0$ we obtain

\be
\alpha = 0 \hspace{0.3cm};\hspace{0.3cm} q = \frac{1 - p + \sqrt{(1 - p)^{2} + 2 l(l+1)p}}{2}.
\label{eq:(6-28)}
\en
In \underline{case A-2} ($\phi^{'}(r \rightarrow 0) \sim a - b r^{\sigma},\sigma>0$) equation (\ref{eq:(6-19)}) becomes

\be
2\beta q \frac{q - \sigma - 1}{b \sigma} r^{q - \sigma} + \frac{l(l+1) - \Gamma r^{2}}{a} \left(\alpha - \beta r^{q}\right) \approx 0.
\label{eq:(6-29)}
\en
If $l = 0$ we must have

\be
\frac{\beta}{\alpha} = \frac{b}{2 a} \frac{\Gamma \sigma}{\sigma + 2} \hspace{0.3cm}; \hspace{0.3cm}
q = \sigma + 2,
\label{eq:(6-30)}
\en
or

\be \alpha=0 \hspace{0.3cm};\hspace{0.3cm} q = \sigma + 1.
\label{eq:(6-30)bis} \en For $l \neq 0$

\be
\frac{\beta}{\alpha} = \frac{b}{2 a} l(l + 1) \hspace{0.3cm};\hspace{0.3cm} q = \sigma,
\label{eq:(6-31)}
\en
or

\be
\alpha = 0 \hspace{0.3cm};\hspace{0.3cm} q = \sigma + 1.
\label{eq:(6-31)bis}
\en

In all these cases the integral of the second order variation of
the energy (\ref{eq:(6-3)bis}) can be shown to converge in the
limit $r \rightarrow 0$.

\subsection{\large Stability of multicomponent scalar solitons.}

The analysis of stability in the scalar case can be extended to
the multicomponent scalar fields. Now we have $N$ integration
constants $\Lambda_{i}$ and a degeneration in energy of the SSS
solutions on the sphere of radius $\Lambda =
\sqrt{\sum_{i=1}^{N}\Lambda_{i}^{2}}$ in $\Re^{N}$. Obviously the
variations of the energy vanish for perturbations which remain
inside this sphere obtained by modifying the constants
$\Lambda_{i}$ in equation (\ref{eq:(4-10)}) in such a way that the
``total mean-square charge" $\Lambda$ remains unchanged. The
asymptotic boundary conditions satisfied by the fields
$\phi_{i}^{'}(r,\Lambda_{j})$ (obtained from the asymptotic
behaviour of the associated one-component scalar field solution
$\phi^{'}(r,\Lambda)$ through Eq.(\ref{eq:(4-10)})) are modified
by these perturbations and the associated charges (defined from
Eq.(\ref{eq:(4-10)bis}) as $\Lambda_{i}$) are modified.
Consequently, charge conservation condition blocks such
perturbations and prevents a soliton from evolving spontaneously
towards another equal-energy configuration in the sphere.

For general perturbations $\delta \phi_{i}(\vec{r})$ the first-order modifications of the scalar charges take the form

\bea
\Delta \Lambda_{i} = \frac{1}{4\pi} \int d_{3}\vec{r} \vec{\nabla}\cdot\left[\stackrel{\bullet}{f}(X_{0}) (\vec{\nabla}\delta\phi_{i}) -
2\stackrel{\bullet\bullet}{f}(X_{0})\sum_{j=1}^{N}\left(\vec{\nabla}\phi_{j}\cdot\vec{\nabla}\delta\phi_{j}\right)
\vec{\nabla}\phi_{i}, \right],
\label{eq:(6-31)ter}
\ena
with $X = -\sum_{i=1}^{N} \phi_{i}^{'2}$. The requirement of charge conservation ($\Delta \Lambda_{i} = 0$) imposes boundary conditions on the perturbing fields which, as in the one-component case, must vanish asymptotically faster than the SSS fields themselves.

The first-order variation of the energy functional is obtained
from the integral of Eq.(\ref{eq:(4-5)}) and reads

\be
\Delta_{1}\epsilon =2\sum_{i=1}^{N} \frac{\Lambda_{i}}{\Lambda}\left(\int d_{3}\vec{r} \nabla\cdot\left[\stackrel{\bullet}{f}(X_{0}) \delta \phi_{i}\vec{\nabla}\phi \right] -
\int d_{3}\vec{r} \delta \phi_{i}
\vec{\nabla}\cdot\left[\stackrel{\bullet}{f}(X_{0})\vec{\nabla}(\phi)\right]\right),
\label{eq:(6-31)q}
\en
where Eq.(\ref{eq:(4-11)}) has been used and a partial integration has been performed. Each term of this sum has the form of Eq.(\ref{eq:(6-2)}) and vanishes because of the same reasons. Thus the first variation of the energy vanishes, which is an extremum condition. The second variation of the energy functional takes the form

\bea
\Delta_{2}\epsilon &=& \int d_{3}\vec{r} \left[\stackrel{\bullet}{f}(X_0) \sum_{i=1}^{N}
\left(\vec{\nabla} \delta\phi_{i}\right) ^{2} - 2\phi^{'2} \stackrel{\bullet\bullet}{f}(X_0) \left(\sum_{i=1}^{N} \frac{\Lambda_{i}}{\Lambda} \frac{\partial \delta\phi_{i}}{\partial r}\right)^{2} \right], \label{eq:(6-31)f}
\ena
and can be written as

\bea
\Delta_{2}\epsilon &=& \int d_{3}\vec{r} \left[\stackrel{\bullet}{f}(X_0) +
2X_{0} \stackrel{\bullet\bullet}{f}(X_0) \cos^{2}(\Omega) \right]
\sum_{i=1}^{N} \left(\frac{\partial \delta\phi_{i}}{\partial r}\right)^{2} + \\
\nonumber
&+& \int d_{3}\vec{r} \stackrel{\bullet}{f}(X_0) \sum_{i=1}^{N}\left[\frac{1}{r^{2}}
\left(\frac{\partial \delta\phi_{i}}{\partial \theta}\right)^{2} + \frac{1}{r^{2}\cos^{2}(\theta)}\left(\frac{\partial \delta\phi_{i}}{\partial \varphi}\right)^{2}\right],
\label{eq:(6-31)s}
\ena
in terms of the angle $\Omega(\vec{r})$ in the internal space between the vector formed by the radial derivatives of the components of the perturbing fields and the direction $\frac{\Lambda_{i}}{\Lambda}$ defined by the SSS solution . The second integral in this equation is always positive. If $\stackrel{\bullet\bullet}{f}(X_0)$ is negative the first integral is also positive. Otherwise we have

\be
\stackrel{\bullet}{f}(X_0) + 2X_{0} \stackrel{\bullet\bullet}{f}(X_0) \cos^{2}(\Omega) \geq
\stackrel{\bullet}{f}(X_0) + 2X_{0} \stackrel{\bullet\bullet}{f}(X_0) \geq 0,
\label{eq:(6-31)sv}
\en
and, owing to Eq.(\ref{eq:(6-4)ter}), this integral is always positive for admissible many-components scalar models with finite-energy SSS solutions. Consequently, all these solutions are statically stable. Moreover, the analysis of the dynamical evolution of small perturbations performed for scalar solitons can be straightforwardly generalized to this multicomponent case. Such an analysis proves the dynamical stability of these solitons.

\subsection{\large Static stability of generalized electromagnetic solitons}

Similar analysis of stability can be performed for generalized abelian gauge fields. We consider a finite-energy ESS solution \footnote{To avoid difficulties related to the gauge determination we work directly with the fields. When the use of the potentials becomes necessary in some step of the calculation we will fix the gauge through appropriate conditions.} of the field equations (\ref{eq:(5-9)}) ($\vec{E}_{0}(r),\vec{H}_{0}=0$) and introduce a small perturbing field ($\vec{E}_{1}(\vec{r})$, $\vec{H}_{1}(\vec{r})$) which does not modify the total electric charge of the soliton. The first-order modification of the charge density is obtained by perturbing the first of the field equations (\ref{eq:(5-9)bis}), which leads to

\be
\vec{\nabla}\cdot\vec{\sigma} = 0,
\label{eq:(6-32)}
\en
where

\be
\vec{\sigma} = \frac{\partial \varphi}{\partial X_{0}}\vec{E}_{1} + 2\frac{\partial^{2} \varphi}{\partial X_{0}^{2}}(\vec{E}_{0}\cdot\vec{E}_{1})\vec{E}_{0}
\label{eq:(6-33)}
\en
and the index $0$ in the derivatives means that they are calculated for the unperturbed solution (note that, owing to the parity invariance, the odd partial derivatives of $\varphi$ with respect to $Y$ vanish in $Y=0$). From the integration of (\ref{eq:(6-32)}) we see that $\vec{\sigma}$ must vanish asymptotically faster than $r^{-2}$. Then, the regular perturbing field $\vert\vec{E}_{1}(\vec{r})\vert$ must be damped faster than $E_{0}(r)$ itself. This boundary condition is similar to the one introduced for scalar models but, owing to the electric charge conservation implicit in the field equations, the physical
meaning becomes here more transparent.

Let us now consider the variations of the energy functional under such charge-preserving perturbations. The first-order variation is obtained from the integration of Eq.(\ref{eq:(5-5)}) and reads

\be
\Delta_{1} \epsilon = -2 \int d_{3}\vec{r} \vec{\nabla}\cdot\left[A^{0}\vec{\sigma}\right] + 2 \int d_{3}\vec{r} A^{0}\vec{\nabla}\cdot\vec{\sigma},
\label{eq:(6-34)}
\en
where we have introduced the time-like component of the four-vector potential for the solution ($\vec{E}_{0} = -\vec{\nabla}A^{0}, \vec{A} = 0$). This expression vanishes due to the boundary conditions and the linearized field equation (\ref{eq:(6-32)}). This is an extremum condition.

In calculating the second variation of the energy let us expand the first of the field equations (\ref{eq:(5-9)bis}) to the second-order. We are lead to

\be
\vec{\nabla}\cdot\left(\vec{\sigma} + \vec{\eta}\right) = 0,
\label{eq:(6-35)}
\en
where now the term

\bea
\vec{\eta} &=& 2 \frac{\partial^{2} \varphi}{\partial X_{0}^{2}}(\vec{E}_{0}\cdot\vec{E}_{1})\vec{E}_{1} +
\frac{\partial^{2} \varphi}{\partial X_{0}^{2}}(\vec{E}_{1}^{2} - \vec{H}_{1}^{2})\vec{E}_{0} +
2 \frac{\partial^{3} \varphi}{\partial X_{0}^{3}}(\vec{E}_{0}\cdot\vec{E}_{1})^{2}\vec{E}_{0} +  \nonumber\\
&+& 2 \frac{\partial^{3} \varphi}{\partial X_{0}\partial Y_{0}^{2}}(\vec{E}_{0}\cdot\vec{H}_{1})^{2}\vec{E}_{0} +
2\frac{\partial^{2} \varphi}{\partial Y_{0}^{2}}(\vec{E}_{0}\cdot\vec{H}_{1})\vec{H}_{1},
\label{eq:(6-36)}
\ena
includes the second-order corrections. Using this equation the second variation of the energy, obtained from
the integration of Eq.(\ref{eq:(5-5)}), becomes

\bea\label{eq:(6-37)} \Delta_{2} \epsilon &=& \int d_{3}\vec{r}
\left[\frac{\partial \varphi}{\partial X_{0}}\vec{E}_{1}^{2} +
2\frac{\partial^{2} \varphi}{\partial X_{0}^{2}}(\vec{E}_{0}\cdot\vec{E}_{1})^{2}\right] +\int d_{3}\vec{r} \left[\frac{\partial \varphi}{\partial X_{0}}\vec{H}_{1}^{2} - 2\frac{\partial^{2} \varphi}{\partial Y_{0}^{2}}(\vec{E}_{0}\cdot\vec{H}_{1})^{2}\right] - \\
\nonumber &-& 2 \int d_{3}\vec{r}
\vec{\nabla}\cdot\left[A^{0}\vec{\eta}\right]. \ena
The last
integral in the r.h.s. of this equation vanishes, owing to the
boundary conditions. The first term is positive if
$\frac{\partial^2 \varphi}{\partial X_0^2}\geq 0$; on the other
hand, if $\frac{\partial^2 \varphi}{\partial X_0^2} < 0$ the
integrand of this term can be written as

\be (\vec{E}_{1})^{2}\left(\frac{\partial \varphi}{\partial X_{0}}
+ 2\frac{\partial^{2} \varphi}{\partial
X_{0}^{2}}(\vec{E}_{0}^{2}\cos^{2}(\theta))\right) \geq
(\vec{E}_{1})^{2}\left(\frac{\partial \varphi}{\partial X_{0}} +
2\frac{\partial^{2} \varphi}{\partial
X_{0}^{2}}\vec{E}_{0}^{2}\right), \label{eq:(6-38)} \en
where
$\theta$ is the angle between $\vec{E}_{0}$ and $\vec{E}_{1}$. By
deriving Eq.(\ref{eq:(5-10)}) with respect to $r$ and taking into
account the monotonicity of $E_{0}(r)$ we see that the first term
in (\ref{eq:(6-37)}) is always positive. Concerning the second
term of (\ref{eq:(6-37)}) it is positive if $\frac{\partial^2
\varphi}{\partial Y_0^2}\leq 0$ while if $\frac{\partial^2
\varphi}{\partial Y_0^2}>0$ we can write its integrand as

\be (\vec{H}_{1})^{2}\left(\frac{\partial \varphi}{\partial X_{0}}
- 2\frac{\partial^{2} \varphi}{\partial
Y_{0}^{2}}(\vec{E}_{0}^{2}\cos^{2}(\theta))\right)\geq
(\vec{H}_{1})^{2}\left(\frac{\partial \varphi}{\partial X_{0}} -
2\frac{\partial^{2} \varphi}{\partial
Y_{0}^{2}}(\vec{E}_{0}^{2})\right). \label{eq:(6-39)} \en
From the
arbitrariness of the perturbing fields, the positivity of this
term and, finally, the positivity of the second variation of the
energy requires the condition

\be
\frac{\partial \varphi}{\partial X} \geq 2X\frac{\partial^{2}
\varphi}{\partial Y^{2}},
\label{eq:(6-40)}
\en
to be fulfilled in the range of values of $X$ ($Y=0$) where the ESS solutions are defined. This is a \textit{necessary and sufficient condition of static stability} to be satisfied by the lagrangian densities of
\textit{admissible} models supporting finite-energy ESS solutions. This stability criterion goes beyond the widely used Derrick's \textit{necessary conditions} \cite{derrick64}.

Let us check, using this criterium, the linear stability of the
electrostatic finite-energy solutions of the B-I model. From Eqs.
(\ref{eq:(5-3)}) and (\ref{eq:(6-40)}) we immediately obtain

\be \frac{\partial \varphi}{\partial X_0}-2X_0\frac{\partial^2
\varphi}{\partial Y_0^2} =
\frac{1}{8\pi}\left(1-\mu^2X_0\right)^{1/2} > 0,
\label{eq:(6-40)bis} \en
and the stability condition
(\ref{eq:(6-40)}) is fulfilled since the ESS field is bounded
everywhere ($X_0<1/{\mu^2}$).

\subsection{\large Dynamic stability of generalized electromagnetic solitons}

Let us now analyze the dynamical evolution of the small perturbations of the ESS solitons. The system of linearized field equations, obtained by expanding (\ref{eq:(5-9)bis}) to first order, is formed by Eq.(\ref{eq:(6-32)}) aside from a vector equation:

\bea
\vec{\nabla}\cdot\vec{\sigma} =\vec{\nabla}\cdot(\boldsymbol{\Sigma}\cdot\vec{E}_{1}) &=& 0
\nonumber \\
\frac{\partial \vec{\sigma}}{\partial t} - \vec{\nabla}\times(\boldsymbol{\Omega}\cdot\vec{H}_{1}) &=& \frac{\partial}{\partial t}(\boldsymbol{\Sigma}\cdot\vec{E}_{1}) - \vec{\nabla}\times(\boldsymbol{\Omega}\cdot\vec{H}_{1}) = 0,
\label{eq:(6-41)}
\ena
where now we have introduced the symmetric tensors

\bea \boldsymbol{\Sigma} &=& \frac{\partial \varphi}{\partial
X_{0}}\mathbb{I}_{3} + 2\frac{\partial^{2} \varphi}{\partial
X_{0}^{2}}(\vec{E}_{0}\otimes\vec{E}_{0})
\nonumber\\
\boldsymbol{\Omega} &=& \frac{\partial \varphi}{\partial
X_{0}}\mathbb{I}_{3} - 2\frac{\partial^{2} \varphi}{\partial
Y_{0}^{2}} (\vec{E}_{0}\otimes\vec{E}_{0}), \label{eq:(6-42)} \ena
which will be useful in simplifying the notations in the sequel.
The perturbing fields must also satisfy the first set of Maxwell
equations. Expanding (\ref{eq:(5-9)q}) up to the first order we
obtain

\bea
\vec{\nabla} \times \vec{E}_{1} &=&  -\frac{\partial \vec{H}_{1}}{\partial t}
\nonumber \\
\vec{\nabla}\cdot\vec{H}_{1} &=& 0.
\label{eq:(6-42)bis}
\ena
Let us look for solutions of these equations which are products of functions of time and space variables for both electric and magnetic fields, of the form

\bea
\vec{E}_{1}(t,\vec{r}) &=& T_{e}(t)\cdot\vec{e}(\vec{r})
\nonumber\\
\vec{H}_{1}(t,\vec{r}) &=& T_{h}(t)\cdot\vec{h}(\vec{r}).
\label{eq:(6-43)}
\ena
In this way we are lead to the first-order equations for the time variables

\bea
\dot{T_{e}} &=& \lambda T_{h}
\nonumber\\
\dot{T_{h}} &=& -\mu T_{e},
\label{eq:(6-43)bis}
\ena
where $\lambda$ and $\mu$ are separation constants. By deriving these equations we are lead to the system

\bea
\ddot{T}_{e}(t) + \Gamma T_{e}(t) &=& 0
\nonumber\\
\ddot{T}_{h}(t) + \Gamma T_{h}(t) &=& 0,
\label{eq:(6-44)}
\ena
with $\Gamma = \lambda\cdot\mu$. The identification $\lambda = \mu = \sqrt{\Gamma}$ can be introduced without loss of generality, as a consequence of the first-order equations and \textit{the positivity of} $\Gamma$, which will be established below. The eigenvalue $\Gamma$ being positive, the solutions (normalized to unity) take the form

\bea
T_{e}(t) &=& \cos(\sqrt{\Gamma} t + \delta)
\nonumber\\
T_{h}(t) &=& \sin(\sqrt{\Gamma} t + \delta),
\label{eq:(6-45)}
\ena
where $\delta$ is the same constant phase for both solutions. In this case the eigenfunctions remain bounded as time evolves and the soliton is dynamically stable. The field equations for the spatial components are

\bea
\vec{\nabla}\cdot(\boldsymbol{\Sigma}\cdot\vec{e}) &=& 0
\nonumber\\
\vec{\nabla}\times(\boldsymbol{\Omega}\cdot\vec{h}) &=& \sqrt{\Gamma} \boldsymbol{\Sigma}\cdot\vec{e},
\label{eq:(6-46)}
\ena
where we have used the definitions (\ref{eq:(6-42)}). Note that the first of Eqs.(\ref{eq:(6-46)}) is an immediate consequence of the second one. Moreover, the first set of Maxwell equations leads to

\bea
\vec{\nabla} \times \vec{e} &=& \sqrt{\Gamma} \vec{h}
\nonumber\\
\vec{\nabla}\cdot\vec{h} &=& 0.
\label{eq:(6-47)}
\ena
We shall now introduce a four-potential $A^{\mu}$ for the perturbing fields, defined in the Hamilton gauge ($A^{0} = 0$) in such a way that \footnote{This gauge-fixing condition is allowed by the gauge invariance of the equations for the perturbing fields (\ref{eq:(6-41)})-(\ref{eq:(6-42)bis}) which are independent from any gauge choice for the unperturbed fields.}

\bea
\vec{E}_{1} &=& -\frac{\partial \vec{A}}{\partial t}
\nonumber\\
\vec{H}_{1} &=& \vec{\nabla}\times\vec{A}.
\label{eq:(6-48)}
\ena
This vector potential is determined up to the gradient of an arbitrary time-independent scalar field. In terms of this vector potential, the first set of field equations (\ref{eq:(6-42)bis}) are identically satisfied while the second set (\ref{eq:(6-41)}) becomes

\bea
&\frac{\partial}{\partial t}& \left(\vec{\nabla}\cdot(\boldsymbol{\Sigma}\cdot\vec{A})\right) = 0,
\nonumber \\
&\frac{\partial^{2}}{\partial t^{2}}& (\boldsymbol{\Sigma}\cdot\vec{A}) +
\vec{\nabla}\times\left(\boldsymbol{\Omega}\cdot(\vec{\nabla}\times\vec{A})\right) = 0,
\label{eq:(6-48)bis}
\ena
If we write equations (\ref{eq:(6-48)}) for the separated functions (\ref{eq:(6-43)}), by integrating in time the first one and using (\ref{eq:(6-43)bis}), we obtain the general form of the vector potential \textit{for these functions} as

\be
\vec{A}(t,\vec{r}) = T_{a}(t) \vec{a}(\vec{r}) + \vec{\nabla}\phi(\vec{r}),
\label{eq:(6-49)}
\en
where $\phi(\vec{r})$ is a time-independent function and

\be
T_{a}(t) = \frac{T_{h}(t)}{\sqrt{\Gamma}} \hspace{0.5cm} ; \hspace{0.5cm} \vec{a}(\vec{r}) = \vec{e}(\vec{r}).
\label{eq:(6-50)}
\en
This vector potential, determined in Eq.(\ref{eq:(6-49)}) up to the gradient of a time-independent scalar field, becomes univocally fixed by requiring its form to separate in time and space variables, taking the form of the first term in the r.h.s of Eq.(\ref{eq:(6-49)}). In terms of this potential, using the definitions (\ref{eq:(6-42)}), the field equations for the spatial part of the perturbation reduce to the unique vector equation

\be
\vec{\nabla}\times\left(\boldsymbol{\Omega}\cdot(\vec{\nabla}\times\vec{a})\right) =
\Gamma \boldsymbol{\Sigma}\cdot\vec{a},
\label{eq:(6-51)}
\en
which outlines the \textit{eigenvalue problem} for the linear oscillations in this electromagnetic case.

As in the scalar case, the standard analysis of this problem can be performed for the ESS solitons of admissible generalized electromagnetic field models without any reference to the explicit form of the lagrangian density. In this way we shall show that two given eigenfunctions $\vec{a}_{1}(\vec{r})$ and $\vec{a}_{2}(\vec{r})$ associated to the eigenvalues $\Gamma_{1}$ and $\Gamma_{2}$, respectively, are orthogonal and finite-norm with respect to the scalar product

\be
<\vec{a}_{1}\cdot\vec{a}_{2}> =\zeta \int d_{3}\vec{r}( \vec{a}_{1}\cdot\boldsymbol{\Sigma}\cdot\vec{a}_{2}),
\label{eq:(6-52)}
\en
where $\zeta$ is a normalizing factor. Indeed, multiplying Eq.(\ref{eq:(6-51)}) (defined for a given eigenfunction $\vec{a}_{1}$) by an eigenfunction $\vec{a}_{2}$, integrating in space, using the identity

\be
\vec{a}_{2}\cdot\vec{\nabla}\times\left(\boldsymbol{\Omega}\cdot(\vec{\nabla}\times\vec{a}_{1})\right)
=
\vec{\nabla}\cdot\left[\vec{a}_{2}\times\left(\boldsymbol{\Omega}\cdot(\vec{\nabla}\times\vec{a}_{1})\right)
\right] +
(\vec{\nabla}\times\vec{a}_{1})\cdot\boldsymbol{\Omega}\cdot(\vec{\nabla}\times\vec{a}_{2}),
\label{eq:(6-53)} \en and taking into account the symmetry of the
tensors $\boldsymbol{\Sigma}$ and $\boldsymbol{\Omega}$, we obtain
the equation

\bea \int d_{3}\vec{r}
\vec{\nabla}\cdot\left[\vec{a}_{2}\times\left(\boldsymbol{\Omega}\cdot(\vec{\nabla}\times\vec{a}_{1})
\right) \right] &+& \int d_{3}\vec{r}
(\vec{\nabla}\times\vec{a}_{1})\cdot\boldsymbol{\Omega}\cdot(\vec{\nabla}\times\vec{a}_{2})=
\nonumber \\ &=& \Gamma_{1} \int
d_{3}\vec{r}(\vec{a}_{1}\cdot\boldsymbol{\Sigma}\cdot\vec{a}_{2}).
\label{eq:(6-54)} \ena Owing to the boundary conditions the first
integral in the l.h.s of this equation vanishes while the second
one converges. By permuting the indices and subtracting we finally
obtain

\be
(\Gamma_{1} - \Gamma_{2})\int d_{3}\vec{r} (\vec{a}_{1}\cdot\boldsymbol{\Sigma}\cdot\vec{a}_{2}) =
(\Gamma_{1} - \Gamma_{2}) <\vec{a}_{1}\cdot\vec{a}_{2}> = 0.
\label{eq:(6-55)}
\en
We see that the eigenfunctions associated to different eigenvalues are orthogonal with respect to the scalar product (\ref{eq:(6-52)}). Moreover, if $\vec{a}_{1} = \vec{a}_{2}$ we obtain from (\ref{eq:(6-54)})

\be
\Gamma \int
d_{3}\vec{r}(\vec{a}\cdot\boldsymbol{\Sigma}\cdot\vec{a}) = \int d_{3}\vec{r}
(\vec{\nabla}\times\vec{a})\cdot\boldsymbol{\Omega}\cdot(\vec{\nabla}\times\vec{a}),
\label{eq:(6-56)}
\en
and both integrals converge. The integrand in the l.h.s. of this equation can be written as

\be
(\vec{a}\cdot\boldsymbol{\Sigma}\cdot\vec{a}) = \frac{\partial \varphi}{\partial X_{0}} \left(\left(\vec{a}\right)^{2} - \left(\vec{a}\cdot\frac{\vec{r}}{r}\right)^{2}\right) +
\left(\frac{\partial \varphi}{\partial X_{0}} + 2\vec{E}_{0}^{2}\frac{\partial^{2} \varphi}{\partial X_{0}^{2}}\right) \left(\vec{a}\cdot\frac{\vec{r}}{r}\right)^{2},
\label{eq:(6-57)}
\en
and is positive for ESS soliton solutions of any admissible electromagnetic model. The integrand of the r.h.s. of (\ref{eq:(6-56)}) takes the form

\bea
(\vec{\nabla}\times\vec{a})\cdot\boldsymbol{\Omega}\cdot(\vec{\nabla}\times\vec{a}) &=&
\frac{\partial \varphi}{\partial X_{0}}\left(\left(\vec{\nabla}\times\vec{a}\right)^{2} -
\left(\frac{\vec{r}}{r}\cdot\vec{\nabla}\times\vec{a}\right)^{2}\right) + \\ \nonumber &+& \left(\frac{\partial \varphi}{\partial X_{0}} -
2\vec{E}_{0}^{2}\frac{\partial^{2} \varphi}{\partial Y_{0}^{2}}\right) \left(\frac{\vec{r}}{r}\cdot\vec{\nabla}\times\vec{a}\right)^{2}.
\label{eq:(6-58)}
\ena
This term is positive \textit{if (and only if)} the condition for static stability (\ref{eq:(6-40)}) is fulfilled. Under this condition the eigenvalues $\Gamma$ are well defined and positive and, consequently, the behaviour of any initially bounded perturbation remains bounded as time evolves. We conclude that the statically stable ESS solitons of admissible electromagnetic models which satisfy (\ref{eq:(6-40)}) are also \textit{dynamically stable}. Moreover, the spectrum of eigenvalues is discrete and the eigenfunctions generate the functional space of the \textit{physical vector potentials}, which can be written as

\be \vec{A}(t,\vec{r}) = \sum_{n} C_{n} \sin(\sqrt{\Gamma_{n}}
t)\vec{a}_{n}(\vec{r}), \label{eq:(6-59)} \en
and are in a
one-to-one correspondence with the physical perturbed states of
the soliton. Indeed, any charge-preserving perturbation of the
soliton field is described, in the Hamilton gauge, by vector
potentials which can be obtained from one (and only one) of the
form (\ref{eq:(6-59)}) by the addition of gradients of
time-independent scalar functions.

The analysis of the spatial structure of the eigenfunctions and physical perturbations can now be performed by separating in radial and angular parts the components of the vector potentials $\vec{a}_{n}(\vec{r})$ in the natural basis of the polar coordinate system. This procedure, which is standard in spherically-symmetric physical problems \cite{bona07}, will determine the asymptotic and central-behaviour of the perturbing fields, as in the scalar case already considered. We shall leave this study for future developments.

\subsection{\large Stability of generalized non-abelian gauge solitons.}

Owing to the essential self-interactions involving the field potentials, the treatment of the static and dynamic stability of the solitons for generalized non-abelian gauge models is more involved than in the abelian case. A detailed study of the stability of some extended static finite-energy solutions for the standard Yang-Mills model has been performed in Ref. \cite{jackiw80}, where similar difficulties arise. Most methods of that work can be generalized to the present situation and we shall follow this way in analyzing the stability behaviour of the ESS solutions (\ref{eq:(5-26bis)}).

Consider a finite-energy ESS solution of the field equations (\ref{eq:(5-21)}) of the form (\ref{eq:(5-26bis)}) and introduce small regular perturbing fields through the definitions

\bea
&\vec{E}_{a}&(r) = -\vec{\nabla}A^{0}_{a}(r) \hspace{0.2cm},\hspace{0.2cm} \vec{H}_{a} = 0
\hspace{0.2cm},\hspace{0.2cm} \vec{A}_{a} = 0
\nonumber\\
&\delta\vec{E}_{a}&(\vec{r})\hspace{0.2cm}, \hspace{0.2cm} \delta\vec{H}_{a}(\vec{r})\hspace{0.2cm}, \hspace{0.2cm} \delta A^{0}_{a}(\vec{r})\hspace{0.2cm}, \hspace{0.2cm} \delta\vec{A}_{a}(\vec{r}).
\label{eq:(6-60)}
\ena
To first order these fields are related through

\bea
&\delta&\vec{E}_{a}(\vec{r}) = -\vec{\nabla}\delta A^{0}_{a} - \frac{\partial \delta\vec{A}_{a}}{\partial t} -
g\sum_{bc} C_{abc} \delta\vec{A}_{b}  A^{0}_{c}
\nonumber\\
&\delta&\vec{H}_{a}(\vec{r}) = \vec{\nabla} \times
\delta\vec{A}_{a},
\label{eq:(6-60)bis}
\ena
and they are assumed to leave invariant the color charges $Q_{a}$ associated to the unperturbed solution. To first-order the modifications of the charge densities are obtained from the perturbation of the time-components of the field equations (\ref{eq:(5-21)}) (the generalized Gauss laws) and read

\be
\vec{\nabla}\cdot\vec{\sigma}_{a} =-g\sum_{bc} C_{abc} \frac{\partial \varphi}{\partial X_{0}} \delta\vec{A}_{b}\cdot\vec{E}_{c},
\label{eq:(6-61)}
\en
where

\be
\vec{\sigma}_{a} = \frac{\partial \varphi}{\partial X_{0}}\delta\vec{E}_{a} + 2\frac{\partial^{2} \varphi}{\partial X_{0}^{2}} \left(\sum_{p}\vec{E}_{p}\cdot\delta\vec{E}_{p}\right)\vec{E}_{a},
\label{eq:(6-62)}
\en
and the modifications of the total charges read

\be \Delta Q_{a} = \frac{1}{4\pi}\int d_{3}\vec{r} \vec{\nabla}\cdot\vec{\sigma}_{a} =
-g\sum_{bc} C_{abc} \int d_{3}\vec{r} \frac{\partial \varphi}{\partial X_{0}} \delta\vec{A}_{b}\cdot\vec{E}_{c} = 0.
\label{eq:(6-62)bis}
\en
The first-order perturbations of the vector equations, given by the spatial components of Eqs.(\ref{eq:(5-21)}) (the generalized Amp\`ere laws) read

\be
-\frac{\partial}{\partial t} \vec{\sigma}_{a} + \vec{\nabla}\times \vec{\omega}_{a} =
-g\sum_{bc} C_{abc} \frac{\partial \varphi}{\partial X_{0}}\left[\delta A^{0}_{b} \vec{E}_{c} +
A^{0}_{b} \delta\vec{E}_{c} \right],
\label{eq:(6-63)}
\en
where

\be \vec{\omega}_{a} = \frac{\partial \varphi}{\partial
X_{0}}\delta\vec{H}_{a} - 2\frac{\partial^{2} \varphi}{\partial
Y_{0}^{2}}
\left(\sum_{p}\vec{E}_{p}\cdot\delta\vec{E}_{p}\right)\vec{E}_{a}.
\label{eq:(6-64)} \en The r.h.s. in Eq.(\ref{eq:(6-61)}) is the
color-charge density carried by the perturbations and its spatial
integral must vanish, according to our initial assumptions. This
requirement restricts the asymptotic behaviour of
$\vec{\sigma}_{a}$ and, as for the other field models already
considered, leads to boundary conditions to be satisfied by the
perturbing fields. Moreover, similarly to the multi-component
scalar case, the color charges of the unperturbed solution
($Q_{a}$) fix a direction in the color-charge space (called in
Ref.\cite{jackiw80} ``electromagnetic" direction, while the
orthogonal directions are termed ``charged"). Owing to the
first-integral equation (\ref{eq:(5-25)bis}), the potentials
$A^{0}_{a}$ and the fields $\vec{E}_{a}$ of the ESS solutions lie
in this direction. For the perturbing fields to remain purely
\textit{electromagnetic} the associated charge densities that are
induced by them must vanish and, owing to Eq.(\ref{eq:(6-61)}),
$\delta\vec{A}_{b}$ must also lie in this direction. In what
follows, we shall prove the stability of the finite-energy ESS
solutions against this kind of non-charged perturbations.

Now let us analyze the variations of the energy functional. The first variation is obtained by perturbing the spatial integral of (\ref{eq:(5-23)}) (this is a gauge-invariant quantity, as well as its variations) around the ESS solutions, which reads

\be
\Delta_{1} \epsilon = -2 \int d_{3}\vec{r} \sum_{a}\vec{\nabla}\cdot\left[A^{0}_{a}\vec{\sigma}_{a}\right] +
2 \int d_{3}\vec{r} \sum_{a} A^{0}_{a}\vec{\nabla}\cdot\vec{\sigma}_{a}.
\label{eq:(6-65)}
\en
The first integral in this expression vanishes, owing to the boundary conditions. Using Eq.(\ref{eq:(6-61)}) the variation becomes

\be
\Delta_{1} \epsilon = -2 g\int d_{3}\vec{r} \sum_{abc} C_{abc} \frac{\partial \varphi}{\partial X_{0}} A^{0}_{a} \delta\vec{A}_{b}\cdot\vec{E}_{c}.
\label{eq:(6-66)}
\en
As expected this expression vanishes, owing to the parallelism of $A^{0}_{a}$ and $\vec{E}_{c}$ in the color space (see Eqs.(\ref{eq:(5-26bis)}) and (\ref{eq:(5-28)})) and the antisymmetry of the structure constants.

In obtaining the second variation of the energy functional we follow the same steps as in the abelian case. First we expand the generalized Gauss law to the second order. After the cancellation of the first-order terms we are lead to

\be
\vec{\nabla}\cdot\vec{\omega}_{a} = -g \sum_{bc} C_{abc} \left[\frac{\partial \varphi}{\partial X_{0}}
\delta\vec{A}_{b}\cdot\delta\vec{E}_{c} +
2\frac{\partial^{2} \varphi}{\partial X_{0}^{2}} \left(\sum_{p}\vec{E}_{p}\cdot\delta\vec{E}_{p}\right)
\delta\vec{A}_{b}\cdot\vec{E}_{c}\right],
\label{eq:(6-67)}
\en
where now

\bea
\vec{\omega}_{a} &=& 2 \frac{\partial^{2} \varphi}{\partial X_{0}^{2}} \left(\sum_{p}\vec{E}_{p}\cdot\delta\vec{E}_{p}\right) \delta\vec{E}_{a} + \frac{\partial^{2} \varphi}{\partial X_{0}^{2}}\sum_{p}\left(\delta\vec{E}_{p}^{2} - \delta\vec{H}_{p}^{2}\right)\vec{E}_{a} +
\nonumber\\
&+& 2 \frac{\partial^{3} \varphi}{\partial X_{0}^{3}}\left(\sum_{p} \vec{E}_{p}\cdot\delta\vec{E}_{p}\right)^{2} \vec{E}_{a} + 2\frac{\partial^{3} \varphi}{\partial X_{0}\partial Y_{0}^{2}}\left(\sum_{p} \vec{E}_{p}\cdot\delta\vec{H}_{p}\right)^{2} \vec{E}_{a} +
\nonumber\\
&+& 2 \frac{\partial^{2} \varphi}{\partial Y_{0}^{2}}\left(\sum_{p} \vec{E}_{p}\cdot\delta\vec{H}_{p}\right)\delta\vec{H}_{a}.
\label{eq:(6-68)}
\ena

By expanding the integral of (\ref{eq:(5-23)}) up to second order and using Eqs.(\ref{eq:(6-61)}) and (\ref{eq:(6-67)}) the second variation of the energy becomes

\bea\label{eq:(6-69)} \Delta_{2} \epsilon &=& \int d_{3}\vec{r}
\left[\frac{\partial \varphi}{\partial X_{0}} \sum_{a}
\delta\vec{E}_{a}^{2} + 2\frac{\partial^{2} \varphi}{\partial
X_{0}^{2}}(\sum_{a} \vec{E}_{a}\cdot\delta\vec{E}_{a})^{2}\right]
+ \nonumber \\ &+& \int d_{3}\vec{r} \left[\frac{\partial
\varphi}{\partial X_{0}}\sum_{a}\delta\vec{H}_{a}^{2} -
2\frac{\partial^{2} \varphi}{\partial
Y_{0}^{2}}(\sum_{a}\vec{E}_{a}\cdot\delta\vec{H}_{a})^{2}\right] -
\\ \nonumber &-&2\int d^3\vec{r}\vec{\nabla} \cdot(A_0^a\cdot
\vec{\omega}^a)-2 g\int d_{3}\vec{r} \frac{\partial
\varphi}{\partial X_{0}} \sum_{abc} C_{abc} A^{0}_{a}
\delta\vec{A}_{b}\cdot\delta\vec{E}_{c}. \ena Once again, the
divergence term in this expression vanishes owing to the boundary
conditions. The integrand of the last term is the scalar product
in color space between the potential $A^{0}_{a}$ of the
unperturbed field and the first component of the second-order
perturbation of the color-charge density in the r.h.s of
Eq.(\ref{eq:(6-67)}). For \textit{electromagnetic} perturbations
this component must satisfy the condition

\be
\frac{\partial \varphi}{\partial X_{0}} \sum_{abc} C_{abc} A^{0}_{a} \delta\vec{A}_{b}.\delta\vec{E}_{c} = 0,
\label{eq:(6-70)}
\en
(note that the remaining component in the r.h.s. of Eq.(\ref{eq:(6-67)}) lies already in the \textit{electromagnetic} direction). Consequently, the last term in Eq.(\ref{eq:(6-69)}) must vanish and the second variation of the energy takes a form similar to that of the abelian case. We can now determine the conditions for stability of the finite-energy ESS solutions in this non-abelian case through a similar argumentation. As easily seen stability requires the lagrangian-density function $\varphi(X,Y)$ to satisfy the
condition

\be
\frac{\partial \varphi}{\partial X} \geq 2X\frac{\partial^{2} \varphi}{\partial Y^{2}},
\label{eq:(6-71)}
\en
in all the range of values of the gauge invariants ($X,Y=0$) defined by the solution. This condition is formally the same as in the abelian case and is also \textit{necessary and sufficient} for the stability of the solitons against \textit{electromagnetic} perturbations. Obviously, it is a gauge-invariant criterion.

The analysis of the dynamical stability of non-abelian solitons should now be performed starting with Eqs.(\ref{eq:(6-61)}) and (\ref{eq:(6-63)}) and following similar steps as in the abelian case. But the presence of the antisymmetric structure constants and the symplectic character of the eigenvalue problem require new qualitative procedures and longer calculations which would lengthen excessively the contents of this work. This issue will be approached elsewhere.

\section{\large Conclusions and perspectives}

In this work we have solved the problem of characterizing a large
class of physically consistent relativistic lagrangian field
theories in three-space dimensions, supporting static spherically
symmetric non-topological soliton solutions. The fields concerned
were one and many-components scalar fields (whose lagrangian
densities depend on the kinetic term alone) and generalized gauge
fields of compact semi-simple Lie groups. This characterization is
exhaustive and leads to the classification of such models into six
types, according to the central and asymptotic behaviours of the
soliton fields. We have performed a broad analysis of the linear
stability of the solutions, obtaining necessary and sufficient
stability conditions which go beyond the usual Derrick criterion.
We also have carried out a general spectral analysis of the linear
perturbations around the soliton solutions, confirming their
dynamical stability and setting grounds for their quantum
extensions.

All these results allow the explicit determination of a large number of examples of such a class of lagrangians, providing a wide panoply of tools for the analysis of diverse physical problems, as those mentioned in the introduction and others. Among these problems let us outline three of particular interest, which we are addressing from the methods developed here.

1) As already discussed in section 3.1., the photon-photon
interaction mediated by the QED vacuum can be classically
described in terms of effective lagrangians which are polynomial
expressions in the gauge invariants $X$ and $Y$ and can be
obtained in a perturbative procedure \cite{Dobado97}, where the
lowest order is the well-known Euler-Heisenberg lagrangian
\cite{Heisenberg36}

\be \varphi(X,Y) \sim \eta X + \xi (4X^{2} + 7Y^{2}),
\label{eq:(7-1)} \en ($\eta$ and $\xi$ being positive constants).
The sequence of these lagrangians exhibits finite-energy ESS
solutions and suggests an interpretation in terms of the screening
effects of the vacuum on the field of point-like charges.
Unfortunately, as already mentioned, the perturbative expansion
involved in this procedure is a low-energy (or a low-intensity
field) approximation and is not accurate to describe the strong
fields present near the center of the ESS solutions. It is thus
necessary to explore this issue with other effective lagrangians,
obtained from the perturbative renormalization of the self-energy
of point-like fields. The analysis of this problem is in progress
\cite{dr08-1}, \cite{dr08-2}.

2) Scalar field models as that of the example treated in section
3.4., which belong to the case B-3, exhibiting short-ranged SSS
solutions (solitons or not), can be extended to generalized gauge
field models supporting similar ESS solutions. This behaviour,
which is related to the form of the lagrangian density around the
vacuum, may arise in effective actions for some fundamental
forces. If we assume the effective dynamics of the non-abelian
gauge fields in electroweak interactions to be described by this
kind of lagrangians, the short range of these forces could be
explained in terms of the non-linear self-couplings among these
fields, coming from the integration of some higher energy degrees
of freedom of a more fundamental theory. In this case the appeal
to any symmetry breaking or Higgs mechanism should become
superfluous. In our sense, this alternative deserves to be
thoroughly explored.

3) A new approach to the phenomenological description of the
hadronic structure can be envisaged, using the results of section
6 on the spectral analysis of the excitations of the (multi-)
scalar or generalized gauge-invariant solitons. In the
phenomenological descriptions based on the Skyrme model the hadron
arises as a topological soliton of a non-linear field theory.
Other models, which are believed to give an effective low-energy
approach to the non-perturbative regime of QCD (as the
Friedberg-Lee model \cite{fried77},\cite{l-p92} and related
theories), describe hadrons as confined states of quarks in
non-topological solitonic bags of non-linear phenomenological
fields. As an alternative to these viewpoints, we are considering
a generalized gauge-invariant lagrangian model for gauge fields
(``gluons") coupled to a quark-like fermionic sector and
implementing properly chosen symmetries. Such a phenomenological
model may be interpreted as an effective lagrangian for QCD or,
alternatively, as a field-theoretical low-energy limit of string
theory. The classical generalized gauge-field lagrangian can
exhibit soliton solutions in absence of other fields. If such
solutions are minima of the functional of energy of the full
action, their small perturbations will involve fermionic and
bosonic modes. The quantization of these modes leads to
``quasi-quarks" and ``quasi-gluons" as quantum excitations of the
soliton field. This quantum extension becomes a model for the
hadron containing these particles. In this picture the confinement
would be a consequence of the fact that quarks are quasi-particles
associated to these quantum excitations and (as the phonons in a
solid) they cannot exist outside the hadron.

Another domain where the results of this work could be useful
concerns the search for self-gravitating (scalar and gauge) field
configurations in General Relativity \cite{v-d-g}. The
classification of the lagrangian field theories considered here,
supporting non-topological solitons in flat space, can be extended
to the static spherically symmetric solutions of the Einstein
equations resulting from the coupling of these fields to
gravitation. Indeed, we have verified that these equations have
first-integrals which have the same form as (or can be closely
related to) the ones obtained from the corresponding field
theories in flat space (of the generic form of
Eqs.(\ref{eq:(2-4)}) or (\ref{eq:(5-10)})). This result opens the
possibility of generalizing to the gravitational case many of the
methods and results obtained here \cite{dr07-3}. We will continue
to address this topic in future work.

It would be also interesting to study the soliton solutions of
generalized non-abelian gauge field theories with other ansatzes
than the ESS one. In fact, such a kind of solutions have been
already found for the SU(2) non-abelian B-I theory within the
``monopole ansatz" \cite{g-k00}. This issue should be a theme for
a future investigation.

Let us conclude with some comments concerning an important
question which has not been considered here. It refers to the
analysis of propagation of wave-like solutions of these models. As
can be easily seen, all these theories exhibit plane wave
solutions propagating with the speed of light. But, owing to the
non-linear self-coupling, they also support other radiative
solutions propagating with more complex dispersion relations. In
most cases such waves evolve towards spatially-singular
configurations. Roughly speaking, the wave fronts travelling with
velocities which are dependent on the values of the fields at
every point tend to cumulate, generating discontinuities and
shocks after a critical time. Regularly evolving wave solutions of
a system of field equations are called \textit{exceptional}. If
all the wave solutions of a given system are exceptional, the
system is called \textit{completely exceptional} \cite{lax57}. A
detailed analysis of the problem of wave propagation for
generalized electromagnetic field models was performed by G.
Boillat \cite{boi70}, who established the complete exceptionality
of the Born-Infeld electrodynamics. Moreover, B-I is the only
\textit{admissible} generalized electromagnetic field theory (with
asymptotically coulombian elementary solutions) exhibiting this
property. Nevertheless, the Boillat analysis considers only models
which satisfy the condition $\frac{\partial \varphi}{\partial X}
(X=0,Y=0) = 1$ (case B-2) and, consequently, excludes the models
belonging to B-1 and B-3 cases. It would be interesting to extend
the Boillat analysis for these cases. However, when one considers
the extensions of B-I electrodynamics to the non-abelian case or
in the Kaluza-Klein context, this exceptionality character is lost
\cite{kern99}.\\

\textbf{\large Acknowledgments}\\

Many parts of this work have been discussed with several
colleagues. We are indebted to Drs. E. Alvarez, L. Bel, S.
Bonazzola, B. Carter, M.A. Cobas, B. Coll, R. Hakim, J. -L.
Jaramillo, J. Larena, T. Lehner, Y. Lozano, J. Madore, L. Mornas,
A. Nieto, M.A.R. Osorio, A.R. Plastino, J.A. Rodriguez-Mendez and
V. Vento for useful discussions and suggestions.

\end{document}